\documentclass[aps, prc, reprint,  amsmath, amssymb, superscriptaddress, floatfix, nofootinbib]{revtex4-1} 

\usepackage[pdftex]{graphicx}
\DeclareGraphicsExtensions{.jpg,.png,.pdf}

\usepackage[utf8]{inputenc}
\usepackage{hyperref}
\usepackage{amsmath}
\usepackage{amssymb}
\usepackage{amsfonts}
\usepackage{tabularx}
\usepackage{booktabs}
\usepackage{graphicx}
\usepackage{color}
\usepackage{multirow}
\usepackage[perpage,symbol]{footmisc}
\usepackage{verbatim}
\usepackage{dcolumn}
\usepackage{bm}
\usepackage[inline]{enumitem}
\definecolor{theblue}{RGB}{0,50,230}
\hypersetup{
  colorlinks=true,
  linkcolor=theblue,
  citecolor=theblue,
  urlcolor=theblue
}

\newcommand{\pt}{\ensuremath{p}_{\rm T}}
\newcommand{\raa}{\ensuremath{R}_{\rm AA}}
\newcommand{\vtwo}{\ensuremath{v}_{\rm 2}}
\newcommand{\vthree}{\ensuremath{v}_{\rm 3}}
\newcommand{\snn}{\sqrt{s_{\rm NN}}}

\usepackage{lineno}

\begin{document}

\title{Langevin dynamics of heavy quarks in a soft-hard factorized approach}

\author{Shuang~Li}
\email{lish@ctgu.edu.cn}
\affiliation{%
College of Science, China Three Gorges University, Yichang 443002, China
}%
\affiliation{%
Key Laboratory of Quark and Lepton Physics (MOE), Central China Normal University, Wuhan 430079, China
}%

\author{Fei Sun}
\email{sunfei@ctgu.edu.cn}
\affiliation{%
College of Science, China Three Gorges University, Yichang 443002, China
}%

\author{Wei Xie}
\email{xiewei@ctgu.edu.cn}
\affiliation{%
College of Science, China Three Gorges University, Yichang 443002, China
}%


\author{Wei Xiong}%
\email{xiongw@ctgu.edu.cn}
\affiliation{%
College of Science, China Three Gorges University, Yichang 443002, China
}%

\date{\today}

\begin{abstract}
By utilizing a soft-hard factorized model,
which combines a thermal perturbative description of soft scatterings
and a perturbative QCD-based calculation for hard collisions,
we study the energy and temperature dependence of the
heavy quark diffusion coefficients in Langevin dynamics.
The adjustable parameters are fixed from the comprehensive model-data comparison.
We find that a small value of the spatial diffusion coefficient at transition temperature is preferred by data $2\pi TD_{s}(T_{c}) \simeq 6$.
With the parameter-optimized model,
we are able to describe simultaneously the prompt $D^{0}$ $\raa$ and $\vtwo$ data at $\pt\le8$ GeV in Pb--Pb collisions at $\snn=2.76$ and $\snn=5.02$ TeV.
We also make predictions for non-prompt $D^{0}$ meson for future experimental tests down to the low momentum region.
\end{abstract}

\maketitle

\section{INTRODUCTION}\label{sec:Intro}
Ultrarelativistic heavy-ion collisions provide a unique opportunity
to create and investigate the properties of strongly
interacting matter in extreme conditions of temperature and
energy density, where the normal matter turns into a new form of
nuclear matter, consisting of deconfined quarks and gluons, namely quark-gluon plasma (QGP~\cite{2019pkr}).
Such collisions allow us to study the properties of the produced hot and dense partonic medium,
which are important for our understanding of the properties of the universe in the first few milliseconds
and the composition of the inner core of neutron stars~\cite{ShuryakPR80,2009zz,2015hba}.
Over the past two decades,
the measurements with heavy-ion collisions have been carried at the
Relativistic Heavy Ion Collider (RHIC) at BNL and the Large Hadron Collider (LHC)
at CERN~\cite{Gyulassy05,Shuryak05,Muller12}, to search and explore the fundamental properties of QGP,
notably its transport coefficients related to the medium interaction of hard probes.

Heavy quarks (HQs), including charm and bottom, provide a unique insight into
the microscopic properties of QGP~\cite{Zhou:2014kka,Tang:2014tga,Andronic:2015wma,GrecoPIPNP19,Dong:2019byy,Zhao:2020jqu}.
Due to large mass, they are mainly produced in initial hard scatterings
and then traverse the QGP and experience elastic and inelastic scatterings with its thermalized constituents~\cite{HQQGPRapp10,RalfSummary16}.
The transport properties of HQ inside QGP are encoded in the HQ transport coefficients,
which are expected to affect the distributions of the corresponding open heavy-flavor hadrons.
The resulting experimental observables like the nuclear modification factor $\raa$ and
elliptic anisotropy $\vtwo$ of various $D$ and $B$-mesons
are therefore sensitive to the HQ transport coefficients, in particular their energy and temperature dependence.
There now exist an extensive set of such measurements, which allow a data-based extraction of these coefficients.
In this work, we make such an attempt by using a soft-hard factorized model (see Sec.~\ref{sec:WeakApproach})
to calculate the diffusion and drag coefficients relevant for heavy quark Langevin dynamics.

A particularly important feature of the QGP transport coefficients is their momentum and temperature dependence,
especially how they change within the temperature region accessed by the RHIC and LHC experiments.
For instance the normalized jet transport coefficient $\hat{q}/T^3$
was predicted to present a rapidly increasing behavior
with decreasing temperature and develop a near-$T_c$ peak structure~\cite{JFLPRL09}.
The subsequent studies ~\cite{JFLCPL15,CUJET3JHEP16,JETCoef14,CUJET3CPC18,CUJET3Arxiv18,Ramamurti:2017zjn} seem to confirm this scenario.
Another important transport property, shear viscosity over entropy density ratio $\eta/s$,
also presents a visible $T$-dependence with a considerable increase above $T_c$~\cite{BassNaturePhy19}.
Concerning the HQ diffusion and drag coefficients,
there are also indications of nontrivial temperature dependence
from both the phenomenological extractions~\cite{CaoPRC15,LBTPRC16,XuPRC17,CTGUHybrid1,CTGUHybrid2,Prado:2019ste,Wang:2020ukj}
and theoretical calculations~\cite{HFModelHee13,PHSDPRC16,POWLANGEPJC11,BAMPS10,CUJET3CPC18,CUJET3Arxiv18}.
The difference among the derived hybrid models is mainly induced by
the treatment of the scale of QCD strong coupling, hadronization and non-perturbative effects~\cite{Gossiaux:2019mjc,Cao:2021ces}.
See Ref.~\cite{Rapp:2018qla,XuCoefficient18,Cao:2018ews} for the recent comparisons.

The paper is organized as follows.
In Sec.~\ref{sec:langevin} we introduce the general setup of the employed Langevin dynamics.
Section~\ref{sec:WeakApproach} is dedicated to the detailed calculation of the heavy quark transport coefficients with the factorization model.
In Sec.~\ref{sec:ParamOpt} we show systematic comparisons between modeling results and data
and optimize the model parameters based on global $\chi^2$ analysis.
With the parameter-optimized model, the energy and temperature dependence of the heavy quark transport coefficients
are presented in Sec.~\ref{sec:result}, as well as the comparisons with data and other theoretical calculations.
Section~\ref{sec:summary} contains the summary and discussion.

\section{Langevin dynamics}\label{sec:langevin}
The classicial Langevin Transport Equation (LTE) of a single HQ reads~\cite{HQQGPRapp10}
\begin{subequations}
\begin{align}
&dx^{i} = \frac{p^{i}}{E} dt \label{eq:LTE_X} \\
&dp^{i} = -\eta_{D}p^{i}dt + C^{ik}\rho^{k}\sqrt{dt} \label{eq:LTE_P}.
\end{align}
\end{subequations}
with $i,k=1,2,3$.
The first term on the right hand side of Eq.~\ref{eq:LTE_P} represents the deterministic drag force, $F^{i}_{drag}=-\eta_{D}p^{i}$,
which is given by the drag coefficient $\eta_{D}(E,T)$ with the HQ energy $E=\sqrt{\vec{p}^{\;2} + m^{2}_{Q}}$
and the underlying medium temperature $T$.
The second term denotes the stochastic thermal force, $F^{i}_{thermal}=C^{ik}\rho^{k}/\sqrt{dt}$,
which is described by the momentum argument of the covariance matrix $C^{ik}$,
\begin{equation}\label{eq:LTEtensor}
C^{ik} \equiv \sqrt{\kappa_{T}}(\delta^{ik}-\frac{p^{i}p^{k}}{\vec{p}^{\;2}}) + \sqrt{\kappa_{L}}\frac{p^{i}p^{k}}{\vec{p}^{\;2}},
\end{equation}
together with a Gaussian-normal distributed random variable $\rho^{k}$,
resulting in the uncorrelated random momentum kicks between two different time scales
\begin{equation}
\begin{aligned}\label{eq:ThermalForceCorre}
&<\vec{F}^{i}_{thermal}(t) \cdot \vec{F}^{j}_{thermal}(t^{\prime})>_{\rho} \\
&=C^{ik}C^{jk}\delta(t-t^{\prime}) \\
&\stackrel{(\ref{eq:LTEtensor})}{=}
\bigr[ \kappa_{T}(\delta^{ij}-\frac{p^{i}p^{j}}{\vec{p}^{\;2}}) + \kappa_{L}\frac{p^{i}p^{j}}{\vec{p}^{\;2}} \bigr] \delta(t-t^{\prime}).
\end{aligned}
\end{equation}
$\kappa_{T}$ and $\kappa_{L}$ are the transverse and longitudinal momentum diffusion coefficients, respectively,
which describe the momentum fluctuations in the direction
that perpendicular (i.e. transverse) and parallel (i.e. longitudinal) to the propagation.
Considering the Einstein relationship,
which enforces the drag coefficient starting from the momentum diffusion coefficients as~\cite{CTGUHybrid4}
\begin{equation}
\begin{aligned}\label{eq:EtaD_DissFluc}
\eta_{D} = &\frac{\kappa_{L}}{2TE} + (\xi-1)\frac{\partial \kappa_{L}}{\partial \vec{p}^{\;2}} + \frac{d-1}{2\vec{p}^{\;2}}  \bigr[ \xi(\sqrt{\kappa_{T}}+\sqrt{\kappa_{L}})^{2} \\
&- (3\xi-1)\kappa_{T} - (\xi+1)\kappa_{L} \bigr].
\end{aligned}
\end{equation}
The parameter $\xi$ denotes the discretization scheme of the stochastic integral,
which typically takes the values $\xi=0,0.5,1$,
representing the pre-point Ito, the mid-point Stratonovic, and the post-point discretization schemes, respectively;
$d=3$ indicates the spatial dimension.
In the framework of LTE the HQ-medium interactions are
conveniently encoded into three transport coefficients, i.e. $\eta_{D}$, $\kappa_{T}$ and $\kappa_{L}$ (Eq.~\ref{eq:EtaD_DissFluc}).
All the problems are therefore reduced to the evaluation of $\kappa_{T/L}(E,T)$,
which will be mainly discussed in this work.

Finally, we introduce a few detailed setups of the numerical implementation.
The space-time evolution of the temperature field and the velocity field
are needed to solve LTE (Eq.~\ref{eq:LTE_X} and \ref{eq:LTE_P}).
Following our previous analysis~\cite{CTGUHybrid1, CTGUHybrid3}, they are obtained in a 3+1 dimensional viscous hydrodynamic calculation~\cite{vhlle},
with the local thermalization started at $\tau_{0}=0.6~{\rm fm}/{\it c}$, the shear viscosity ${\eta/s=1/(4\pi)}$
and the critical temperature $T_{c}=165~{\rm MeV}$ (see details in Ref.~\cite{vhlle}).
When the medium temperature drops below $T_{c}$,
heavy quark will hadronize into the heavy-flavor hadrons via a fragmentation-coalescence approach.
The Braaten-like fragmentation functions are employed for both charm and bottom quarks~\cite{FragBraaten93,FragFONLLPRL}.
An instantaneous approach is utilized to characterize the coalescence process for
the formation of heavy-flavor mesons from the heavy and (anti-)light quark pairs.
The relevant coalescence probability is quantified by the
overlap integral of the Winger functions for the meson and partons,
which are defined through a harmonic oscillator and the Gaussian wave-function~\cite{CoalOriginalDover91}, respectively.
See Ref.~\cite{CTGUHybrid4} for more details.

\section{Momentum diffusion coefficients in a soft-hard factorized approach}\label{sec:WeakApproach}
When propagating throughout the QGP, the HQ scattering off the gluons and (anti-)partons of the thermal deconfined medium,
can be characterized as the two-body elementary processes,
\begin{equation}\label{eq:Process}
Q~(p_{1})+i~(p_{2})\rightarrow Q~(p_{3})+i~(p_{4}),
\end{equation}
with $p_{1}=(E_{1},\vec{p}_{1})$ and $p_{2}$ are the four-momentum
of the injected HQ ($Q$) and the incident medium partons $i=q, g$, respectively,
while $p_{3}$ and $p_{4}$ are for the ones after scattering.
Note that the medium partons are massless ($m_{2}=m_{4}\sim0$) in particular comparing with the massive HQ ($m_{1}=m_{3}=m_{Q}$ in a few times $\rm GeV$).
The corresponding four-momentum transfer is $(\omega,~\vec{q}^{\;}) = (\omega,~\vec{q}_{T},~q_{L})$.
The Mandelstam invariants read
\begin{equation}
\begin{aligned}\label{eq:MandVar}
&s \equiv (p_{1}+p_{2})^{2} \\
&t \equiv (p_{1}-p_{3})^{2} = \omega^{2} - q^{2} \\
&u \equiv (p_{1}-p_{4})^{2}
\end{aligned}
\end{equation}
with $q\equiv |\vec{q}\;| \ll E_{1}$ for small momentum exchange.
The transverse and longitudinal momentum diffusion coefficients can be determined by
weighting the differential interaction rate with the squared transverse and longitudinal momentum transfer, respectively.
It yields
\begin{equation}
\begin{aligned}\label{eq:KappaT}
\kappa_{T} &=\frac{1}{2}\int d{\Gamma} \; \vec{q}_{T}^{\;2}
=\frac{1}{2}\int d{\Gamma} \bigr[ \omega^{2} - t - \bigr( \frac{2\omega E_{1}-t}{2|\vec{p}_{1}|} \bigr)^{2} \bigr]
\end{aligned}
\end{equation}
and
\begin{equation}                                                                                                                                                             
\begin{aligned}\label{eq:KappaL}
\kappa_{L} &=\int d{\Gamma} \; q_{L}^{2}
=\frac{1}{4\vec{p}_{1}^{\;2}} \int d{\Gamma} \; (2\omega E_{1}-t)^{2}.
\end{aligned}
\end{equation}

As the momentum transfer vanishes ($|t|\rightarrow0$),
the gluon propagator in the $t$-channel of the elastic process
causes an infrared divergence in the
squared amplitude $\overline{|\mathcal{M}^{2}}|_{t-channel}\propto 1/t^{2}$,
which is usually regulated by a Debye screening mass, i.e. $t\rightarrow t-\lambda m^{2}_{D}$
with an adjustable parameter $\lambda$~\cite{Benjamin88,PBGPRC08}.
Alternatively, it can be overcome by utilizing a soft-hard factorized approach~\cite{HQSteph08QED,HQSteph08QCD},
which starts with the assumptions that the medium is thermal and weakly coupled,
and then the interactions between the heavy quarks and the medium can be computed in thermal perturbation theory.
Finally, this approach allows to decompose the soft HQ-medium interactions with $t>t^{\ast}$, from the hard ones with $t<t^{\ast}$.
For soft collisions the gluon propagator should be replaced by the hard-thermal loop (HTL) propagator~\cite{Braaten91PRL,JeanPR02},
while for hard collisions the hard gluon exchange is considered and the Born approximation is appropriate.
Therefore, the final results of $\kappa_{T/L}$ include the contributions from both soft and hard components.

As discussed in the QED case~\cite{HQSteph08QED}, $\mu+\gamma\rightarrow \mu+\gamma$,
the complete calculation for the energy loss of the energic incident heavy-fermion
is independent of the intermediate scale $t^{\ast}$ in high energy limit.
However, in the QCD case, there is the complication that the challenge of the validity of the HTL scenario,
$m_{D}^{2}\ll T^{2}$~\cite{HQSteph08QCD}, due to the temperatures reached at RHIC and LHC energies.
Consequently, in the QCD case, the soft-hard approach is in fact
not independent of the intermediate scale $t^{\ast}$~\cite{PBGPRC08,POWLANGEPJC11}.
In this analysis we have checked that the calculations for $\kappa_{T/L}$
are not sensitive to the choice of the artificial cutoff $t^{\ast}\sim m_{D}^{2}$.

In the next parts of this section, we will focus on the energy and temperature dependence
of the interaction rate ${\Gamma}$ at leading order in $g$ for the elastic process,
as well as the momentum diffusion coefficients (Eq.~\ref{eq:KappaT} and~\ref{eq:KappaL})
in soft and hard collisions, respectively.

\subsection{$\kappa_{T/L}$ in soft region $t^{\ast}<t<0$}\label{subsec:soft}
In soft collisions the exchanged four-momentum is soft, $\sqrt{-t}\sim gT$ ($\lambda_{mfp}\sim 1/g^{2}T$~\cite{JetMediumJHEP16}),
and the $t$-channel long-wavelength gluons are screened by the mediums,
thus, they feel the presence of the medium and require the resummation.
Here we just show the final results,
and the details are relegated to~\ref{app:appendixA}.
The transverse and longitudinal momentum diffusion coefficients can be expressed as
\begin{equation}
\begin{aligned}\label{eq:KappaT_Soft}
\kappa_{T}(E_{1},T)=&\frac{C_{F}g^{2}}{16\pi^{2}v_{1}^{3}} \int^{0}_{t^{\ast}} dt \; (-t)^{3/2}
\int_{0}^{v_{1}}dx \frac{v_{1}^{2}-x^{2}}{(1-x^{2})^{5/2}} \\
&\bigr[ \rho_{L}(t,x) + (v_{1}^{2}-x^{2})\rho_{T}(t,x) \bigr]
coth\bigr(\frac{x}{2T}\sqrt{\frac{-t}{1-x^{2}}} \bigr)
\end{aligned}
\end{equation}
and
\begin{equation}
\begin{aligned}\label{eq:KappaL_Soft}
\kappa_{L}(E_{1},T)=&\frac{C_{F}g^{2}}{8\pi^{2}v_{1}^{3}} \int^{0}_{t^{\ast}} dt \; (-t)^{3/2}
\int_{0}^{v_{1}}dx \frac{x^{2}}{(1-x^{2})^{5/2}} \\
&\bigr[ \rho_{L}(t,x) + (v_{1}^{2}-x^{2})\rho_{T}(t,x) \bigr]
coth\bigr(\frac{x}{2T}\sqrt{\frac{-t}{1-x^{2}}} \bigr),
\end{aligned}
\end{equation}
respectively, with the HQ velocity $v_{1}=|\vec{p}|/E_{1}$
and the transverse and longitudinal parts of the HTL gluon spectral functions\footnote[4]{
The spectral function involves only the low frequency excitations, namely the Landau cut,
while the quasiparticle excitations is irrelevant in this regime.
See Eq.\ref{eq:GammaRhoTL1} for more details.}
are given by
\begin{equation}
\begin{aligned}\label{eq:GammaRhoT_Soft}
\rho_{T}(t,x)=&\frac{\pi m_{D}^{2}}{2} x(1-x^{2}) \biggr\{
\bigr[-t+\frac{m_{D}^{2}}{2} x^{2}(1+\frac{1-x^{2}}{2x}ln\frac{1+x}{1-x}) \bigr]^{2} \\
&+ \bigr[ \frac{\pi m_{D}^{2}}{4}x(1-x^{2}) \bigr]^{2} \biggr\}^{-1}
\end{aligned}
\end{equation}
and
\begin{equation}
\begin{aligned}\label{eq:GammaRhoL_Soft}
\rho_{L}(t,x)=&\pi m_{D}^{2} x \biggr\{
\bigr[ \frac{-t}{1-x^{2}}+m_{D}^{2}(1-\frac{x}{2}ln\frac{1+x}{1-x}) \bigr]^{2} \\
&+ \bigr( \frac{\pi m_{D}^{2}}{2} x \bigr)^{2} \biggr\}^{-1}.
\end{aligned}
\end{equation}

\subsection{$\kappa_{T/L}$ in hard region $t_{min}<t<t^{\ast}$}\label{subsec:hard}
In hard collisions the exchanged four-momentum is hard, $\sqrt{-t}\gtrsim T$ ($\lambda_{mfp}\sim 1/g^{4}T$~\cite{JetMediumJHEP16}),
and the pQCD Born approximation is valid in this regime.
In analogy with the previous part we give the $\kappa_{T/L}$ results directly,
and the detailed aspects of the calculations can be found in~\ref{app:appendixB}.
The momentum diffusion coefficients reads
\begin{equation}
\begin{aligned}\label{eq:KappaT_Hard}
\kappa^{Qi}_{T}(E_{1},T) =& \frac{1}{256\pi^{3}|\vec{p}_{1}|E_{1}} \int_{|\vec{p}_{2}|_{min}}^{\infty}d|\vec{p}_{2}| E_{2} n_{2}(E_{2}) \\
&\int_{-1}^{cos\psi|_{max}} d(cos\psi) \int_{t_{min}}^{t^{\ast}}dt
\frac{1}{a} \bigr[ -\frac{m_{1}^{2}(D+2b^{2})}{8\vec{p}_{1}^{\;2}a^{4}} \\
&+ \frac{E_{1}tb}{2\vec{p}_{1}^{\;2}a^{2}} - t(1+\frac{t}{4\vec{p}_{1}^{\;2}}) \bigr] \; \overline{|\mathcal{M}^{2}}|^{Qi}
\end{aligned}
\end{equation}
and
\begin{equation}
\begin{aligned}\label{eq:KappaL_Hard}
\kappa^{Qi}_{L}(E_{1},T) =& \frac{1}{256\pi^{3}|\vec{p}_{1}|^{3}E_{1}} \int_{|\vec{p}_{2}|_{min}}^{\infty}d|\vec{p}_{2}| E_{2} n_{2}(E_{2}) \\
&\int_{-1}^{cos\psi|_{max}} d(cos\psi) \int_{t_{min}}^{t^{\ast}} dt
\frac{1}{a} \bigr[ \frac{E_{1}^{2}(D+2b^{2})}{4a^{4}} \\
&-\frac{E_{1}tb}{a^{2}} + \frac{t^{2}}{2} \bigr] \; \overline{|\mathcal{M}^{2}}|^{Qi}.
\end{aligned}
\end{equation}
The integrations limits and the short notations are shown in Eq.~\ref{eq:P2Min}-\ref{eq:cVal}.

\subsection{Complete results in soft-hard scenario}\label{subsec:all}
Combining the soft and hard contributions to the momentum diffusion coefficients via
\begin{equation}
\begin{aligned}\label{eq:KappaAll}
\kappa_{T/L}(E,T)&=\kappa_{T/L}^{soft}(E,T) + \kappa_{T/L}^{hard}(E,T) \\
&=\kappa_{T/L}^{soft}(E,T) + \sum_{i=q,g} \kappa_{T/L}^{hard-Qi}
\end{aligned}
\end{equation}
while $\kappa_{T/L}^{soft}(E\equiv E_{1},T)$ is given by Eq.~\ref{eq:KappaT_Soft} and \ref{eq:KappaL_Soft},
and $\kappa_{T/L}^{hard-Qi}$ is expressed in Eq.~\ref{eq:KappaT_Hard} and \ref{eq:KappaL_Hard} for a given incident medium parton $i=q,g$.
Adopting the post-point discretization scheme of the stochastic integral, i.e. $\xi=1$ in Eq.~\ref{eq:EtaD_DissFluc},
the drag coefficient $\eta_{D}(E,T)$ can be obtained by inserting Eq.~\ref{eq:KappaAll} into Eq.~\ref{eq:EtaD_DissFluc}.

\section{Data-based parameter optimization}\label{sec:ParamOpt}
Following the strategies utilized in our previous work~\cite{CTGUHybrid4, CTGUHybrid5},
the two key parameters in this study, the intermediate cutoff $t^{\ast}$
and the scale $\mu$ of running coupling (Eq.~\ref{eq:TwoLoopG}),
are tested within a wide range of possibility and drawn constrains
by comparing the relevant charm meson data with model results.
We calculate the corresponding final observable {\bf y} for the desired species of D-meson.
Then, a $\chi^{2}$ analysis can be performed by comparing the model predictions with experimental data
\begin{equation}
\begin{aligned}\label{eq:ChiSqu}
&\chi^{2}=\sum_{i=1}^{N} \biggr( \frac{\bf y^{\rm Data}_{i}-y^{\rm Model}_{i}}{\sigma_{i}} \biggr)^{2}.
\end{aligned}
\end{equation}
In the above $\sigma_i$ is the total uncertainty in data points,
including the statistic and systematic components which are added in quadrature.
$n=N-1$ denotes the degree of freedom ($d.o.f$) when there are $N$ data points used in the comparison.
In this study, we use an extensive set of LHC data in the range $\pt\le8~\rm{GeV}$:
$D^{0}$, $D^{+}$, $D^{\ast +}$ and $D_{s}^{+}$ $\raa$ data collected at mid-rapidity ($|y|<0.5$) in the most central ($0-10\%$)
and semi-central ($30-50\%$) Pb--Pb collisions at $\snn=2.76~{\rm TeV}$~\cite{ALICEDesonPbPb2760RAA,ALICEDsPbPb2760RAA}
and $\snn=5.02~{\rm TeV}$~\cite{ALICEDesonPbPb5020RAA},
as well as the $\vtwo$ data in semi-central ($30-50\%$) collisions~\cite{ALICEDesonPbPb2760V2,ALICEDesonPbPb5020V2,CMSd05020V2}.

\begin{figure}[!htbp]
\centering
\setlength{\abovecaptionskip}{-0.1mm}
\includegraphics[width=.48\textwidth]{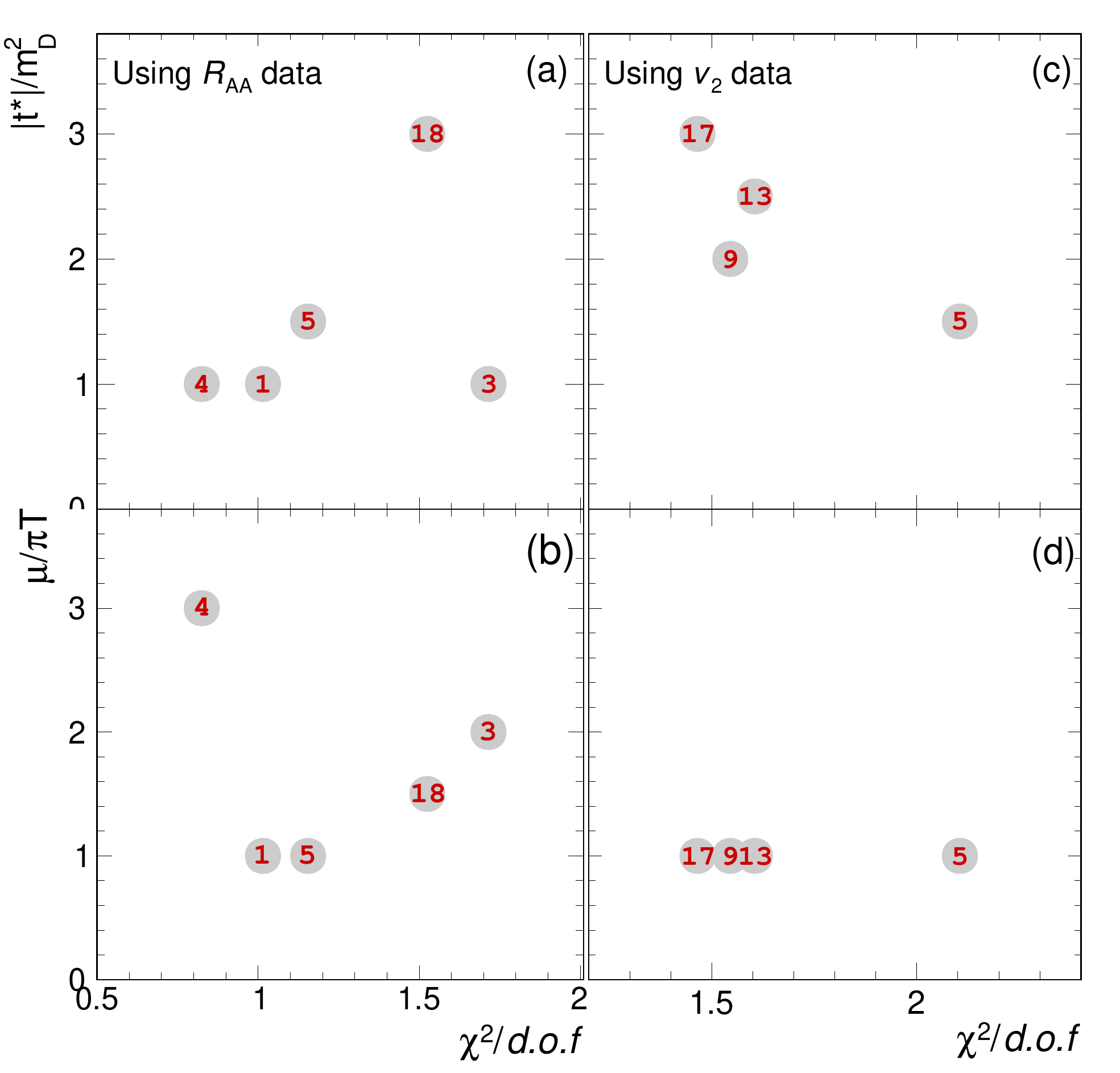}
\caption{Comparision of $\chi^{2}/d.o.f$ based on the experimental data of $\raa$ (left) and $\vtwo$ (right).
The model predictions are calculated by using various combinations of parameter $|t^{\ast}|/m_{D}^{2}$ (upper) and $\mu/\pi T$ (lower),
which are represented as $\chi^{2}/d.o.f$ (x-axis) and the desired parameter (y-axis), respectively.
See the legend and text for details.}
\label{fig:Param_RAAV2}
\end{figure}

We scan a wide range of valus for ($t^{\ast}$, $\mu$): $1\le\frac{|t^{\ast}|}{m_{D}^{2}}\le3$ and $1\le\frac{\mu}{\pi T}\le3$.
A total of 20 different combinations were computed and compared with the experimental data.
The obtained results are summarized in Tab.~\ref{tab:OptimizedParam}.
The $\chi^2$ values are computed separately for $\raa$ and $\vtwo$ as well as for all data combined.
To better visualize the results, we also show them in Fig.~\ref{fig:Param_RAAV2},
with left panels for $\raa$ analysis and right panels for  $v_{2}$ analysis.
In both panels, the y-axis labels the desired parameters, $\frac{|t^{\ast}|}{m_{D}^{2}}$ (upper) and $\frac{\mu}{\pi T}$ (lower),
and x-axis labels the $\chi^2/d.o.f$
within the selected ranges\footnote[5]{$\chi^2/d.o.f$ is shown in the range $0.5<\chi^2/d.o.f<2.5$ for better visualization.}.
The different points (filled gry circles) represent the different combinations of parameters ($\frac{|t^{\ast}|}{m_{D}^{2}}$, $\frac{\mu}{\pi T}$)
in Tab.~\ref{tab:OptimizedParam}, with the number on top of each point to display the relevant ``$Model~ID$" for that model.
A number of observations can be drawn from the comprehensive model-data comparison.
For the $\raa$, several models achieve $\chi^2/d.o.f \simeq 1$ with $\frac{|t^{\ast}|}{m_{D}^{2}}\lesssim1.5$
and widespread values of $\mu$: $1\le\frac{\mu}{\pi T}\le3$.
This suggests that $\raa$ appears to be more sensitive to the intermediate cutoff
while insensitive to the scale of coupling constant.
For the $\vtwo$, it clearly shows a stronger sensitivity to $\mu$,
which seems to give a better description ($\chi^2/d.o.f \simeq 1.5-2.0$) of the data with $\frac{\mu}{\pi T}=1$.
It is interesting to see that $\raa$ data is more powerful to constrain $\frac{|t^{\ast}|}{m_{D}^{2}}$,
while $\vtwo$ data is more efficient to nail down $\frac{\mu}{\pi T}$.
Taken all together, we can identify a particular model that
outperforms others in describing both $\raa$ and $\vtwo$ data simultaneously with $\chi^2/d.o.f = 1.3$.
This one will be the parameter-optimized model in this work: $|t^{\ast}|=1.5m_{D}^{2}$ and $\mu=\pi T$.

\begin{table}[!htbp]
\centering
\begin{tabular}{c|c|c|c|c|c}
\hline
\multicolumn{1}{c}{\multirow{2}{*}{$\it Model~ID$}}
 & \multicolumn{1}{c}{\multirow{1}{*}{\centering $|t^{\ast}|/m_{D}^{2}$}}
 & \multicolumn{1}{c}{\multirow{1}{*}{\centering $\mu/\pi T$}}
 & \multicolumn{1}{c}{\multirow{1}{*}{\centering $\chi^{2}/d.o.f$}}
 & \multicolumn{1}{c}{\multirow{1}{*}{\centering $\chi^{2}/d.o.f$}}
 & \multicolumn{1}{c}{\multirow{2}{*}{\centering Total}}
  \\
\multicolumn{1}{c}{\centering }
 & \multicolumn{1}{c}{\centering (Cutoff)}
 & \multicolumn{1}{c}{\centering (Scale)}
 & \multicolumn{1}{c}{\centering ($\raa$)}
 & \multicolumn{1}{c}{\centering ($\vtwo$)}
 & \multicolumn{1}{c}{\centering }
 \\
\cline{1-6}
\multicolumn{1}{c}{\centering 1}
 & \multicolumn{1}{c}{\centering 1.00}
 & \multicolumn{1}{c}{\centering 1.00}
 & \multicolumn{1}{c}{\centering 1.02}
 & \multicolumn{1}{c}{\centering 5.08}
 & \multicolumn{1}{c}{\centering 1.61}
 \\
\multicolumn{1}{c}{\centering 2}
 & \multicolumn{1}{c}{\centering 1.00}
 & \multicolumn{1}{c}{\centering 1.50}
 & \multicolumn{1}{c}{\centering 2.46}
 & \multicolumn{1}{c}{\centering 6.08}
 & \multicolumn{1}{c}{\centering 2.98}
 \\
\multicolumn{1}{c}{\centering 3}
 & \multicolumn{1}{c}{\centering 1.00}
 & \multicolumn{1}{c}{\centering 2.00}
 & \multicolumn{1}{c}{\centering 1.71}
 & \multicolumn{1}{c}{\centering 6.11}
 & \multicolumn{1}{c}{\centering 2.35}
 \\
\multicolumn{1}{c}{\centering 4}
 & \multicolumn{1}{c}{\centering 1.00}
 & \multicolumn{1}{c}{\centering 3.00}
 & \multicolumn{1}{c}{\centering 0.83}
 & \multicolumn{1}{c}{\centering 4.20}
 & \multicolumn{1}{c}{\centering 1.32}
 \\
\multicolumn{1}{c}{\centering \textbf{5}}
 & \multicolumn{1}{c}{\centering \textbf{1.50}}
 & \multicolumn{1}{c}{\centering \textbf{1.00}}
 & \multicolumn{1}{c}{\centering \textbf{1.16}}
 & \multicolumn{1}{c}{\centering \textbf{2.10}}
 & \multicolumn{1}{c}{\centering \textbf{1.30}}
 \\
\multicolumn{1}{c}{\centering 6}
 & \multicolumn{1}{c}{\centering 1.50}
 & \multicolumn{1}{c}{\centering 1.50}
 & \multicolumn{1}{c}{\centering 3.90}
 & \multicolumn{1}{c}{\centering 9.82}
 & \multicolumn{1}{c}{\centering 4.76}
 \\
\multicolumn{1}{c}{\centering 7}
 & \multicolumn{1}{c}{\centering 1.50}
 & \multicolumn{1}{c}{\centering 2.00}
 & \multicolumn{1}{c}{\centering 5.38}
 & \multicolumn{1}{c}{\centering 10.90}
 & \multicolumn{1}{c}{\centering 6.18}
 \\
\multicolumn{1}{c}{\centering 8}
 & \multicolumn{1}{c}{\centering 1.50}
 & \multicolumn{1}{c}{\centering 3.00}
 & \multicolumn{1}{c}{\centering 4.85}
 & \multicolumn{1}{c}{\centering 9.12}
 & \multicolumn{1}{c}{\centering 5.47}
 \\
\multicolumn{1}{c}{\centering 9}
 & \multicolumn{1}{c}{\centering 2.00}
 & \multicolumn{1}{c}{\centering 1.00}
 & \multicolumn{1}{c}{\centering 2.44}
 & \multicolumn{1}{c}{\centering 1.54}
 & \multicolumn{1}{c}{\centering 2.31}
 \\
\multicolumn{1}{c}{\centering 10}
 & \multicolumn{1}{c}{\centering 2.00}
 & \multicolumn{1}{c}{\centering 1.50}
 & \multicolumn{1}{c}{\centering 2.99}
 & \multicolumn{1}{c}{\centering 7.11}
 & \multicolumn{1}{c}{\centering 3.58}
 \\
\multicolumn{1}{c}{\centering 11}
 & \multicolumn{1}{c}{\centering 2.00}
 & \multicolumn{1}{c}{\centering 2.00}
 & \multicolumn{1}{c}{\centering 6.81}
 & \multicolumn{1}{c}{\centering 10.19}
 & \multicolumn{1}{c}{\centering 7.30}
 \\
\multicolumn{1}{c}{\centering 12}
 & \multicolumn{1}{c}{\centering 2.00}
 & \multicolumn{1}{c}{\centering 3.00}
 & \multicolumn{1}{c}{\centering 8.91}
 & \multicolumn{1}{c}{\centering 8.19}
 & \multicolumn{1}{c}{\centering 8.80}
 \\
\multicolumn{1}{c}{\centering 13}
 & \multicolumn{1}{c}{\centering 2.50}
 & \multicolumn{1}{c}{\centering 1.00}
 & \multicolumn{1}{c}{\centering 4.04}
 & \multicolumn{1}{c}{\centering 1.61}
 & \multicolumn{1}{c}{\centering 3.68}
\\
\multicolumn{1}{c}{\centering 14}
 & \multicolumn{1}{c}{\centering 2.50}
 & \multicolumn{1}{c}{\centering 1.50}
 & \multicolumn{1}{c}{\centering 2.08}
 & \multicolumn{1}{c}{\centering 6.78}
 & \multicolumn{1}{c}{\centering 2.77}
 \\
\multicolumn{1}{c}{\centering 15}
 & \multicolumn{1}{c}{\centering 2.50}
 & \multicolumn{1}{c}{\centering 2.00}
 & \multicolumn{1}{c}{\centering 6.59}
 & \multicolumn{1}{c}{\centering 8.51}
 & \multicolumn{1}{c}{\centering 6.87}
 \\
\multicolumn{1}{c}{\centering 16}
 & \multicolumn{1}{c}{\centering 2.50}
 & \multicolumn{1}{c}{\centering 3.00}
 & \multicolumn{1}{c}{\centering 11.80}
 & \multicolumn{1}{c}{\centering 12.04}
 & \multicolumn{1}{c}{\centering 11.83}
 \\
\multicolumn{1}{c}{\centering 17}
 & \multicolumn{1}{c}{\centering 3.00}
 & \multicolumn{1}{c}{\centering 1.00}
 & \multicolumn{1}{c}{\centering 5.46}
 & \multicolumn{1}{c}{\centering 1.46}
 & \multicolumn{1}{c}{\centering 4.88}
 \\
\multicolumn{1}{c}{\centering 18}
 & \multicolumn{1}{c}{\centering 3.00}
 & \multicolumn{1}{c}{\centering 1.50}
 & \multicolumn{1}{c}{\centering 1.52}
 & \multicolumn{1}{c}{\centering 8.05}
 & \multicolumn{1}{c}{\centering 2.47}
 \\
\multicolumn{1}{c}{\centering 19}
 & \multicolumn{1}{c}{\centering 3.00}
 & \multicolumn{1}{c}{\centering 2.00}
 & \multicolumn{1}{c}{\centering 6.27}
 & \multicolumn{1}{c}{\centering 9.88}
 & \multicolumn{1}{c}{\centering 6.79}
 \\
\multicolumn{1}{c}{\centering 20}
 & \multicolumn{1}{c}{\centering 3.00}
 & \multicolumn{1}{c}{\centering 3.00}
 & \multicolumn{1}{c}{\centering 12.94}
 & \multicolumn{1}{c}{\centering 15.21}
 & \multicolumn{1}{c}{\centering 13.27}
 \\
\hline
\end{tabular}
\caption{Summary of the adjustable parameters in this work,
together with the relevant $\chi^{2}/d.o.f$ obtained for $\raa$ and $\vtwo$.}
\label{tab:OptimizedParam}
\end{table}

\section{Results}\label{sec:result}
In this section we will first examine the $t^{\ast}$ dependence of
$\kappa_{T/L}$ (Eq.~\ref{eq:KappaAll}) for both charm and bottom quark.
Then, the relevant energy and temperature dependence of $\kappa_{T/L}(E,T)$
will be discussed with the optimized parameters.
For the desired observables we will perform the comparisons with the results
from lattice QCD at zero momentum limit,
as well as the ones from experimental data in the low to intermediate $\pt$ region.

\subsection{Energy and temperature dependence of the transport coefficients}\label{subsec:DiffCoef_ETDep}
\begin{figure*}[!htbp]
\centering
\setlength{\abovecaptionskip}{-0.1mm}
\includegraphics[width=.35\textwidth]{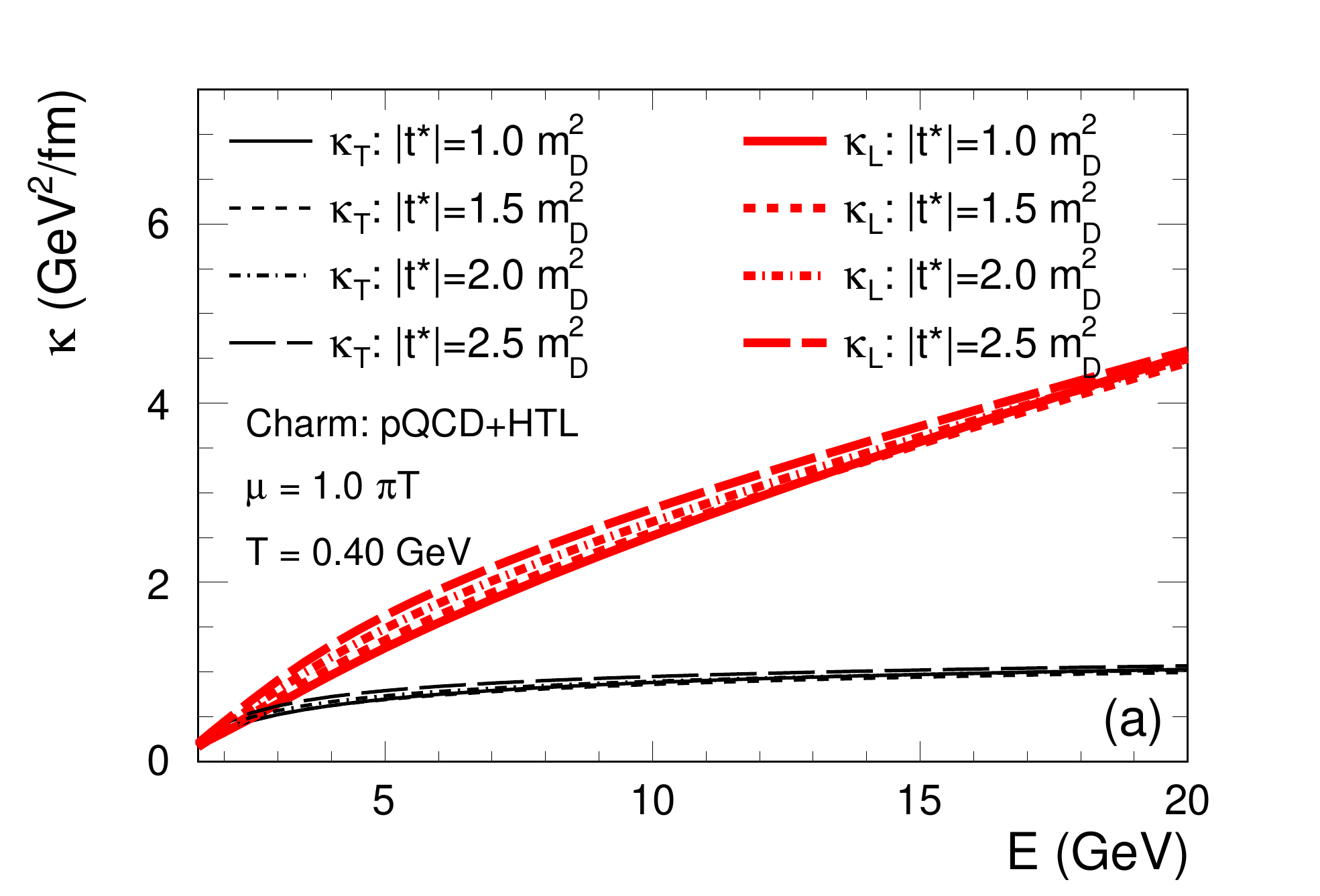}
\includegraphics[width=.35\textwidth]{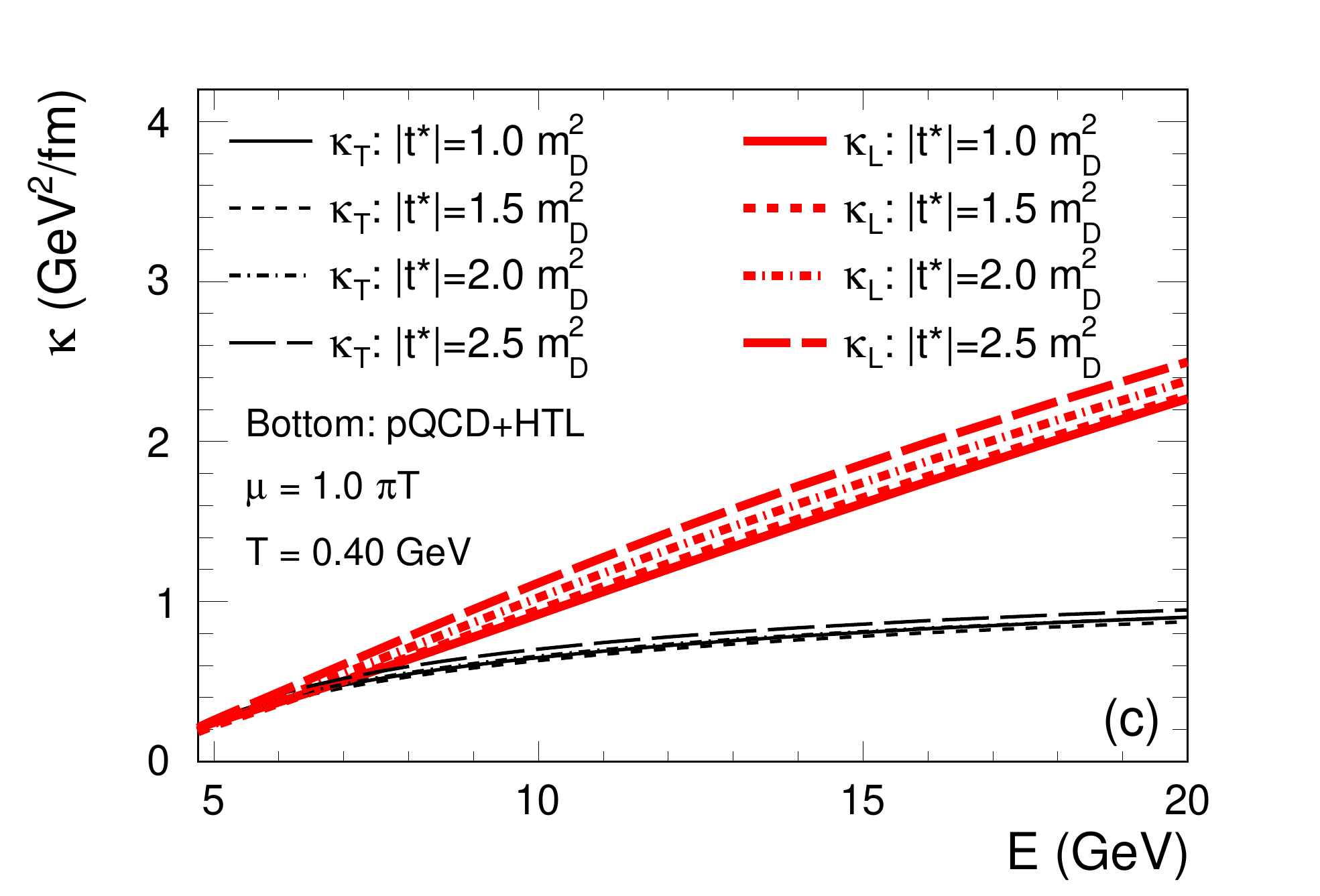}
\includegraphics[width=.35\textwidth]{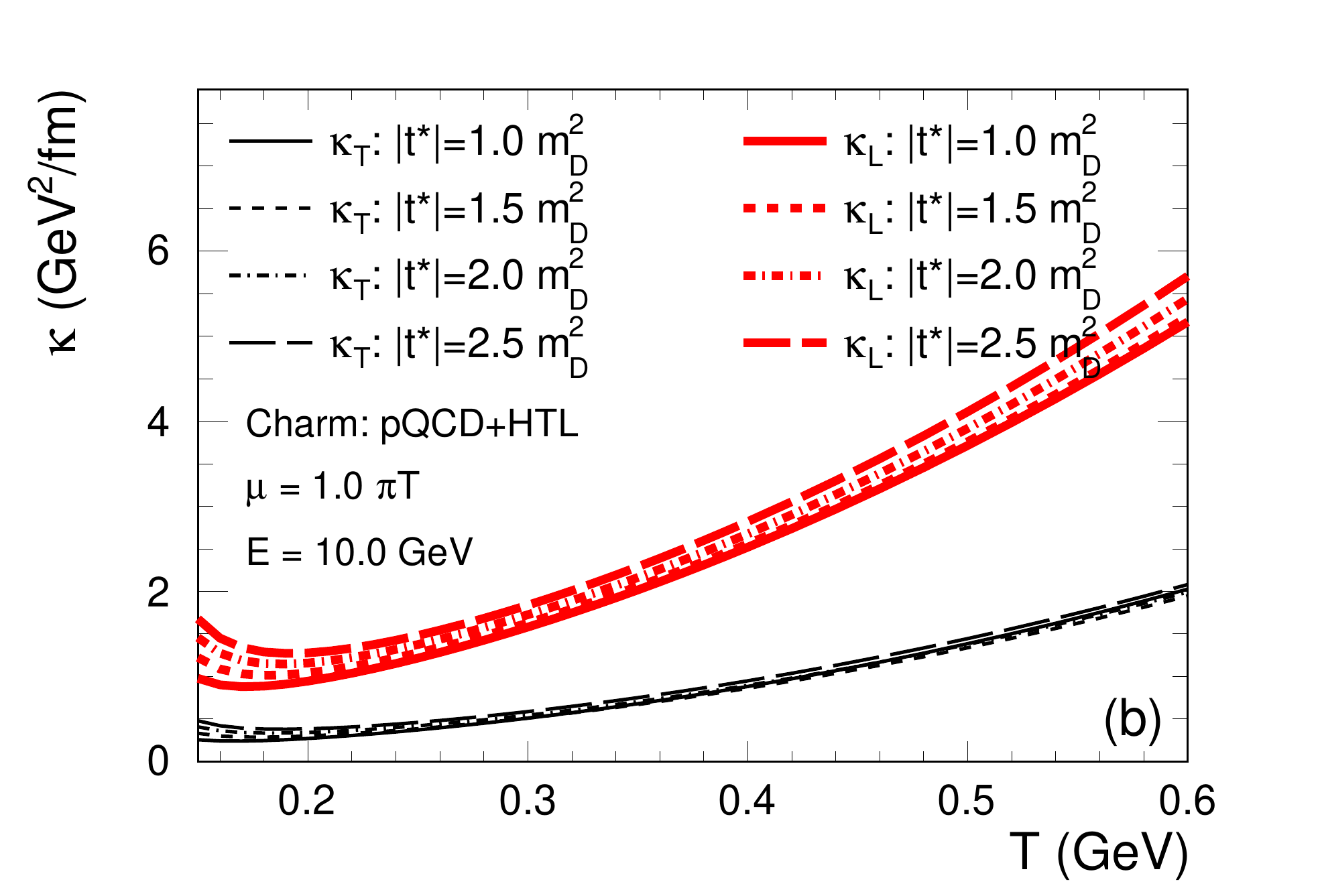}
\includegraphics[width=.35\textwidth]{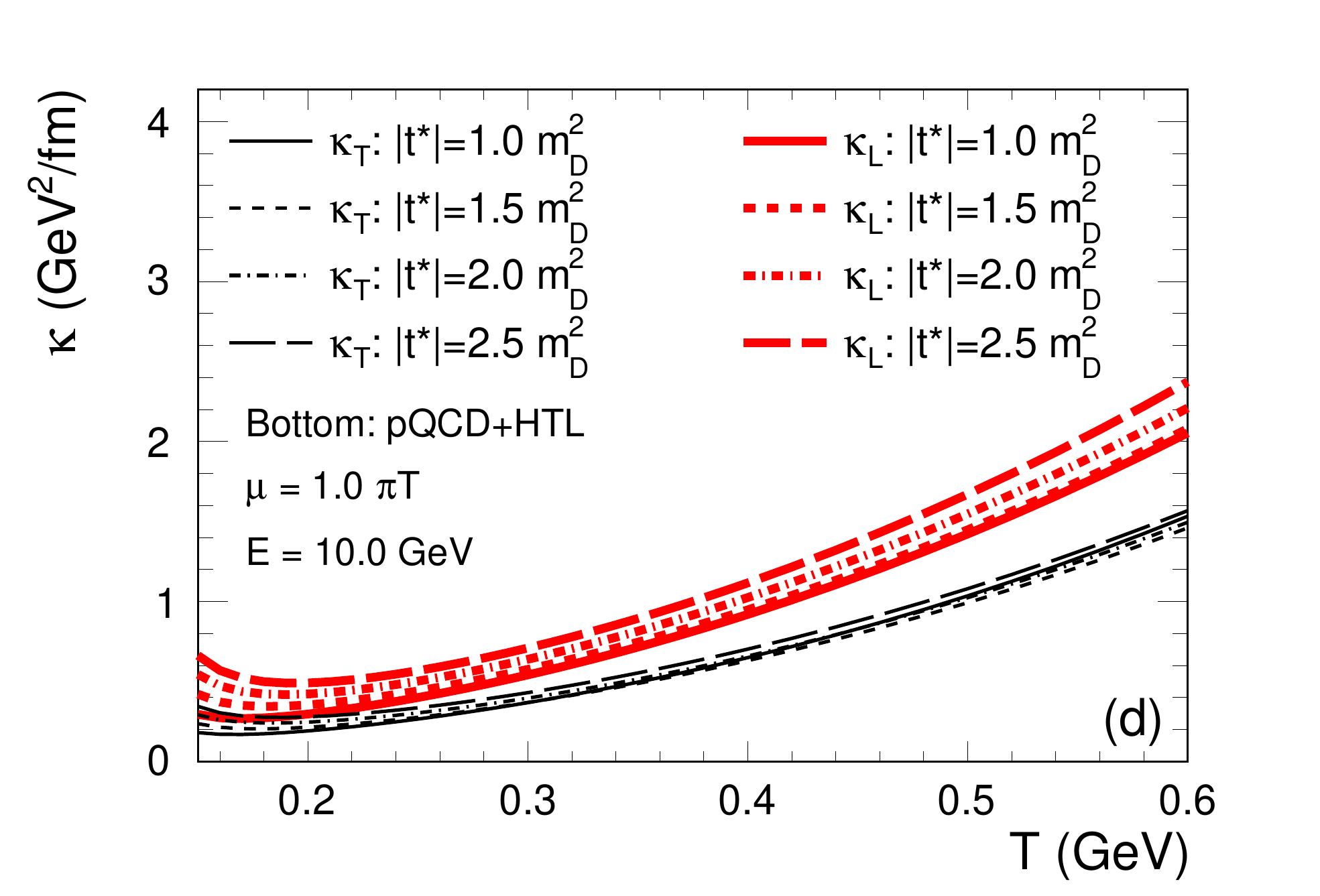}
\caption{(Color online) Comparison of the momentum diffusion coefficients $\kappa_{T}$ (thin black curves) and $\kappa_{L}$ (thick red curves),
for charm (left) and bottom quarks (right), displaying separately the results
based various testing parameters: $m_{D}^{2}\le{|t^{\ast}|}\le2.5m_{D}^{2}$ and $\mu=\pi T$.}
\label{fig:Kappa_Cutoff}
\end{figure*}

In Fig.~\ref{fig:Kappa_Cutoff}, charm (left) and bottom quark (right) $\kappa_{T}$ (thin curves) and $\kappa_{L}$ (thick curves) are calculated,
with $m_{D}^{2}\le{|t^{\ast}|}\le2.5m_{D}^{2}$ and $\mu=\pi T$,
at the temperature $T=0.40$ GeV (upper) and the energy $E=10.0$ GeV (lower).
$\kappa_{T}$ and $\kappa_{L}$ are, as expected, identical at zero momentum limit ($E=m_{Q}$),
while the latter one has a much stronger energy dependence at larger momentum.
Furthermore, $\kappa_{T/L}(E,T)$ behave a mild sensitivity to the intermediate cutoff $t^{\ast}$~\cite{POWLANGEPJC11}.
Because the soft-hard approach is strictly speaking valid when the coupling is small $m_{D}^{2}\ll T^{2}$~\cite{HQSteph08QCD},
thus, the above observations support the validity of this approach within the temperature regions even though the coupling is not small.

\begin{figure*}[!htbp]
\centering
\setlength{\abovecaptionskip}{-0.1mm}
\includegraphics[width=.32\textwidth]{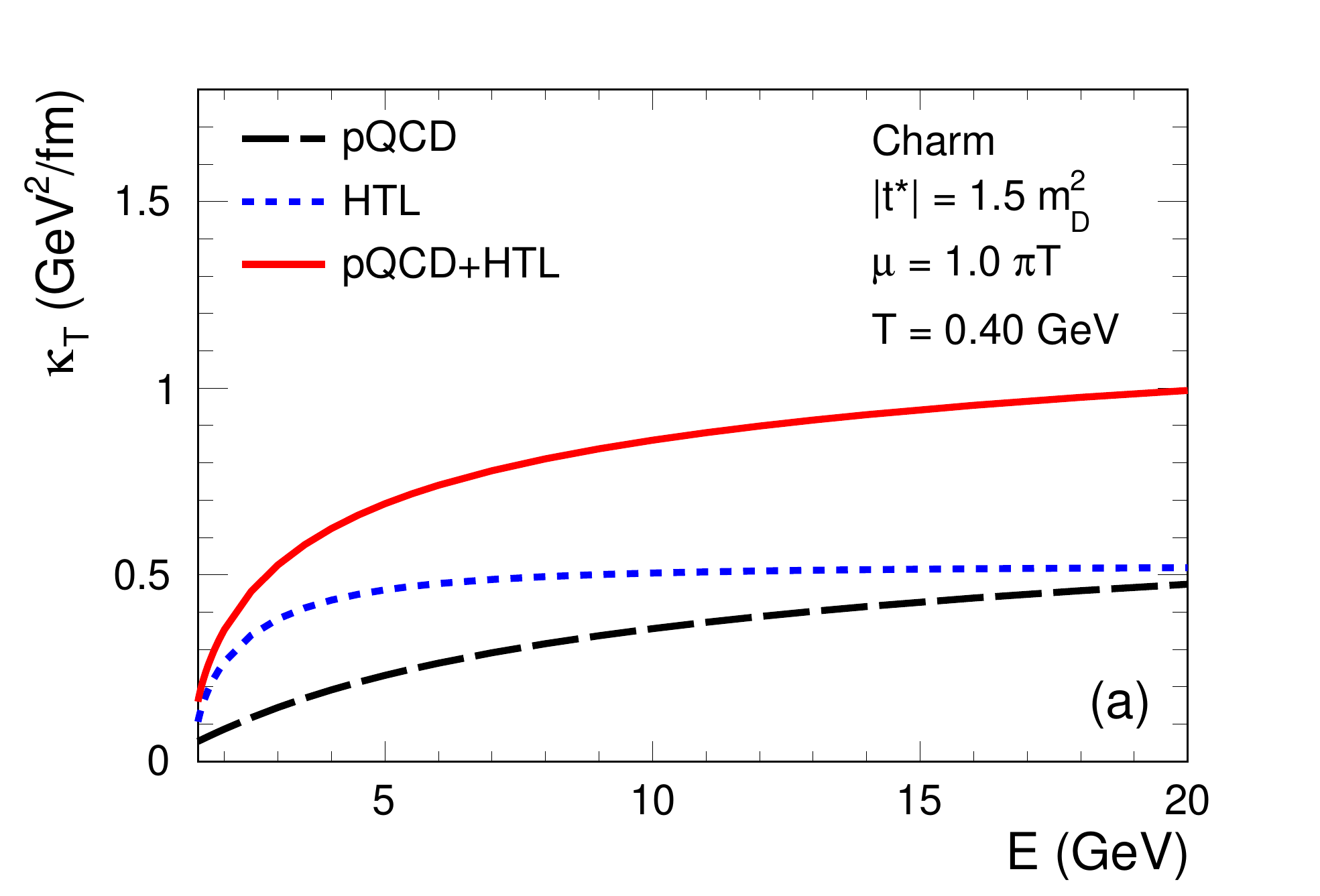}
\includegraphics[width=.32\textwidth]{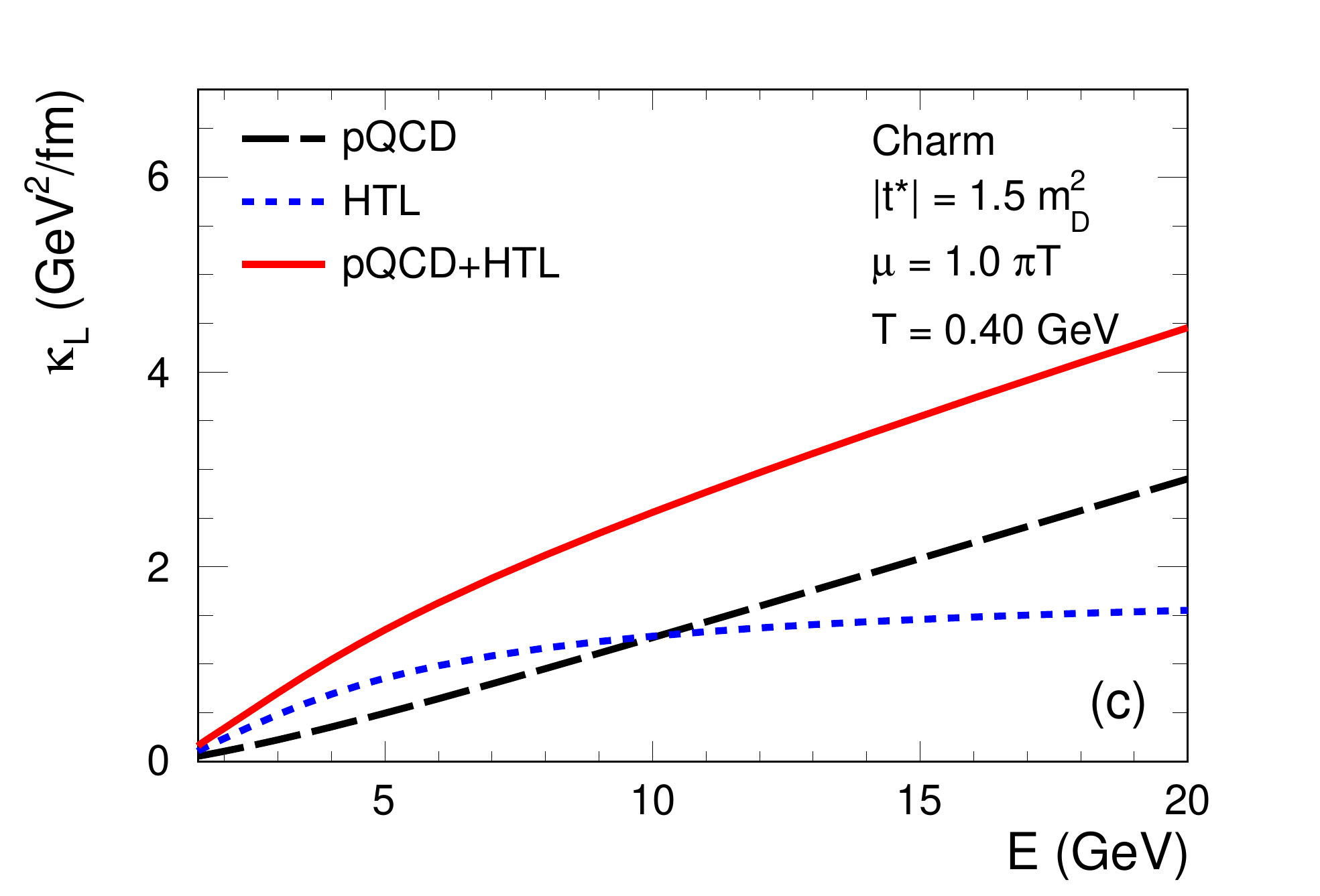}
\includegraphics[width=.32\textwidth]{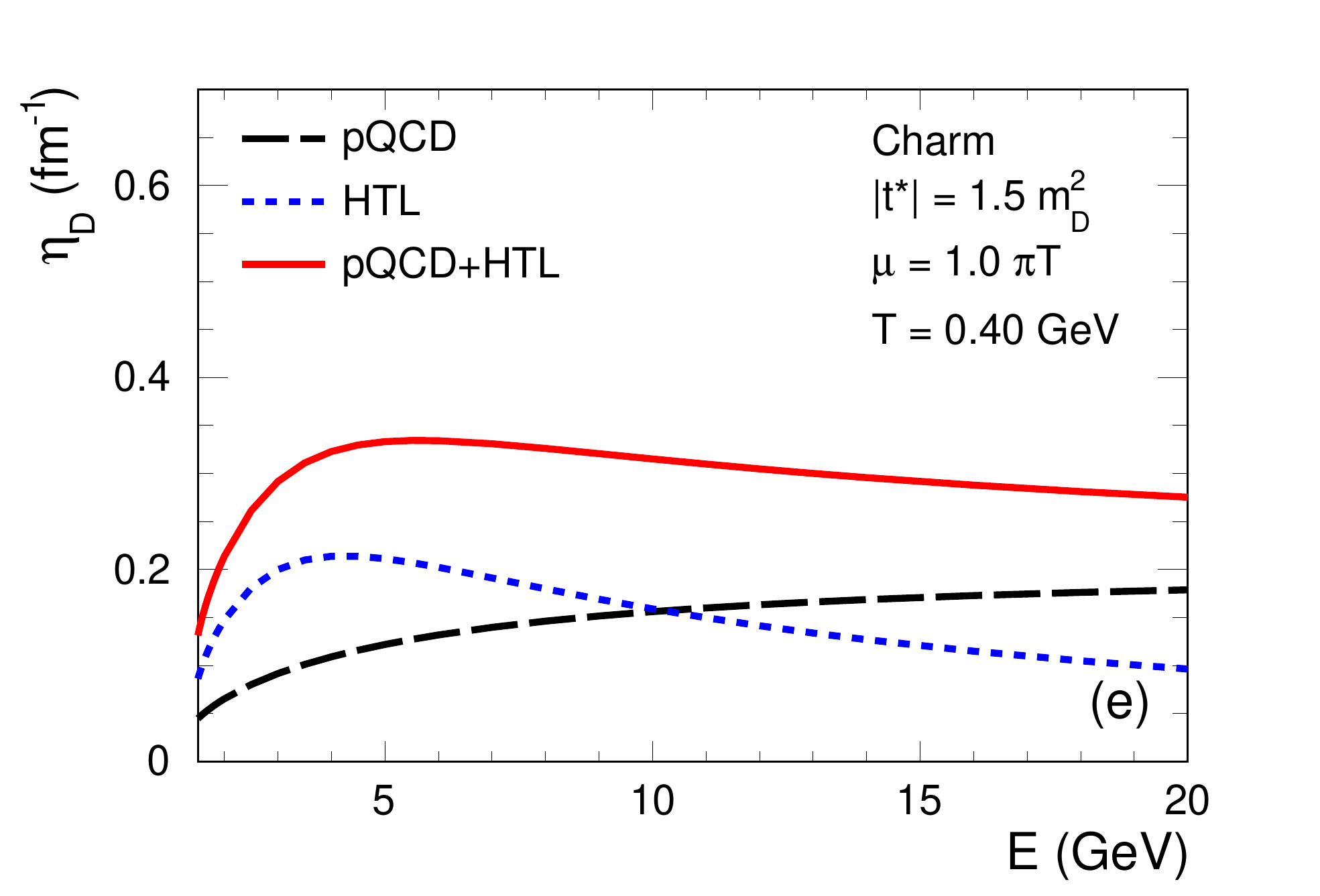}
\includegraphics[width=.32\textwidth]{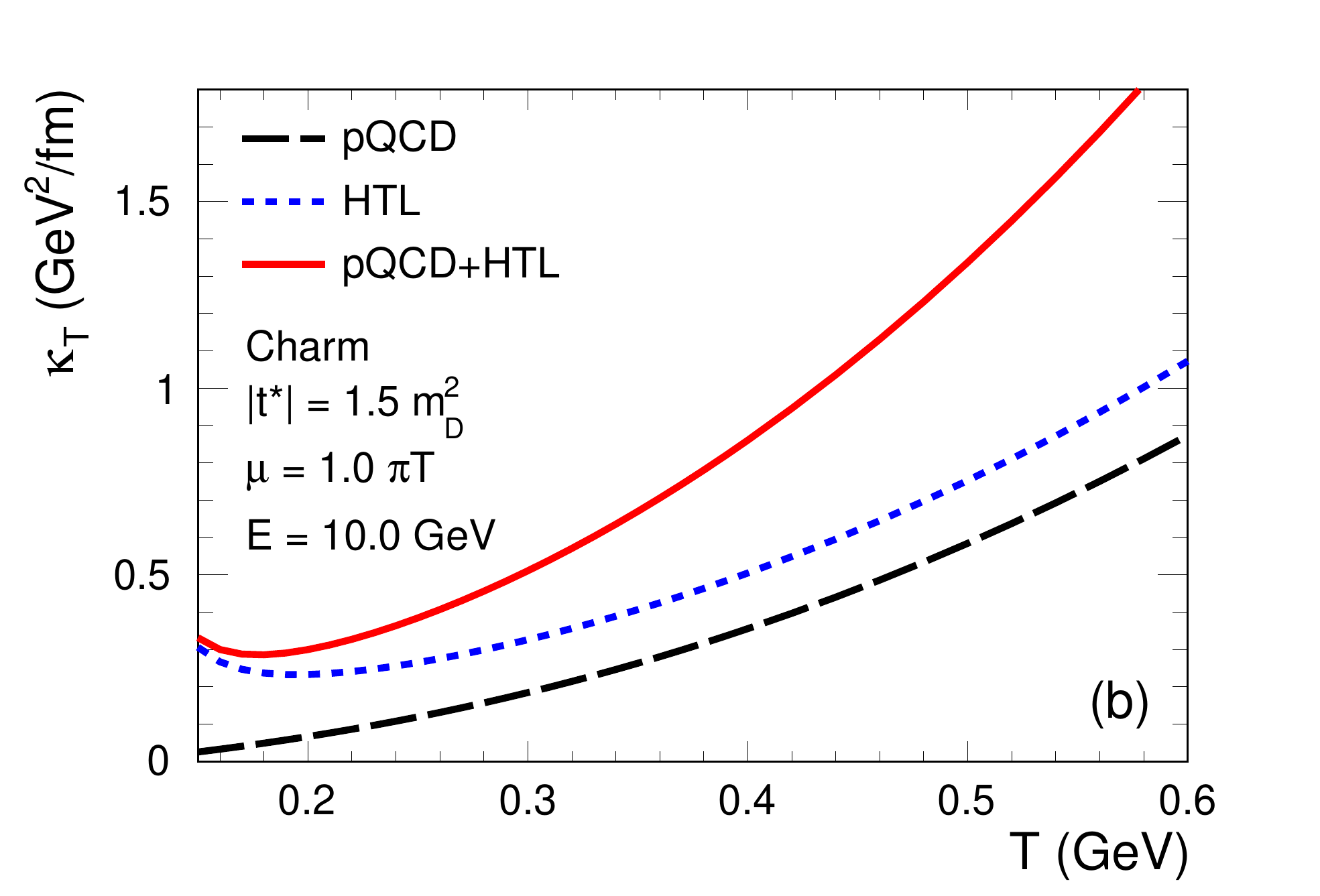}
\includegraphics[width=.32\textwidth]{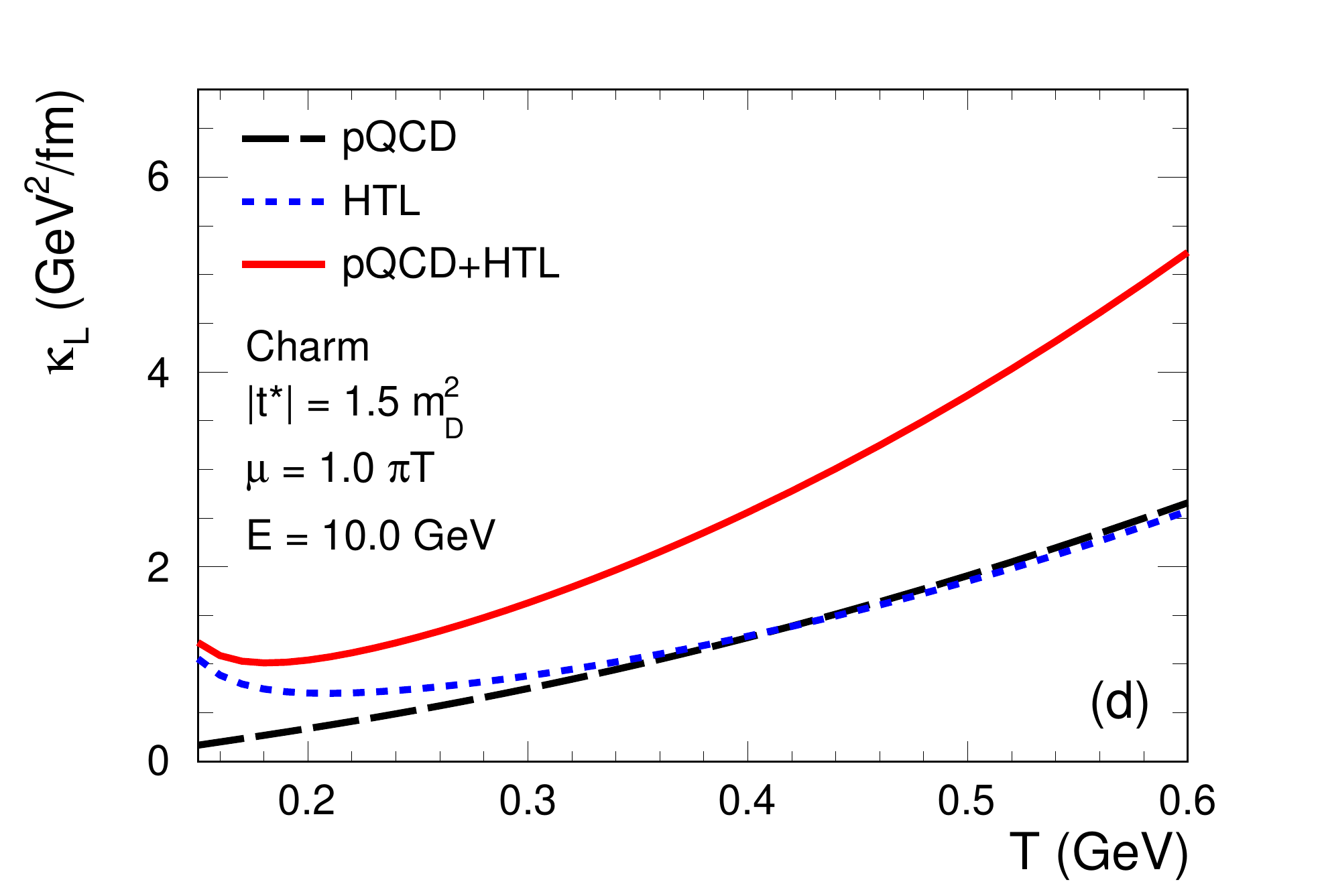}
\includegraphics[width=.32\textwidth]{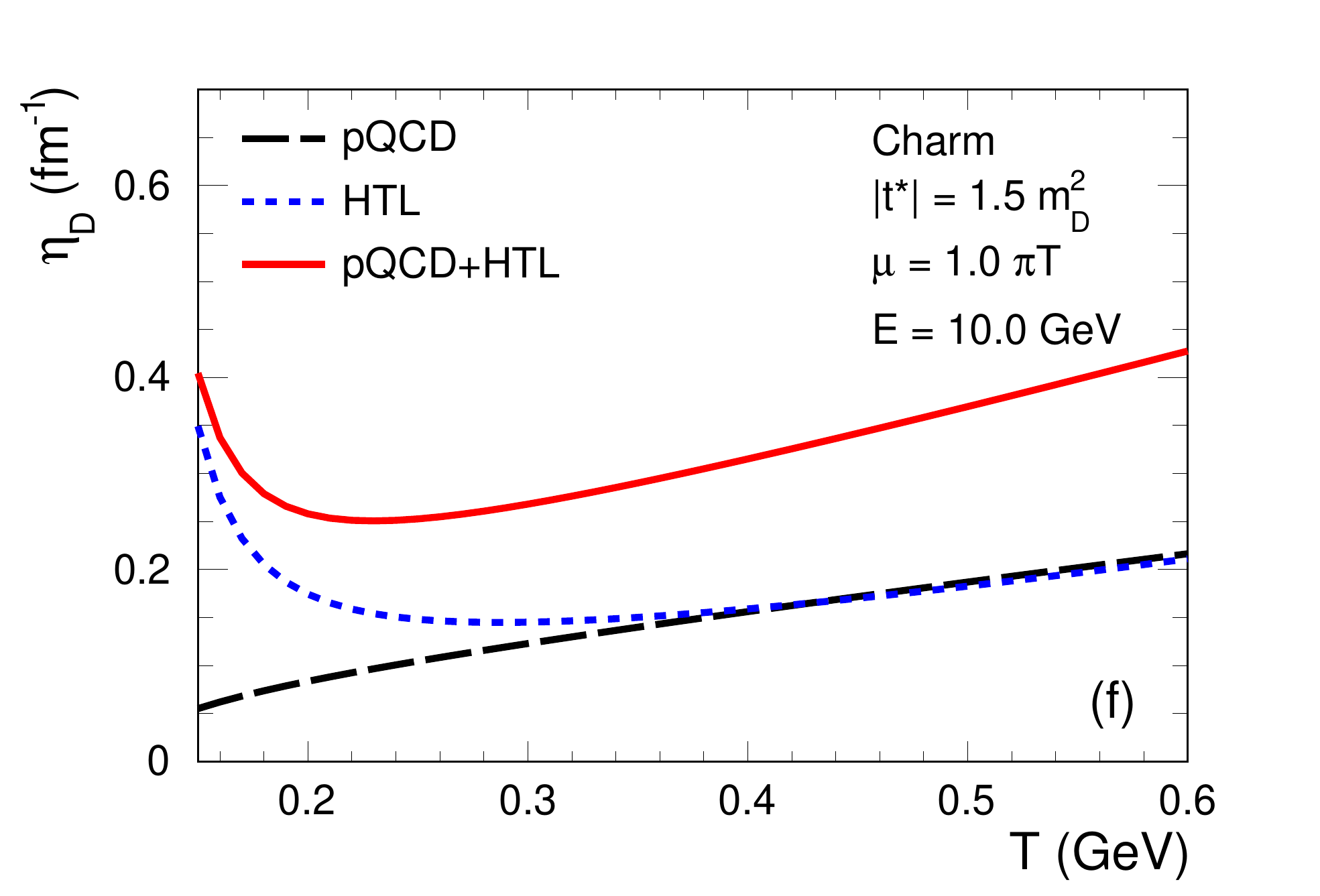}
\caption{(Color online) Charm quark $\kappa_{T}$ (left) and $\kappa_{L}$ (middle) are
shown at fixed temperature $T=0.40$ GeV (upper) and at fixed energy $E=10.0$ GeV (lower),
contributed by the soft (dotted blue curves) and hard collisions (dashed black curves).
The combined results (solid red curves) are shown as well for comparison.
The derived drag coefficient $\eta_{D}$ (right; Eq.~\ref{eq:EtaD_DissFluc}) are obtained with the post-point scenario.}
\label{fig:Charm_Coef_vsE_vsT}
\end{figure*}

\begin{figure*}[!htbp]
\centering
\setlength{\abovecaptionskip}{-0.1mm}
\includegraphics[width=.32\textwidth]{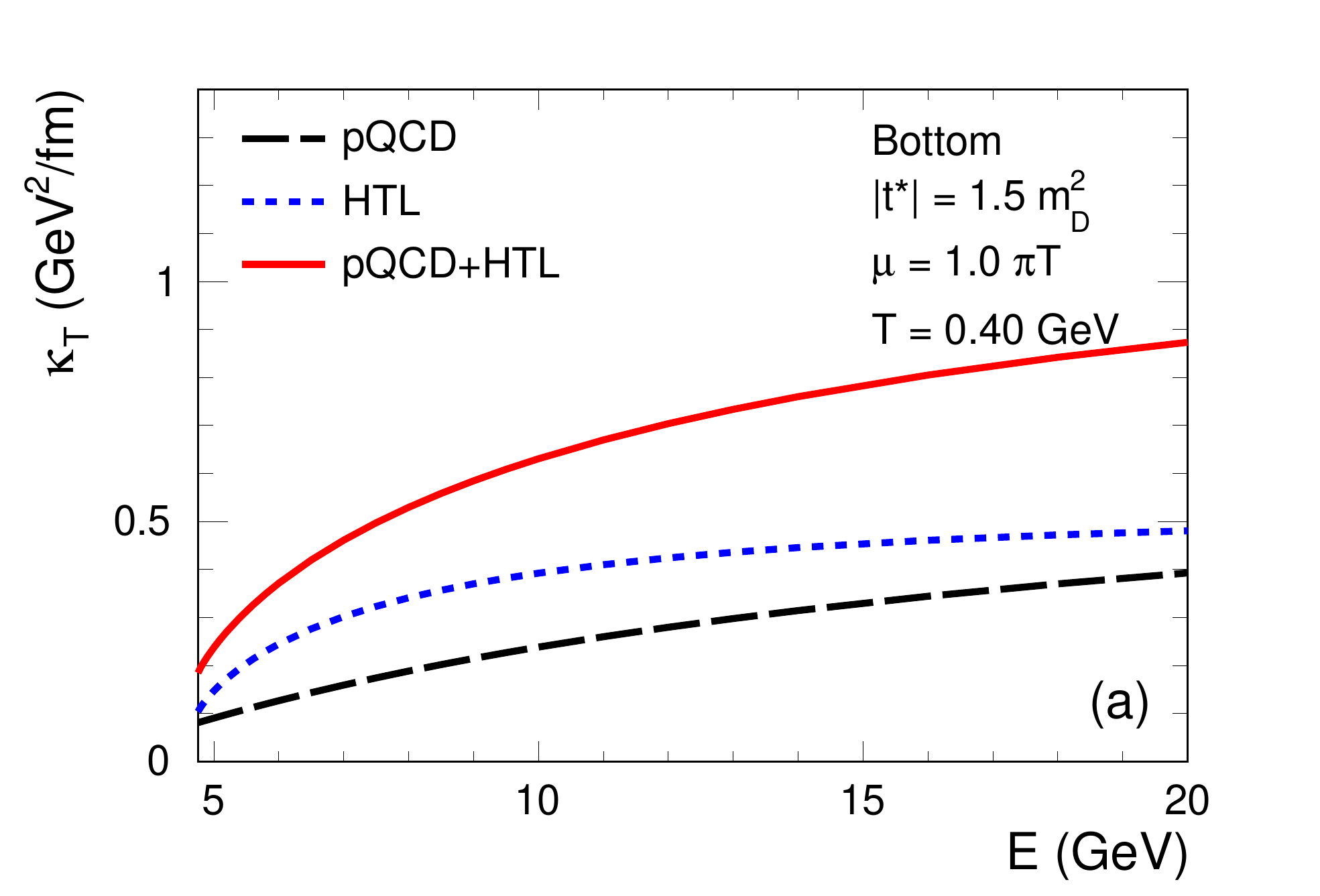}
\includegraphics[width=.32\textwidth]{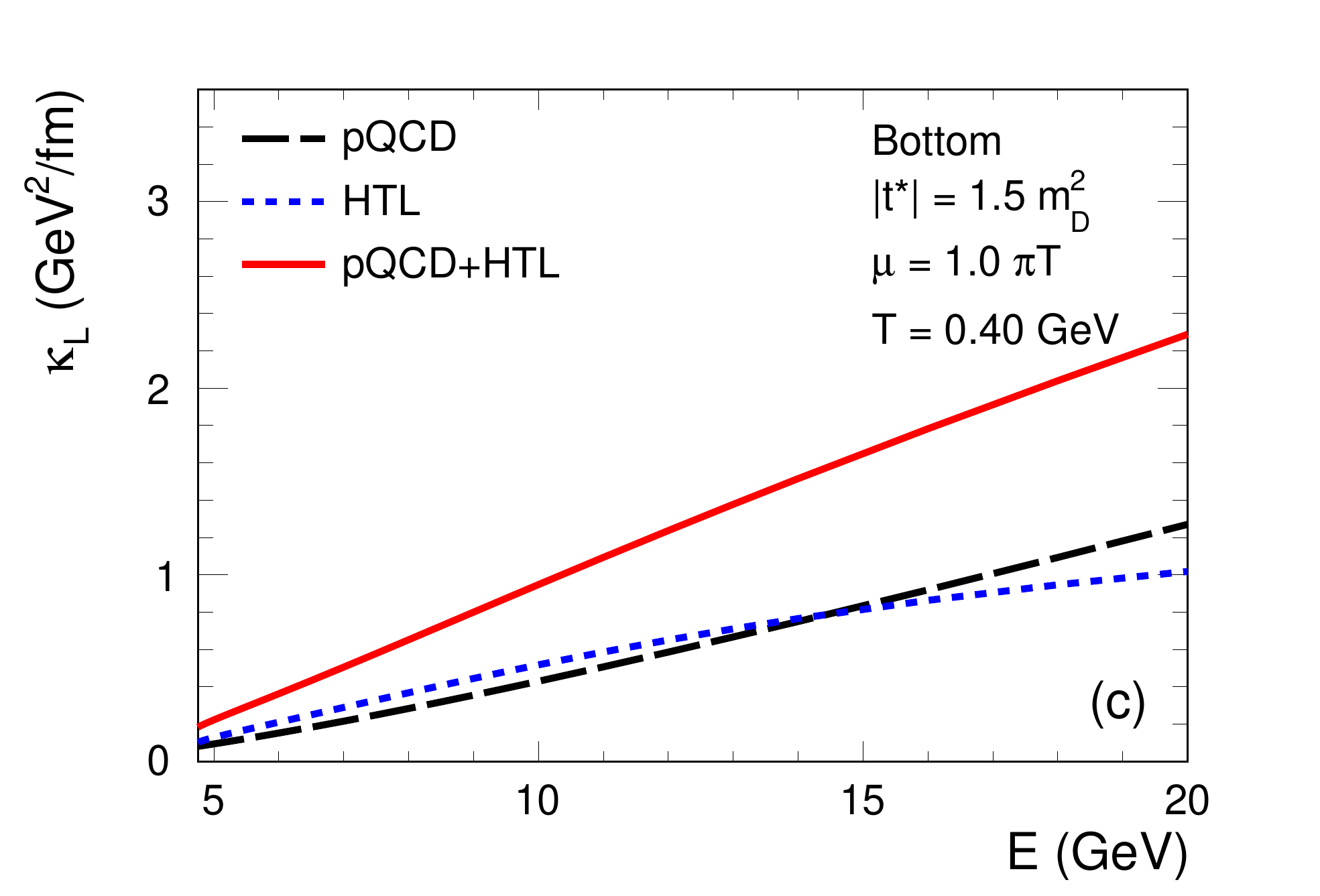}
\includegraphics[width=.32\textwidth]{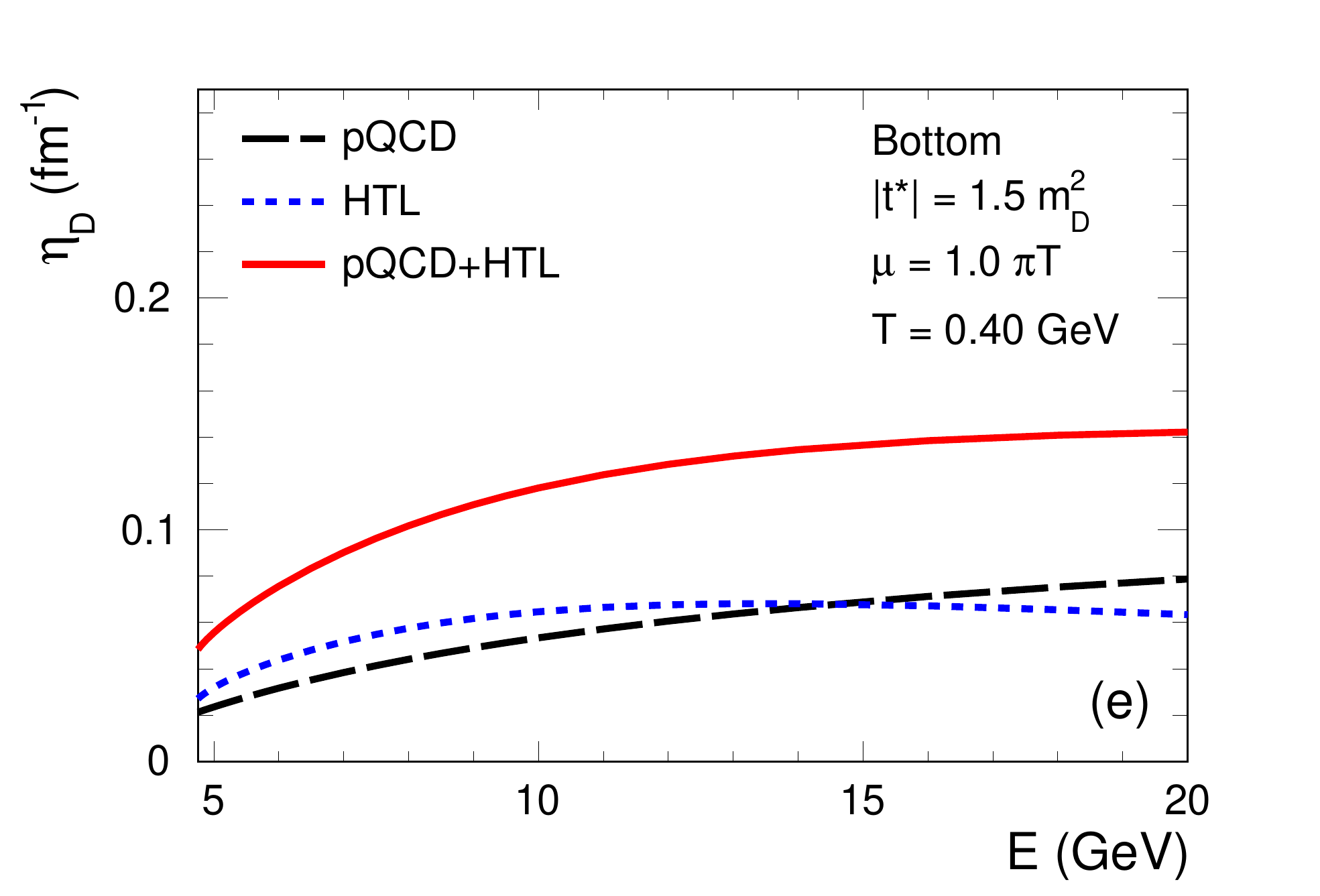}
\includegraphics[width=.32\textwidth]{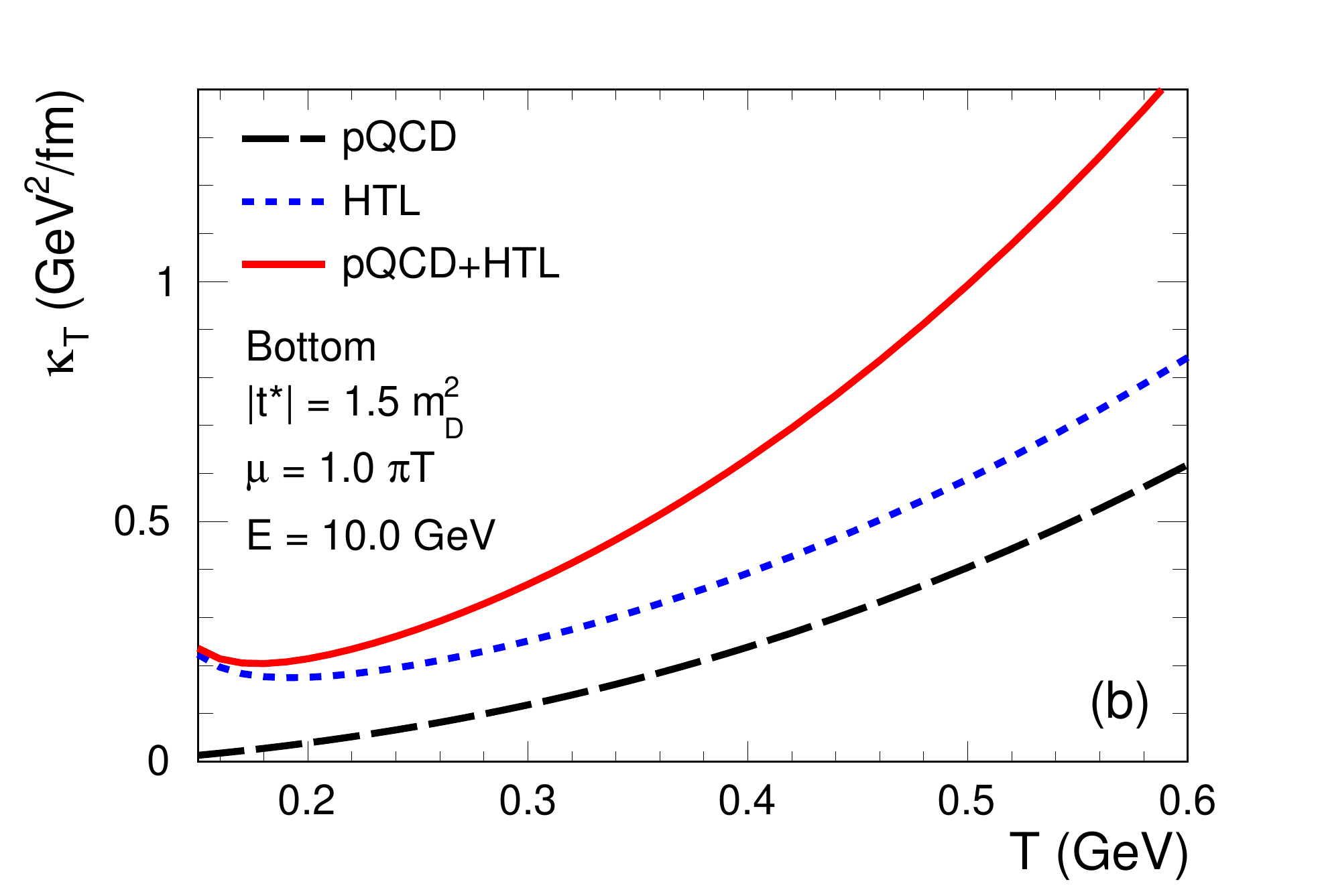}
\includegraphics[width=.32\textwidth]{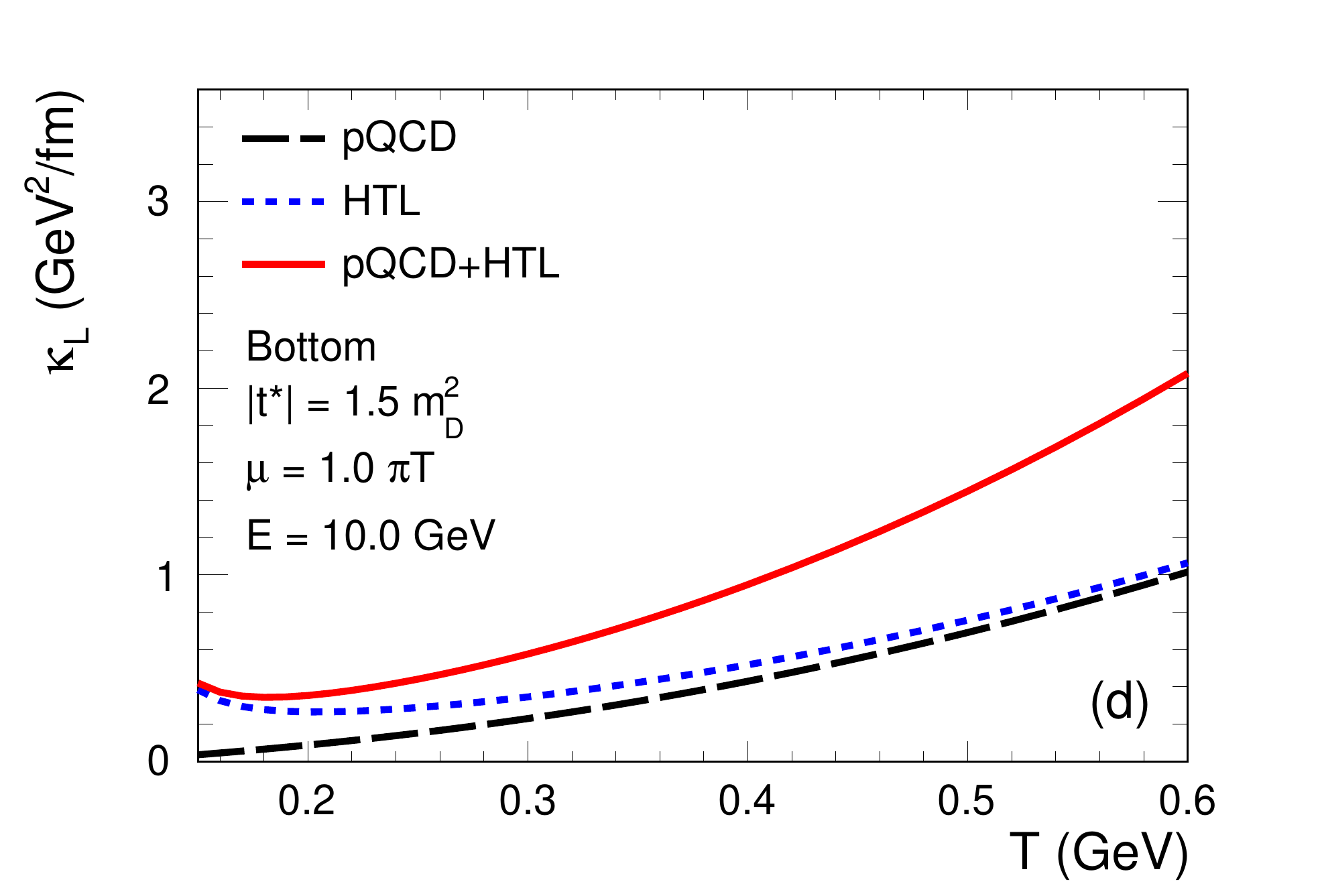}
\includegraphics[width=.32\textwidth]{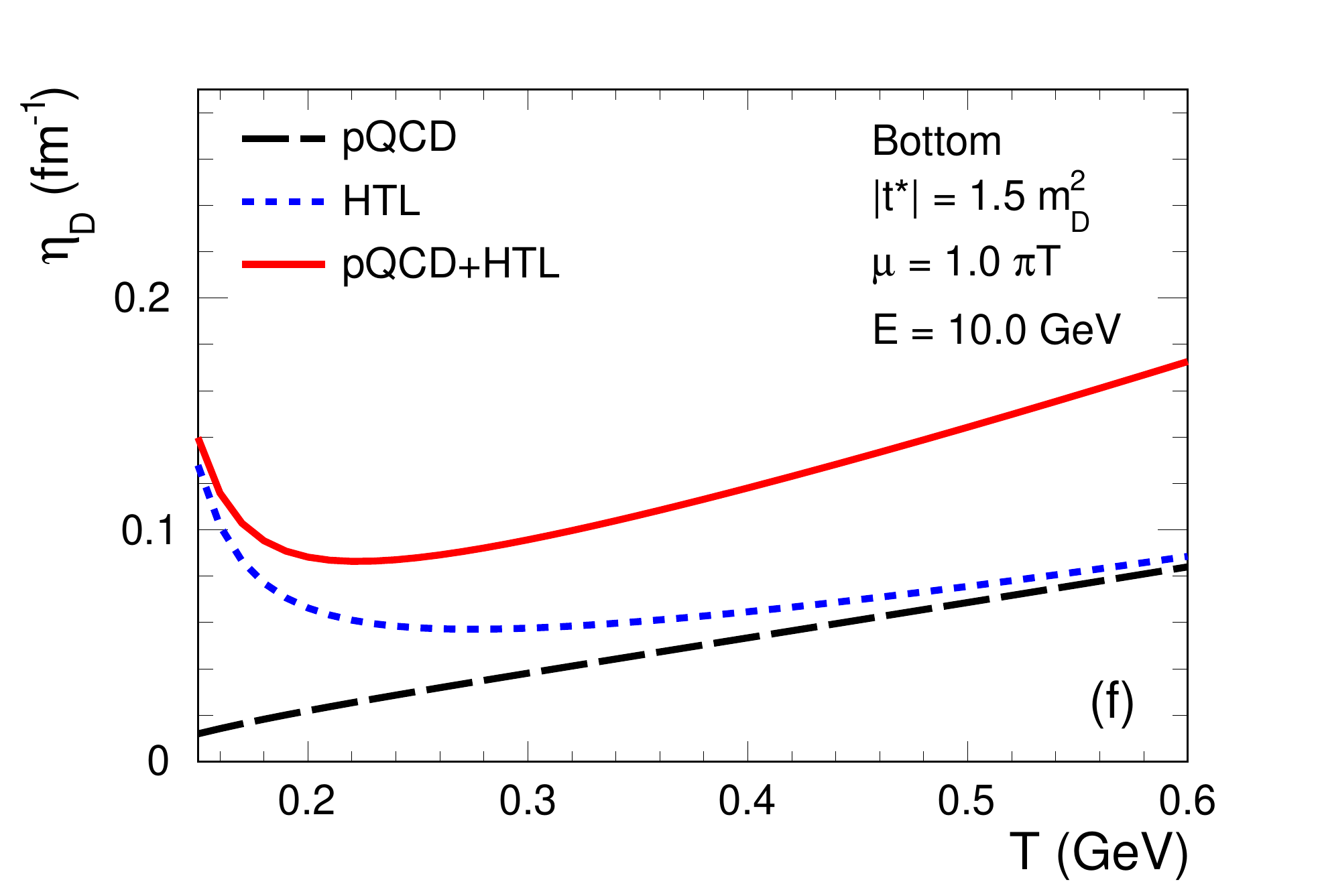}
\caption{Same as Fig.~\ref{fig:Charm_Coef_vsE_vsT} but for bottom quark.}
\label{fig:Bottom_Coef_vsE_vsT}
\end{figure*}

In Fig.~\ref{fig:Charm_Coef_vsE_vsT}, charm quark $\kappa_{T}$ (left) and $\kappa_{L}$ (middle) are evaluated
with the optimized parameters, $|t^{\ast}|=1.5m_{D}^{2}$ and $\mu=\pi T$,
including both the soft (dotted blue curves) and hard contributions (dashed black curves),
at fixed temperature $T=0.40$ GeV (upper) and at fixed energy $E=10.0$ GeV (lower).
It is found that the soft components are significant at low energy/temperature,
while they are compatible at larger values.
The combined results (solid red curves) are presented as well for comparison.
With the post-point scheme ($\xi=1$ in Eq.~\ref{eq:KappaAll}),
the drag coefficients (right) behave
(1) a nonmonotonic dependence, in particular on the temperature,
which is in part due to the improved treatment of the screening in soft collisions,
and in part due to the procedure of inferring $\eta_{D}$ from $\kappa_{T/L}$~\cite{RalfSummary16};
(2) a weak energy dependence at $E\gtrsim2m_{c}=3$ GeV.
Similar conclusions can be drawn for bottom quark, as shown in Fig.~\ref{fig:Bottom_Coef_vsE_vsT}.

\begin{figure}[!htbp]
\centering
\setlength{\abovecaptionskip}{-0.1mm}
\includegraphics[width=.40\textwidth]{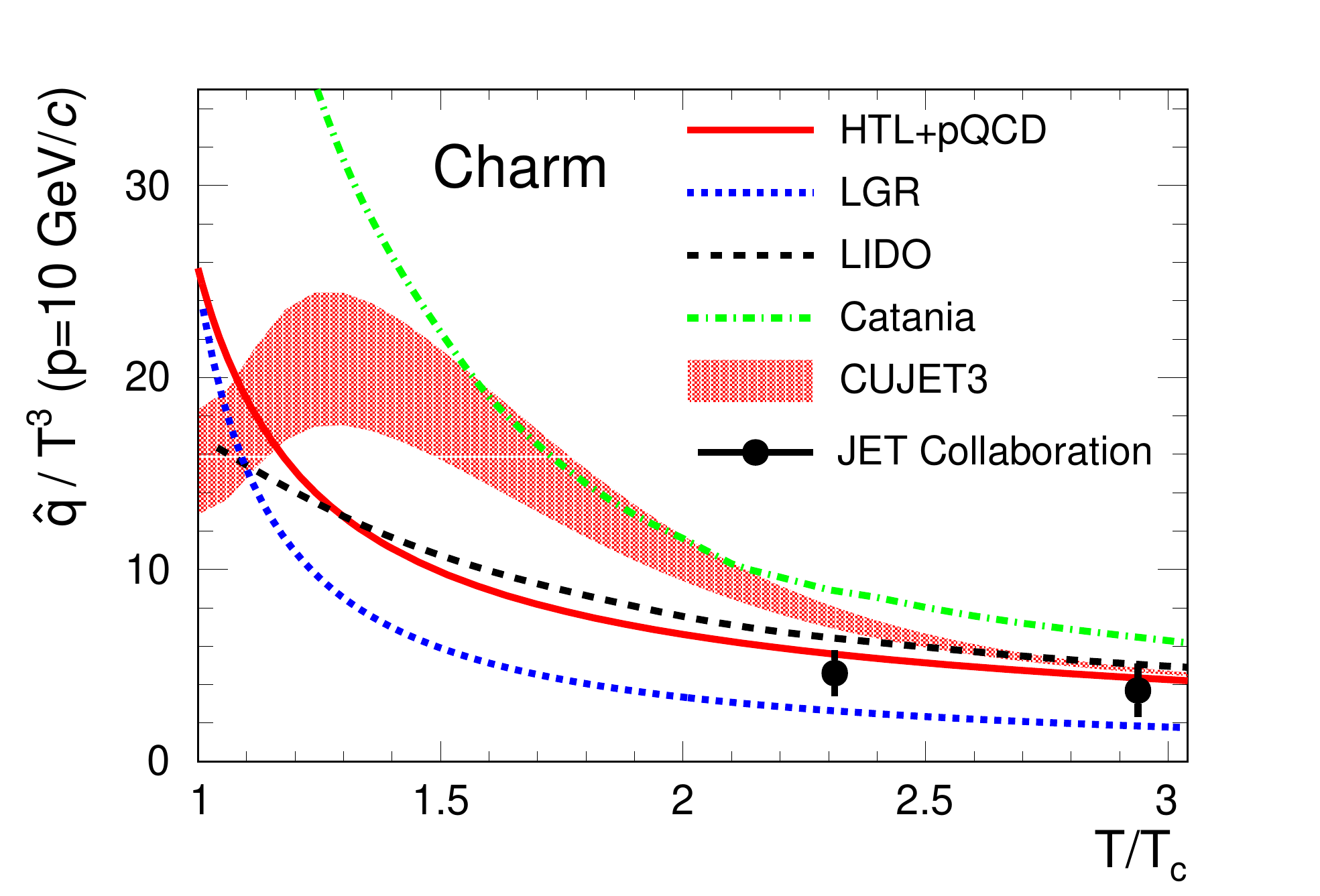}
\caption{(Color online) Transport coefficient, $\hat{q}/T^{3}(T)$, of charm quark from the various calculations,
including: the soft-hard factorized approach (solid red curve), the LGR model with data optimized parameters (dotted blue curve~\cite{CTGUHybrid4}),
a Bayesian ayalysis from LIDO (dashed black curve~\cite{Lido18}),
a quasi-particle model from Catania (dot-dashed green curve~\cite{Catania_Qhat}),
CUJET3 (shadowed red band~\cite{CUJET3JHEP16, CUJET3Arxiv18, CUJET3CPC18})
and JET Collaboration (black circle points~\cite{JETCoef14}) at $p=10$ GeV.}
\label{fig:Qhat_p10}
\end{figure}

Figure~\ref{fig:Qhat_p10} presents the transport coefficient of charm quark, $\hat{q}=2\kappa_{T}$,
at fix momentum $p=10$ GeV (solid red curve).
It can be seen that $\hat{q}/T^{3}$ reaches the maximum near the critical temperature,
and then followed by a decreasing trend with $T$,
providing a good description of the light quark transport parameter (black circle points).
The results from various phenomenological extractions and theoretical calculations,
including a phenomenological fitting analysis with the Langevin-transport with Gluon Radiation (LGR; dotted blue curve~\cite{CTGUHybrid4}),
a LO calculation with a Linearized Boltzmann Diffusion Model
(LIDO\footnote[3]{LIDO results are shown with only elastic scattering channels.}; dashed black curve~\cite{Lido18}),
a nonperturbative treatment with Quasi-Particle Model (QPM in Catania; dot-dashed green curve~\cite{Catania_2PiTDs}),
a novel confinement with semi-quark-gluon-monopole plasma approach (CUJET3; shadowed red band~\cite{CUJET3JHEP16, CUJET3Arxiv18, CUJET3CPC18}),
are displayed as well for comparison.
Similar temperature dependence can be observed except the CUJET3 approach,
which shows a strong enhancement near $T_{c}$ regime.

\begin{figure}[!htbp]
\centering
\setlength{\abovecaptionskip}{-0.1mm}
\includegraphics[width=.40\textwidth]{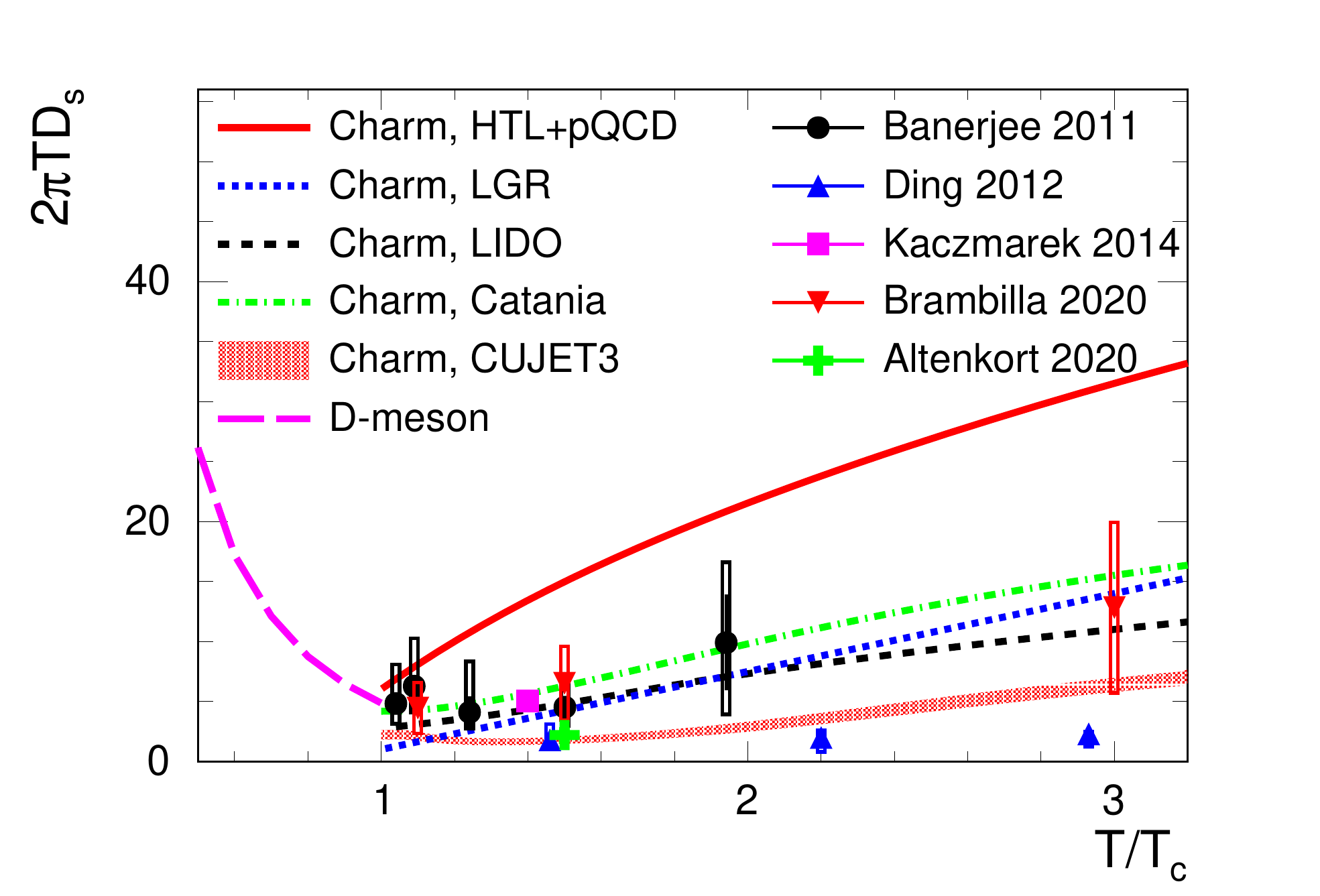}
\caption{(Color online) Spatial diffusion constant $2\pi TD_{s}(T)$ of charm quark from
the sof-hard scenario (solid red curve), LGR (dotted blue curve~\cite{CTGUHybrid4}, LIDO (dashed black curve~\cite{Lido18}),
Catania (dot-dashed green curve~\cite{Catania_2PiTDs}), CUJET3 (shadowed red band~\cite{CUJET3JHEP16, CUJET3Arxiv18, CUJET3CPC18})
and lattice QCD calculations (black circle~\cite{LQCDbanerjee12}, blue triangle~\cite{LQCDding12},
pink square~\cite{LQCDolaf14}, red inverted triangle~\cite{LQCDBrambilla20} and green plus~\cite{LQCDAltenkort20}).
The result for D-meson (long dashed pink curve~\cite{2PiTDs4Dmeson}) in the hadronic phase is shown for comparison.}
\label{fig:2PiTDs_C}
\end{figure}

The scaled spatial diffusion coefficient describes the low energy interaction strength of HQ in medium~\cite{Moore04},
\begin{equation}\label{eq:SpatialDiff}
2\pi TD_{s}= \lim_{E\rightarrow m_{Q}}\frac{2\pi T^2}{m_{Q} \cdot \eta_{D}(E,T)},
\end{equation}
and it can be calculated by substituting Eq.~\ref{eq:EtaD_DissFluc} into Eq.~\ref{eq:SpatialDiff}.
The obtained result for charm quark ($m_{c}=1.5$ GeV) is displayed as the solid red curve in Fig.~\ref{fig:2PiTDs_C}.
It is found that a relatively strong increase of $2\pi TD_{s}(T)$ from crossover temperature $T_c$ toward high temperature.
Meanwhile, the $\vtwo$ data prefers a small value of $2\pi TD_s$ near $T_c$,
$2\pi TD_{s}(T_{c})\simeq3-6$\footnote[4]{This range is estimated from the testing models as displayed in the right panels of Fig.~\ref{fig:Param_RAAV2}.},
which is close to the lattice QCD calculations~\cite{LQCDbanerjee12, LQCDding12, LQCDolaf14, LQCDBrambilla20, LQCDAltenkort20}.
The relevant results from other theoretical analyses,
such as LGR~\cite{CTGUHybrid4}, LIDO~\cite{Lido18}, Catania~\cite{Catania_2PiTDs} and
CUJET3\footnote[5]{CUJET3 results are obtained by performing the energy interpolation down to $E=m_{Q}$.}~\cite{CUJET3JHEP16, CUJET3Arxiv18, CUJET3CPC18},
show a similar trend but with much weaker temperature dependence.

\subsection{Comparison with experimental data: $\raa$ and $\vtwo$}\label{subsec:CmpData}
\begin{figure}[!htbp]
\centering
\setlength{\abovecaptionskip}{-0.1mm}
\includegraphics[width=.40\textwidth]{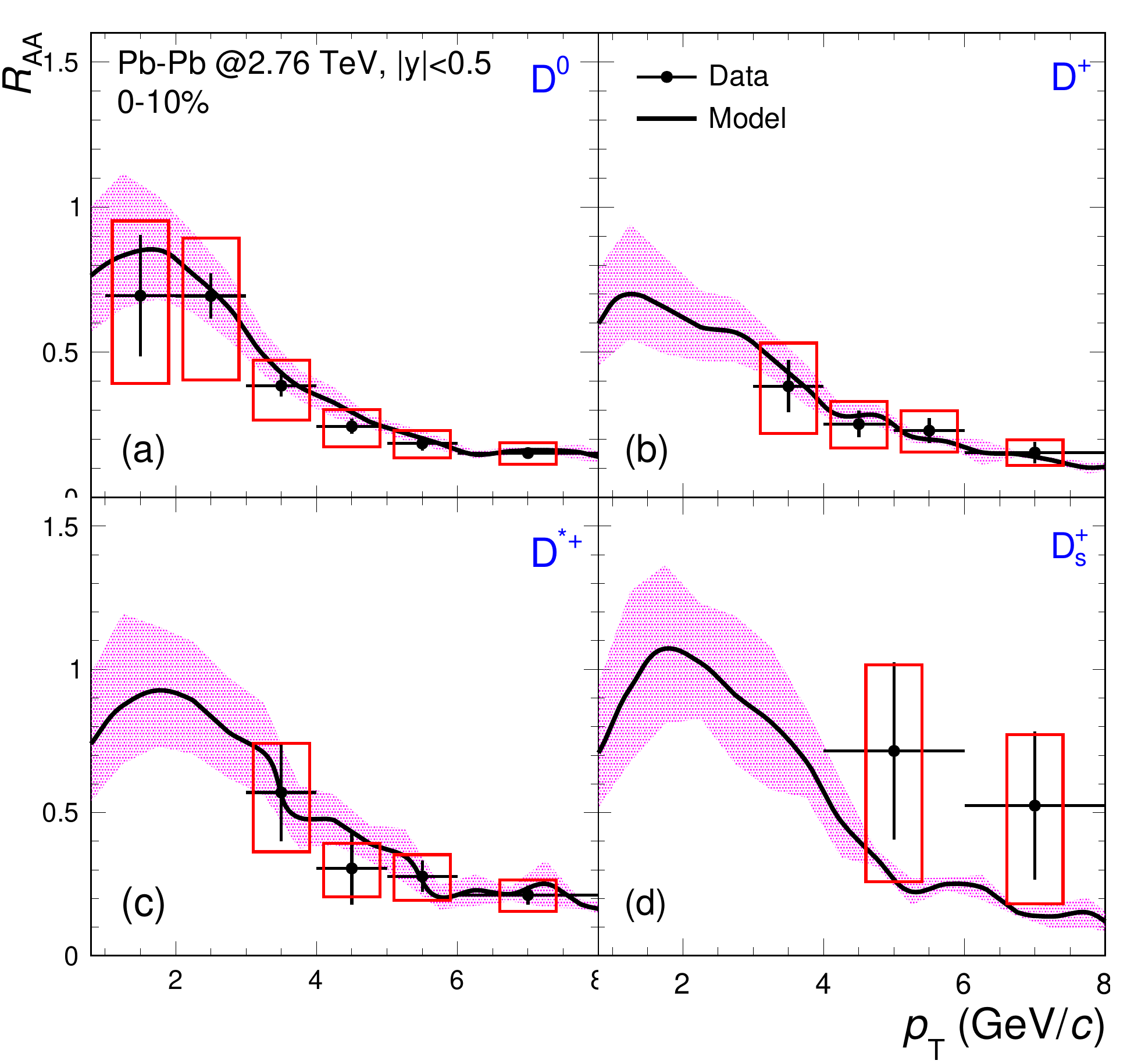}
\caption{(Color online) Comparison between experimental data (red box~\cite{ALICEDesonPbPb2760RAA, ALICEDsPbPb2760RAA})
and soft-hard factorized model calculations (solid black curve with pink uncertainty band)
for the nuclear modification factor $\raa$, of $D^{0}$ (a), $D^{+}$ (b), $D^{\ast +}$ (c) and $D_{s}^{+}$ (d)
at mid-rapidity ($|y|<0.5$) in central ($0-10\%$) Pb--Pb collisions at $\snn=2.76$ TeV.}
\label{fig:RAA_PbPb2760_C0}
\end{figure}

\begin{figure}[!htbp]
\centering
\setlength{\abovecaptionskip}{-0.1mm}
\includegraphics[width=.40\textwidth]{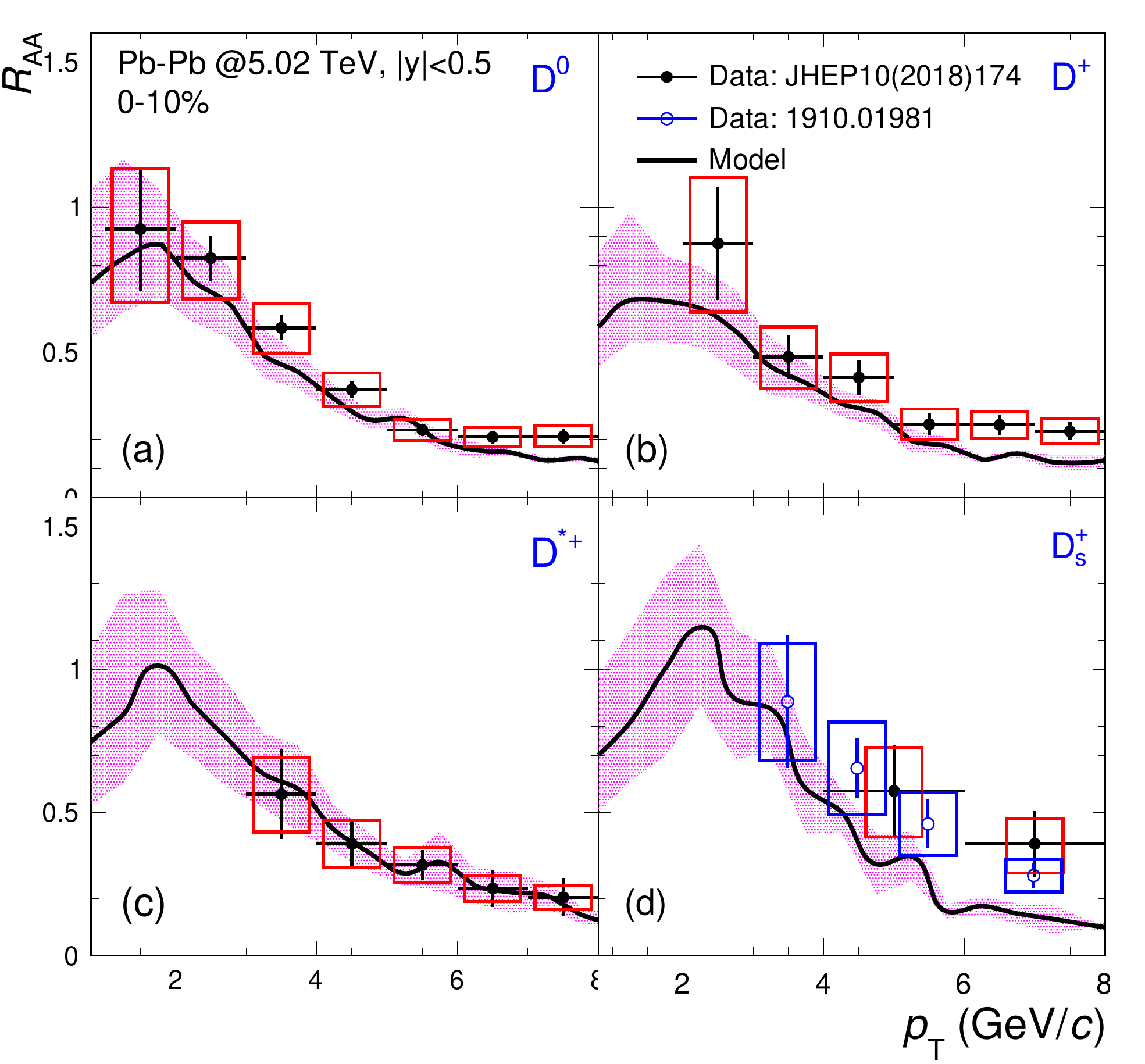}
\caption{Same as Fig.~\ref{fig:RAA_PbPb2760_C0} but for Pb--Pb collisions at $\snn=5.02$ TeV.
The data (solid~\cite{ALICEDesonPbPb5020RAA}, open~\cite{Vertesi:2019awk}) are shown for comparison.}
\label{fig:RAA_PbPb5020_C0}
\end{figure}

Figure~\ref{fig:RAA_PbPb2760_C0} shows the $\raa$
of $D^{0}$ (a), $D^{+}$ (b), $D^{\ast +}$ (c) and $D_{s}^{+}$ (d)
in the most central ($0-10\%$) Pb--Pb collisions at $\snn=2.76$ TeV, respectively.
The calculations are done with FONLL initial charm quark spectra and EPS09 NLO parametrization for the nPDF in Pb~\cite{CTGUHybrid1},
and the pink band reflects the theoretical uncertainties coming from these inputs.
It can be seen that, within the experimental uncertainties,
the model calculations provide a very good description of the measured $\pt$-dependent $\raa$ data for various charm mesons.
Concerning the results in Pb--Pb collisions at $\snn=5.02$ TeV, as shown in Fig.~\ref{fig:RAA_PbPb5020_C0},
a good agreement is found between the model and the measurement at $\pt\lesssim6~{\rm GeV/\it{c}}$,
while a slightly larger discrepancy observed at larger $\pt$.

\begin{figure}[!btbp]
\centering
\setlength{\abovecaptionskip}{-0.1mm}
\includegraphics[width=.40\textwidth]{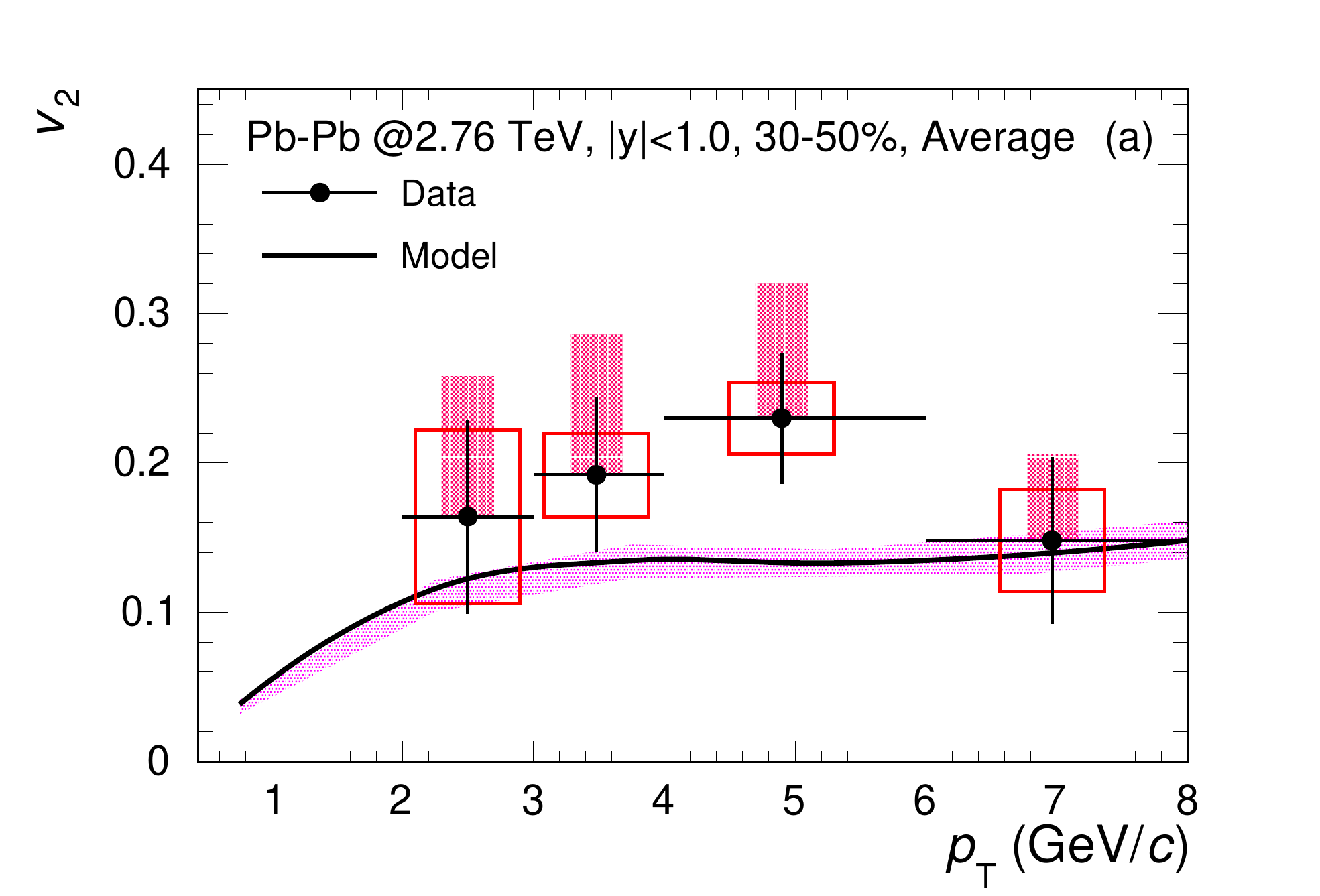}
\includegraphics[width=.40\textwidth]{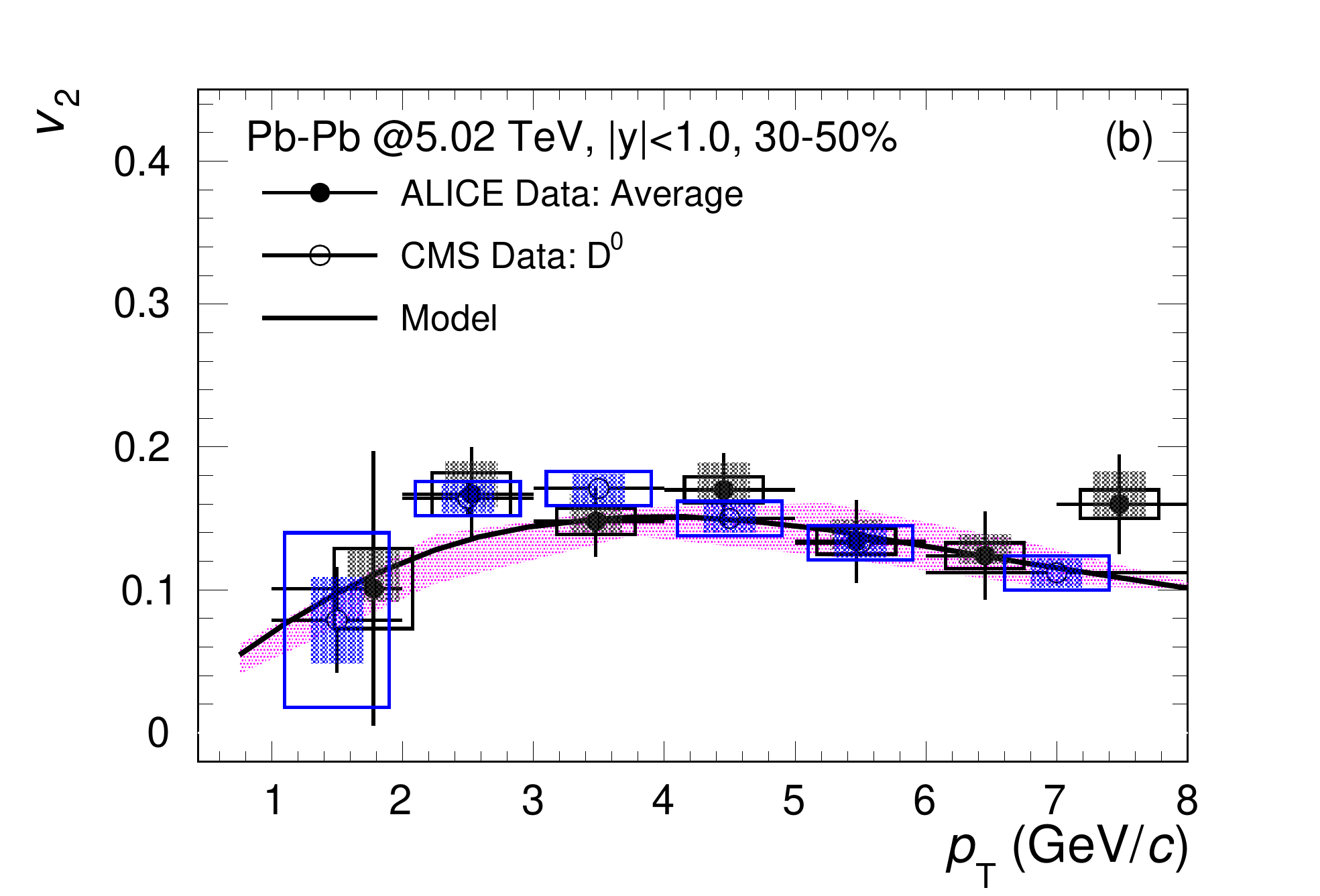}
\caption{(Color online) Comparison between experimental data (red~\cite{ALICEDesonPbPb2760V2},
black~\cite{ALICEDesonPbPb5020V2} and blue boxes~\cite{CMSD0PbPb5020V2})
and model calculations (solid black curve with pink uncertainty band)
for the elliptic flow $\vtwo$ of non-strange D-meson at mid-rapidity ($|y|<0.5$)
in semi-central ($30-50\%$) Pb--Pb collisions at $\snn=2.76$ TeV (a) and $\snn=5.02$ TeV (b).}
\label{fig:V2Dmeson2760_5020_C1}
\end{figure}

Figure~\ref{fig:V2Dmeson2760_5020_C1} presents the elliptic flow coefficient $\vtwo$ of
non-strange D-meson (averaged $D^{0}$, $D^{+}$, and $D^{\ast +}$)
in semi-central ($30-50\%$) Pb--Pb collisions at $\snn=2.76$ TeV (a) and $\snn=5.02$ TeV (b).
Within the uncertainties of the experimental data,
our model calculations describe well the anisotropy of the
transverse momentum distribution of the non-strange D-meson.
The sizable $\vtwo$ of these charm mesons, in particular at intermediate $\pt\simeq3-5$ GeV,
suggests that charm quarks actively participate in the collective expansion of the fireball.

\begin{figure}[!htbp]
\centering
\setlength{\abovecaptionskip}{-0.1mm}
\includegraphics[width=.4\textwidth]{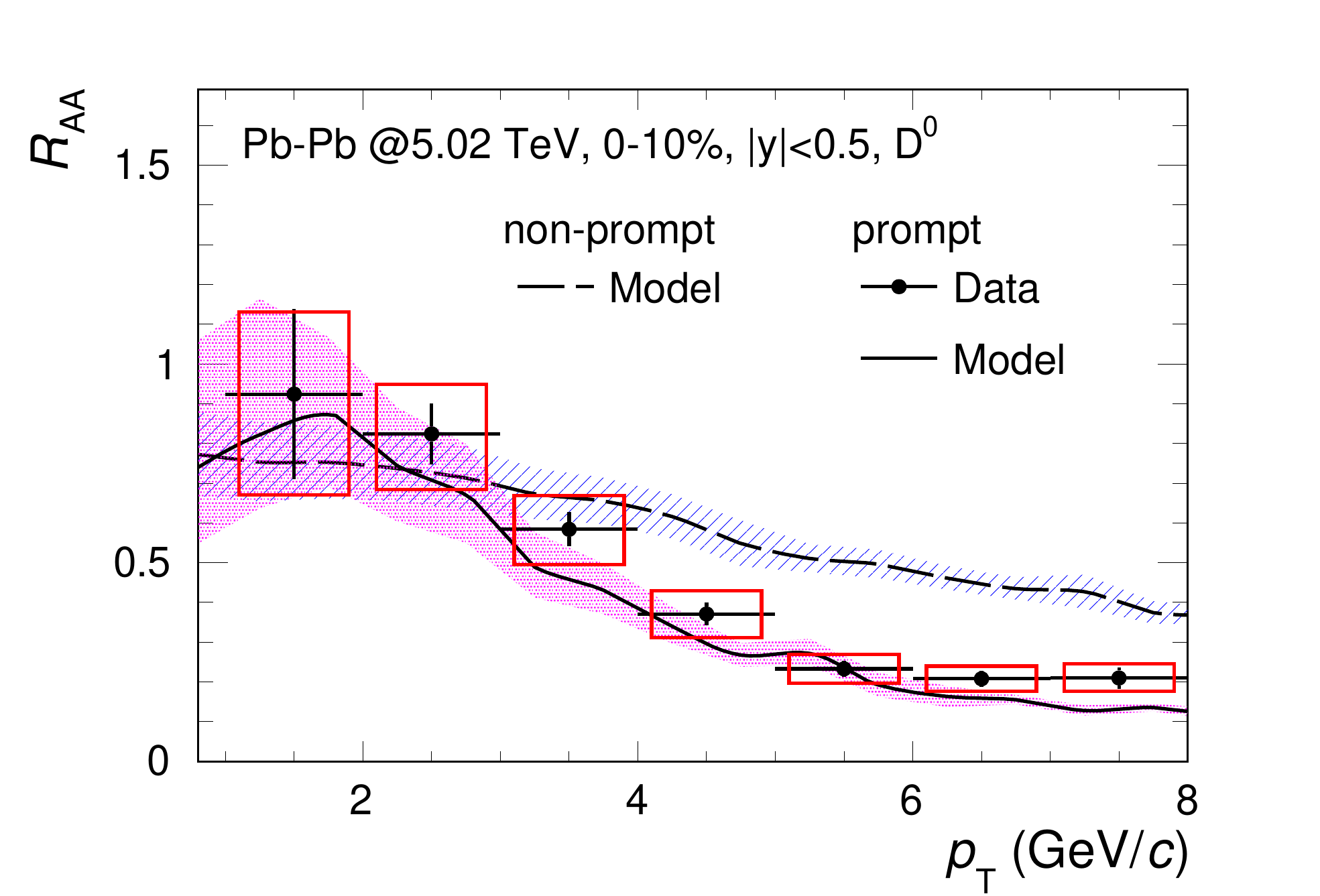}
\caption{Comparison of non-prompt (dashed blue curve) and prompt $D^{0}$ (solid black curve)
$\raa$ calculations as a function of $\pt$ with the measured values (point~\cite{ALICEDesonPbPb5020RAA}),
in the central $0-10\%$ Pb--Pb collisions at $\snn=5.02$ TeV. See legend and text for details.}
\label{fig:RAA_5020}
\end{figure}

In Fig.~\ref{fig:RAA_5020}, the $\pt$-differential $\raa$ of non-prompt $D^{0}$ mesons
are predicted with the parameter-optimized model,
and shown as a function of $\pt$ in central ($0-10\%$) Pb--Pb collisions at $\snn=5.02$ TeV.
Comparing with prompt $D^{0}$ $\raa$ (solid curve), at moderate $\pt$ ($\pt\sim5-7$),
a less suppression behavior is found for $D^{0}$ from $B$-hadron decays (dashed curve),
reflecting a weaker in-medium energy loss effect of bottom quark ($m_{b}=4.75$ GeV),
which has larger mass with respect to that of charm quark ($m_{c}=1.5$ GeV).
We note that the future measurements performed for the non-prompt $D^{0}$ $\raa$ and $\vtwo$,
are powerful in nailing down the varying ranges of the model parameters (see Sec.~\ref{sec:ParamOpt}),
which will largely improve and extend the current understanding of the in-medium effects.

\section{Summary}\label{sec:summary}
In this work we have used a soft-hard factorized model to
investigate the heavy quark momentum diffusion coefficients $\kappa_{T/L}$ in the quark-gluon plasma in a data-driven approach.
In particular we've examined the validity of this scenario by systematically scanning a wide range of possibilities.
The global $\chi^2$ analysis using an extensive set of LHC data on charm meson $\raa$ and $\vtwo$
has allowed us to constrain the preferred range of the two parameters:
soft-hard intermediate cutoff $|t^{\ast}|=1.5m_{D}^{2}$ and the scale of QCD coupling constant $\mu=\pi T$.
It is found that $\kappa_{T/L}$ have a mild sensitivity to $t^{\ast}$,
supporting the validity of the soft-hard approach when the coupling is not small.
With this factorization model, we have calculated the transport coefficient $\hat{q}$,
drag coefficient $\eta_{D}$, spatial diffusion coefficient $2\pi TD_{s}$,
and then compared with other theoretical calculations and phenomenological extractions.
Our analysis suggests that a small value $2\pi TD_{s} \simeq 6$ appears to be much preferred near $T_c$.
Finally we've demonstrated a simultaneous description of charm meson $\raa$ and $\vtwo$ observables in the range $\pt\le8$ GeV.
We've further made predictions for bottom meson observables in the same model.

We end with discussions on a few important caveats in the present study that call for future studies:
\begin{itemize}
\item[$\bullet$]
In the framework of Langevin approach,
the heavy flavor dynamics is encoded into three coefficient, $\kappa_{T}$, $\kappa_{L}$ and $\eta_{D}$,
satisfying Einstein's relationship.
It means that two of them are independent while the third one can be obtained accordingly.
The final results therefore depend on the arbitrary choice of which of the two
coefficients are calculated with the employed model~\cite{RalfSummary16,XuCoefficient18}.
For consistency, we calculate independently the momentum diffusion coefficients $\kappa_{T/L}$ with the factorization approach,
and obtain the drag coefficient via $\eta_{D}=\eta_{D}(\kappa_{T}, \kappa_{L})$ (see Eq.~\ref{eq:EtaD_DissFluc}).
A systematic study among the different options may help remedy this situation.
\item[$\bullet$]
According to the present modeling, the resulting spatial diffusion coefficient
exhibits a relatively strong increase of temperature comparing with lattice QCD calculations,
in particular at large $T$, as shown in Tab.~\ref{tab:2PiTDs_Cmp}.
It corresponds to a larger relaxation time and thus weaker HQ-medium coupling strength.
This comparison can be improved by
(1) determining the key parameters based on the upcoming measurements
on high precision observables (such as $\raa$, $\vtwo$ and $\vthree$) of both $D$ and $B$-mesons at low $\pt$;
(2) replacing the current $\chi^2$ analysis with a state-of-the-art deep learning technology.

\begin{table}[!htbp]
\centering
\begin{tabular}{c|c|c|c}
\hline
\multicolumn{1}{c}{\multirow{1}{*}{ }}
 & \multicolumn{1}{c}{\multirow{1}{*}{\centering $T=T_{c}$}}
 & \multicolumn{1}{c}{\multirow{1}{*}{\centering $T=2T_{c}$}}
 & \multicolumn{1}{c}{\multirow{1}{*}{\centering $T=3T_{c}$}}
  \\
\hline
\multicolumn{1}{c}{\centering This work}
 & \multicolumn{1}{c}{\centering $\sim$6}
 & \multicolumn{1}{c}{\centering $\sim$22}
 & \multicolumn{1}{c}{\centering $\sim$31}
 \\
\hline
\multicolumn{1}{c}{\centering LQCD (median value)~\cite{LQCDbanerjee12, LQCDBrambilla20}}
 & \multicolumn{1}{c}{\centering $\sim$5}
 & \multicolumn{1}{c}{\centering $\sim$10}
 & \multicolumn{1}{c}{\centering $\sim$13}
 \\
\hline
\end{tabular}
\caption{Summary of the different models for $2\pi TD_{s}$ at desired temperature values.}
\label{tab:2PiTDs_Cmp}
\end{table}

\item[$\bullet$]
It is realized~\cite{Cao:2019iqs} that the heavy quark hadrochemistry, the abundance of various heavy flavor hadrons,
provides special sensitivity to the heavy-light coalescence mechanism
and thus plays an important role to understand the observables like the baryon production and the baryon-to-meson ratio.
A systematic comparison including the charmed baryons over a broad momentum region is therefore
crucial for a better constraining of the model parameters,
as well as a better extraction of the heavy quark transport coefficients
in a model-to-data approach.
\item[$\bullet$]
The elastic scattering ($2\rightarrow2$) processes between heavy quark and QGP constituents are dominated for heavy quark with low to moderate transverse momentum~\cite{CTGUHybrid3}.
Thus, here we consider only the elastic energy loss mechanisms to study the observables at $\pt\le8$ GeV.
The missing radiative ($2\leftrightarrow3$) effects may help to reduce the discrepancy with $\raa$ data
in the vicinity of $\pt=8$ GeV, as mentioned above (see Fig.~\ref{fig:RAA_PbPb5020_C0}).
With the soft-hard factorized approach,
it would be interesting to include both elastic and radiative contributions
in a simultaneous best fit to data in the whole $\pt$ region.
We also plan to explore this idea in the future.

\end{itemize}

\begin{acknowledgements}
The authors are grateful to Andrea Beraudo and Jinfeng Liao for helpful discussions and communications.
We also thank Weiyao Ke and Gabriele Coci for providing the inputs as shown in Fig.~\ref{fig:Qhat_p10} and \ref{fig:2PiTDs_C}.
This work is supported by the Hubei Provincial Natural Science Foundation under Grant No.2020CFB163,
the National Science Foundation of China (NSFC) under Grant Nos.12005114, 11847014 and 11875178,
and the Key Laboratory of Quark and Lepton Physics Contracts Nos.QLPL2018P01 and QLPL201905.
\end{acknowledgements}

\appendix

\section{Derivation of the interaction rate and momentum diffusion coefficients in soft collisions}\label{app:appendixA}
In the QED case, $\mu+\gamma\rightarrow \mu+\gamma$,
one can calculate relevant interaction rate with small momentum transfer
by using the imaginary part of the muon self-energy $\Sigma(p_{1})$~\cite{WeldonPRD83}
\begin{equation}
\begin{aligned}\label{eq:Gamma_Soft1}
{\Gamma}(E_{1},T) &= -\frac{1}{2E_{1}} \bar{n}_{F}(E_{1}) {\rm Tr}\bigr[ ({p\! \! \! /}_{1} + m_{1}){\rm Im}\Sigma (p_{1}) \bigr].
\end{aligned}
\end{equation}
where, $p_{1}=(E_{1},\vec{p}_{1})$ and $m_{1}$ are the four-momentum and mass of the injected muon, respectively
\footnote[4]{The notations for the injected muon are same with the ones for the injected HQ in the elastic process.}.
The trace term in Eq.~\ref{eq:Gamma_Soft1} was calculated with a resummed photon propagator,
which is very similar with the one in QCD
\footnote[5]{The structure of the QED HTL-propagator follows Eq.~\ref{eq:PropTL} but with opposite signs~\cite{HQSteph08QED}.}.
It is realized~\cite{HQQGPBraaten91QED, HQQGPBraaten91QCD} that, in the QCD case, the contributions to ${\Gamma}$
can be obtained from the corresponding QED calculations by simply substitution: $e^{2} \rightarrow C_{F}g^{2}$,
where, $e$ ($g$) is the QED (QCD) coupling constant and $C_{F}=4/3$ is the quark Casimir factor.
It yields~\cite{HQQGPBraaten91QED}
\begin{equation}
\begin{aligned}\label{eq:Trace_Soft1}
&{\rm Tr}\bigr[ ({p\! \! \! /}_{1} + m_{1}){\rm Im}\Sigma (p_{1}) \bigr] \\
&= -2C_{F}g^{2} (1+e^{-E_{1}/T}) \int_{q} \int d\omega \; \bar{n}_{B}(\omega) \frac{E_{1}^{2}}{E_{3}} \mathcal{A}\mathcal{B},
\end{aligned}
\end{equation}
with
\begin{equation}
\begin{aligned}\label{eq:Trace_Soft2}
&\mathcal{A} \equiv (1-\frac{\omega+\vec{v}_{1}\cdot\vec{q}}{2E_{1}}) \rho_{L}(\omega,q) + \bigr[ \vec{v}_{1}^{\;2}(1-(\hat{v}_{1}\cdot\hat{q})^{2}) \\
&\qquad - \frac{\omega-\vec{v}_{1}\cdot\vec{q}^{\;}}{E_{1}} \bigr] \rho_{T}(\omega,q)
\end{aligned}
\end{equation}
\begin{equation}
\begin{aligned}\label{eq:Trace_Soft3}
&\mathcal{B} \equiv \bar{n}_{F}(E_{3}) \delta(E_{1}-E_{3}-\omega) - n_{F}(E_{3}) \delta(E_{1}+E_{3}-\omega)
\end{aligned}
\end{equation}
where, $\vec{v}_{1}=\vec{p}_{1}/E_{1}$ denotes the HQ velocity;
$n_{B/F}(E)=(e^{E/T}\mp1)^{-1}$ indicates the thermal distributions for Bosons/Fermions
and $\bar{n}_{B/F}\equiv 1 \pm n_{B/F}$ accounts for the Bose-enhancement or Pauli-blocking effect.
Here we have used the short notation $\int_{q}\equiv \int\frac{d^{3}\vec{q}}{(2\pi)^{3}}$ for phase space integrals.
Taking $|\vec{p}_{1}|$ and $m_{1/3}=m_{Q}$ are both much greater than the underlying medium temperature, i.e. $|\vec{p}_{1}|,m_{1/3}\gg T$,
thus, $E_{1/3}\gg T$ and $n_{F}(E_{3})$ is exponentially suppressed and can be dropped.
Moreover, the first $\delta$ funciton in Eq.~\ref{eq:Trace_Soft3}
can be simplified since $E_{3}=\sqrt{(\vec{p}_{1}-\vec{q})^{2}+m_{1}^{2}}\approx E_{1}-\vec{v}_{1}\cdot\vec{q}$.
Concerning the second $\delta$ funciton, it cannot contribute for $\omega$
less than or on the order of $T$~\cite{HQQGPBraaten91QED}, which will be deleted in this work.
Finally, Eq.~\ref{eq:Trace_Soft2} and~\ref{eq:Trace_Soft3} can be reduced to
\begin{subequations}
\begin{align}
&\mathcal{A} \equiv \rho_{L}(\omega,q) + \vec{v}_{1}^{\;2}\bigr[ 1-(\hat{v}_{1}\cdot\hat{q})^{2} \bigr] \rho_{T}(\omega,q) \label{eq:Trace_Soft4} \\
&\mathcal{B} \equiv \delta(\omega - \vec{v}_{1}\cdot\vec{q}) \label{eq:Trace_Soft5}.
\end{align}
\end{subequations}
By substituting Eq.~\ref{eq:Trace_Soft4} and~\ref{eq:Trace_Soft5} into Eq.~\ref{eq:Trace_Soft1}, one gets
\begin{equation}
\begin{aligned}\label{eq:Trace_Soft6}
&{\rm Tr}\bigr[ ({p\! \! \! /}_{1} + m_{1}){\rm Im}\Sigma (p_{1}) \bigr] \\
&= -2C_{F}g^{2}E_{1} \int_{q} \int d\omega \; \bar{n}_{B}(\omega) \delta(\omega - \vec{v}_{1}\cdot\vec{q}\;) \\
&\qquad \biggr\{ \rho_{L}(\omega,q) + \vec{v}_{1}^{\;2}\bigr[ 1-(\hat{v}_{1}\cdot\hat{q})^{2} \bigr] \rho_{T}(\omega,q) \biggr\}.
\end{aligned}
\end{equation}
Equation~\ref{eq:Gamma_Soft1} can be rewritten as
\begin{equation}
\begin{aligned}\label{eq:Trace_Soft7}
&{\Gamma}(E_{1},T) \\
&= C_{F}g^{2} \int_{q} \int d\omega \; \bar{n}_{B}(\omega) \delta(\omega-\vec{v}_{1}\cdot\vec{q}\;) \\
&\qquad \biggr\{ \rho_{L}(\omega,q) + \vec{v}_{1}^{\;2} \bigr[ 1-(\hat{v}_{1}\cdot\hat{q})^{2} \bigr]\rho_{T}(\omega,q) \biggr\} \\
\end{aligned}
\end{equation}
with the transverse and longitudinal spectral functions
are given by the imaginary part of the retarded propagator
\begin{equation}
\begin{aligned}\label{eq:GammaRhoTL1}
\rho_{T/L}(\omega,q) \equiv 2\cdot ImD_{T/L}^{R}(\omega,q).
\end{aligned}
\end{equation}

We note that, in the weak coupling limit, a consistent method is to use the HTL resummed propagators,
which is contributed by the quasiparticle poles and the Landau damping cuts~\cite{JeanPR02}.
In this analysis, we mainly focus on the low frequency excitation ($|\omega|<q$),
where the Landau damping is dominant and the quasiparticle excitation is irrelevant.
The resulting spectral function is therefore denoted by $\rho$ in Eq.~\ref{eq:GammaRhoTL1}.

The retarded propagator in Eq.\ref{eq:GammaRhoTL1} reads
\begin{equation}
\begin{aligned}\label{eq:RetardSpecFunc}
&D_{T/L}^{R}(\omega,q) \equiv {\Delta}_{T/L}(\omega+i\eta,q),
\end{aligned}
\end{equation}
which is defined by setting $q^{0}=\omega+i\eta$ $(\eta\rightarrow 0_{+})$, i.e. the real energy,
for the dressed gluon propagator ${\Delta}_{T/L}(q^{0},q)$~\cite{JeanPR02}
\begin{equation}
\begin{aligned}\label{eq:PropTL}
&{\Delta}_{T}(q^{0},q)=\frac{-1}{(q^{0})^{2}-q^{2}-{\Pi}_{T}(x)} \\
&{\Delta}_{L}(q^{0},q)=\frac{-1}{q^{2}+{\Pi}_{L}(x)}.
\end{aligned}
\end{equation}
The medium effects are embedded in the HTL gluon self-energy
\begin{equation}
\begin{aligned}\label{eq:PiTL}
&{\Pi}_{T}(x)=\frac{m_{D}^{2}}{2} \bigr[ x^{2}+(1-x^{2})Q(x) \bigr] \\
&{\Pi}_{L}(x)=m_{D}^{2} \bigr[ 1-Q(x) \bigr]
\end{aligned}
\end{equation}
where, $x=q^{0}/q$; $Q(x)$ is the Legendre polynomial of second kind
\begin{equation}
\begin{aligned}\label{eq:Leg2nd}
Q(x)=\frac{x}{2}ln\frac{x+1}{x-1}
\end{aligned}
\end{equation}
and $m_{D}^{2}$ is the Debye screening mass squared for gluon
\begin{equation}
\begin{aligned}\label{eq:DebMas}
m_{D}^{2}=g^{2} T^{2} \bigr( 1 + \frac{N_{f}}{6} \bigr).
\end{aligned}
\end{equation}
The coupling constant, $g$, is quantified by the two-loop QCD beta-function~\cite{TwoLoopGPRD05}
\begin{subequations}
\begin{align}
&g^{-2}(\mu)=2\beta_{0} ln(\frac{\mu}{\Lambda_{QCD}}) + \frac{\beta_{1}}{\beta_{0}} ln \bigr[ 2ln(\frac{\mu}{\Lambda_{QCD}}) \bigr] \label{eq:TwoLoopG} \\
&\qquad \beta_{0}=\frac{1}{16\pi^{2}}(11-\frac{2}{3}N_{f}) \label{eq:TwoLoopBeta0} \\
&\qquad\beta_{1}=\frac{1}{(16\pi^{2})^{2}}(102-\frac{38}{3}N_{f}) \label{eq:TwoLoopBeta1}
\end{align}
\end{subequations}
where, $\pi T \le\mu\le 3\pi T$ and $\Lambda_{QCD}=261~{\rm MeV}$.
$N_{f}$ is the number of active flavors in the QGP.
Finally, for space-like momentum, Eq.~\ref{eq:GammaRhoTL1} can be expressed as
\begin{equation}
\begin{aligned}\label{eq:GammaRhoT2}
\rho_{T}(\omega,q) = &\frac{\pi \omega m_{D}^{2}}{2q^{3}} (q^{2}-\omega^{2})
\biggr\{ \bigr[ q^{2}-\omega^{2} \\
& + \frac{\omega^{2}m_{D}^{2}}{2q^{2}} (1+\frac{q^{2}-\omega^{2}}{2\omega q} ln\frac{q+\omega}{q-\omega}) \bigr]^{2} \\
& + \bigr[ \frac{\pi \omega m_{D}^{2}}{4q^{3}} (q^{2}-\omega^{2}) \bigr]^{2} \biggr\}^{-1} \\
\end{aligned}
\end{equation}

\begin{equation}
\begin{aligned}\label{eq:GammaRhoL2}
\rho_{L}(\omega,q) = &\frac{\pi \omega m_{D}^{2}}{q}
\biggr\{ \bigr[ q^{2}+m_{D}^{2}(1-\frac{\omega}{2q} ln\frac{q+\omega}{q-\omega}) \bigr]^{2} \\
& + \bigr( \frac{\pi \omega m_{D}^{2}}{2q} \bigr)^{2} \biggr\}^{-1},
\end{aligned}
\end{equation}
with which Eq.~\ref{eq:Trace_Soft7} is computable.

The momentum diffusion coefficients can be calculated by subsituting
Eqs.~\ref{eq:Trace_Soft7}, \ref{eq:GammaRhoT2} and \ref{eq:GammaRhoL2}
back into Eq.~\ref{eq:KappaT} and \ref{eq:KappaL}, respectively,
and then performing the angular integral,
yielding\footnote[5]{
Note that the spectral functions (Eq.~\ref{eq:GammaRhoT2} and \ref{eq:GammaRhoL2}) are odd,
and the resulting $\rho_{T/L}(-\omega,q)=-\rho_{T/L}(\omega,q)$
are used to obtain Eq.~\ref{eq:KappaT1} and \ref{eq:KappaL1}.}
\begin{equation}
\begin{aligned}\label{eq:KappaT1}
\kappa_{T}(E_{1},T)=&\frac{C_{F}g^{2}}{8\pi^{2}v_{1}} \int_{0}^{q_{max}}dq \; q^{3}
\int_{0}^{v_{1}q}d\omega \; (1-\frac{\omega^{2}}{v_{1}^{2}q^{2}}) \\
&\bigr[ \rho_{L}(\omega,q) + (v_{1}^{2}-\frac{\omega^{2}}{q^{2}})\rho_{T}(\omega,q) \bigr] coth\frac{\omega}{2T}
\end{aligned}
\end{equation}
and
\begin{equation}
\begin{aligned}\label{eq:KappaL1}
\kappa_{L}(E_{1},T)=&\frac{C_{F}g^{2}}{4\pi^{2}v_{1}} \int_{0}^{q_{max}}dq \; q
\int_{0}^{v_{1}q}d\omega \; \frac{\omega^{2}}{v_{1}^{2}} \\
&\bigr[ \rho_{L}(\omega,q) + (v_{1}^{2}-\frac{\omega^{2}}{q^{2}})\rho_{T}(\omega,q) \bigr] coth\frac{\omega}{2T}
\end{aligned}
\end{equation}
with $v_{1}\equiv |\vec{v}_{1}|$.
The maximum momentum exchange is $q_{max}=\sqrt{4E_{1}T}$ in the high-energy limit~\cite{BjorkenFerlab82}.

Next, we implement the further calculations by performing a simple change of variables,
\begin{equation}
\begin{aligned}\label{eq:OmegaQ2TX}
&t=\omega^{2}-q^{2}<0  \quad \quad q^{2}=\frac{-t}{1-x^{2}} \\
&x=\frac{\omega}{q}<v_{1} ~~~\qquad \quad \omega=x\sqrt{\frac{-t}{1-x^{2}}},
\end{aligned}
\end{equation}
resulting in
\begin{equation}
\begin{aligned}\label{eq:Jacobin}
dtdx= \biggr| \frac{\partial(t,x)}{\partial(q, \omega)}  \biggr| dqd\omega=2(1-x^{2})dqd\omega.
\end{aligned}
\end{equation}
Using Eq.~\ref{eq:OmegaQ2TX} and \ref{eq:Jacobin},
one arrives at Eq.~\ref{eq:KappaT_Soft}-\ref{eq:GammaRhoL_Soft}.
Similar results can be found in Ref.~\cite{POWLANGEPJC11, POWLANG09}.

\section{Derivation of interaction rate and momentum diffusion coefficients in hard collisions}\label{app:appendixB}
For two-boday scatterings, the transition rate is defined as the rate of collisions with medium parton $i$,
which changes the momentum of the HQ (parton $i$) from $\vec{p}_{1}$ ($\vec{p}_{2}$)
to $\vec{p}_{3}=\vec{p}_{1}-\vec{q}$ ($\vec{p}_{4}=\vec{p}_{2}+\vec{q}$),
\begin{equation}
\begin{aligned}\label{eq:IndiTraRate}
\omega^{Qi}(\vec{p}_{1},\vec{q},T)=&\int_{p_{2}} n(E_{2}) \bar{n}(E_{3}) \bar{n}(E_{4}) v_{rel}d\sigma^{Qi}(\vec{p}_{1},\vec{p}_{2}\rightarrow \vec{p}_{3},\vec{p}_{4}).
\end{aligned}
\end{equation}
Typically, one can assume $\bar{n}(E_{3})=1$ by neglecting the thermal effects on the HQ after scattering.
The differential cross section summed over the spin/polarization and color of the final partons and averaged over those of incident partons,
\begin{equation}
\begin{aligned}\label{eq:DiffCroSec}
&v_{rel}d\sigma^{Qi}(\vec{p}_{1},\vec{p}_{2}\rightarrow \vec{p}_{3},\vec{p}_{4}) \\
&=\frac{1}{2E_{1}} \frac{1}{2E_{2}} \frac{d^{3}\vec{p}_{3}}{(2\pi)^{3}2E_{3}} \frac{d^{3}\vec{p}_{4}}{(2\pi)^{3}2E_{4}} \overline{|\mathcal{M}^{2}}|^{Qi} (2\pi)^{4} \\
&\qquad \delta^{(4)}(p_{1}+p_{2}-p_{3}-p_{4})
\end{aligned}
\end{equation}
where, $v_{rel}=(\sqrt{(p_{1}\cdot p_{2})^{2}-(m_{1}m_{2})^{2}})/(E_{1}E_{2})$
is the relative velocity between the projectile HQ and the target parton.
The interaction rate for a given elastic process reads
\begin{equation}
\begin{aligned}\label{eq:IndiGammaHard}
{\Gamma}^{Qi}(E_{1},T) = &\int d^{3}\vec{q}~\omega^{Qi}(\vec{p}_{1},\vec{q}, T) \\ 
& \hspace{-2.3em} \stackrel{(\ref{eq:IndiTraRate},\ref{eq:DiffCroSec})}{=} \frac{1}{2E_{1}} \int_{p_{2}} \frac{n(E_{2})}{2E_{2}}
\int_{p_{3}} \frac{1}{2E_{3}} \int_{p_{4}} \frac{\bar{n}(E_{4})}{2E_{4}} \\
&\quad \overline{|\mathcal{M}^{2}}|^{Qi} (2\pi)^{4} \delta^{(4)}(p_{1}+p_{2}-p_{3}-p_{4})
\end{aligned}
\end{equation}
The momentum diffusion coefficients can be obtained by inserting Eq.~\ref{eq:IndiGammaHard} into Eq.~\ref{eq:KappaT} and \ref{eq:KappaL}, yielding
\begin{equation}
\begin{aligned}\label{eq:KappaT3}
&\kappa^{Qi}_{T}(E_{1},T) \\
&= \frac{1}{2E_{1}} \int_{p_{2}} \frac{n(E_{2})}{2E_{2}} \int_{p_{3}} \frac{1}{2E_{3}} \int_{p_{4}} \frac{\bar{n}(E_{4})}{2E_{4}}
\frac{\vec{q}_{T}^{\;2}}{2} \\
&\quad \theta(|t|-|t^{\ast}|) \overline{|\mathcal{M}^{2}}|^{Qi} (2\pi)^{4} \delta^{(4)}(p_{1}+p_{2}-p_{3}-p_{4}) \\
&= \frac{1}{256\pi^{4}|\vec{p}_{1}|E_{1}} \int_{|\vec{p}_{2}|_{min}}^{\infty}d|\vec{p}_{2}| E_{2} n(E_{2})
\int_{-1}^{cos\psi|_{max}} d(cos\psi) \\
&\quad \int_{t_{min}}^{t^{\ast}}dt \int_{\omega_{min}}^{\omega_{max}}d\omega \frac{\bar{n}({\omega+E_{2}})}{\sqrt{G(\omega)}}
\bigr[ \omega^{2}-t-\frac{(2E_{1}\omega-t)^{2}}{4\vec{p}_{1}^{2}} \bigr] \overline{|\mathcal{M}^{2}}|^{Qi}
\end{aligned}
\end{equation}
and
\begin{equation}
\begin{aligned}\label{eq:KappaL3}
&\kappa^{Qi}_{L}(E_{1},T) \\
&= \frac{1}{2E_{1}} \int_{p_{2}} \frac{n(\vec{p}_{2})}{2E_{2}} \int_{p_{3}} \frac{1}{2E_{3}} \int_{p_{4}} \frac{\bar{n}(\vec{p}_{4})}{2E_{4}}
\bigr( \frac{2E_{1}\omega-t}{2|\vec{p}_{1}|} \bigr)^{2} \\
&\quad \theta(|t|-|t^{\ast}|) \overline{|\mathcal{M}^{2}}|^{Qi} (2\pi)^{4} \delta^{(4)}(p_{1}+p_{2}-p_{3}-p_{4}) \\
&= \frac{1}{512\pi^{4}|\vec{p}_{1}|^{3}E_{1}} \int_{|\vec{p}_{2}|_{min}}^{\infty}d|\vec{p}_{2}| E_{2} n(E_{2})
\int_{-1}^{cos\psi|_{max}} d(cos\psi) \\
&\quad \int_{t_{min}}^{t^{\ast}} dt \int_{\omega_{min}}^{\omega_{max}}d\omega \frac{\bar{n}({\omega+E_{2}})}{\sqrt{G(\omega)}}
(2E_{1}\omega-t)^{2} \overline{|\mathcal{M}^{2}}|^{Qi}
\end{aligned}
\end{equation}
Note that,
\begin{itemize}
\item[(1)] for hard collisions the momentum exchange is constrained
by imposing $\theta(|t|-|t^{\ast}|)$ in the first equality of Eq.~\ref{eq:KappaT3} and~\ref{eq:KappaL3};
\item[(2)] the tree level matrix elements squared includes the contributions from the various channels,
which are given in Ref.~\cite{Combridge79}:
           \begin{itemize}
                   \item[$\bullet$] for $Q+q\rightarrow Q+q$ ($t$-channel only)
                           \begin{equation}
                                   \begin{aligned}\label{eq:MatrixHQq}
                                           &\overline{|\mathcal{M}^{2}}|^{Qq}(s,t) \\ &= 2N_{f} \cdot 2N_{c} \cdot g^{4} \frac{4}{9}
                                           \frac{(m_{1}^{2}-u)^{2}+(s-m_{1}^{2})^{2}+2m_{1}^{2}t}{t^{2}}
                                   \end{aligned}
                           \end{equation}
                   \item[$\bullet$] for $Q+g\rightarrow Q+g$ ($t$, $s$ and $u$-channel combined)
                           \begin{equation}
                                   \begin{aligned}\label{eq:MatrixHQg}
                                           &\overline{|\mathcal{M}^{2}}|^{Qg}(s,t) \\ &= 2(N_{c}^{2}-1) g^{4} \biggr[ 2\frac{(s-m_{1}^{2})(m_{1}^{2}-u)}{t^{2}} \\
					   &\quad + \frac{4}{9}\frac{(s-m_{1}^{2})(m_{1}^{2}-u) + 2m_{1}^{2}(s+m_{1}^{2})}{(s-m_{1}^{2})^{2}}  \\
					   &\quad + \frac{4}{9}\frac{(s-m_{1}^{2})(m_{1}^{2}-u) + 2m_{1}^{2}(m_{1}^{2}+u)}{(m_{1}^{2}-u)^{2}} \\
					   &\quad + \frac{1}{9}\frac{m_{1}^{2}(4m_{1}^{2}-t)}{(s-m_{1}^{2})(m_{1}^{2}-u)} + \frac{(s-m_{1}^{2})(m_{1}^{2}-u) + m_{1}^{2}(s-u)}{t(s-m_{1}^{2})} \\
					   &\quad - \frac{(s-m_{1}^{2})(m_{1}^{2}-u) - m_{1}^{2}(s-u)}{t(m_{1}^{2}-u)} \biggr]
                                   \end{aligned}
                           \end{equation}
           \end{itemize}
Concerning the degeneracy factors, in Eq.~\ref{eq:MatrixHQq}, $2N_{f}$ reflects the identical contribution from all light quark and anti-quark flavors,
and $2N_{c}$ indicates the summing, rather than averaging, over the helicities and colors of the incident light quark,
while in Eq.~\ref{eq:MatrixHQg}, the factor $2(N_{c}^{2}-1)$ denotes the summing over the polarization and colors of the incident gluon.
The running coupling constant takes $g(\mu)^{2}=4\pi\alpha_{s}(\mu)$,
which is given by Eq.~\ref{eq:TwoLoopG} with the scale $\mu=\sqrt{-t}$.
\item[(3)] with the help of $\delta$-function, we can reduce the integral in Eq.~\ref{eq:KappaT3}
and~\ref{eq:KappaL3} from 9-dimension (9D) to 4D in the numerical calculations,
by transforming the integration variables from $(\vec{p}_{2}, \vec{p}_{3}, \vec{p}_{4})$ to $(|\vec{p}_{2}|, cos\psi, t)$,
where $\psi$ is the polar angle of $\vec{p}_{2}$; it yields the resutls as shown in the second equality of Eq.~\ref{eq:KappaT3} and~\ref{eq:KappaL3};
the relevant limits of integration together with the additional notations are summarized below:
\begin{equation}
\begin{aligned}\label{eq:P2Min}
|\vec{p}_{2}|_{min}=\frac{ |t^{\ast}|+\sqrt{(t^{\ast})^{2} + 4m_{1}^{2} |t^{\ast}|} }{4(E_{1}+|\vec{p}_{1}|)}
\end{aligned}
\end{equation}

\begin{equation}
\begin{aligned}\label{eq:PhiMax}
cos\psi|_{max}=min \biggr\{ 1, \frac{E_{1}}{|\vec{p}_{1}|} - \frac{|t^{\ast}| + \sqrt{(t^{\ast})^{2} +
4m_{1}^{2} |t^{\ast}|}}{4|\vec{p}_{1}| \cdot |\vec{p}_{2}|} \biggr\}
\end{aligned}
\end{equation}

\begin{equation}
\begin{aligned}\label{eq:tMin}
t_{min}=-\frac{(s-m_{1}^{2})^{2}}{s}
\end{aligned}
\end{equation}

\begin{equation}
\begin{aligned}\label{eq:OmegaMaxMin}
\omega_{max/min}=\frac{b\pm \sqrt{D}}{2a^{2}}~with
\end{aligned}
\end{equation}

\begin{equation}
\begin{aligned}\label{eq:aVal}
\quad\quad a= \frac{s-m_{1}^{2}}{|\vec{p}_{1}|}
\end{aligned}
\end{equation}

\begin{equation}
\begin{aligned}\label{eq:bVal}
\quad\quad b=-\frac{2t}{\vec{p}_{1}^{\;2}} \bigr[ E_{1}(s-m_{1}^{2})-E_{2}(s+m_{1}^{2}) \bigr]
\end{aligned}
\end{equation}

\begin{equation}
\begin{aligned}\label{eq:cVal}
\quad\quad c=-\frac{t}{\vec{p}_{1}^{\;2}} \biggr\{ t\bigr[ (E_{1}+E_{2})^{2}-s \bigr] +
4\vec{p}_{1}^{\;2}\vec{p}_{2}^{\;2}sin^{2}\psi \biggr\}
\end{aligned}
\end{equation}

\begin{equation}
\begin{aligned}\label{eq:DVal}
\quad\quad D= b^{2}+4a^{2}c=-t \biggr[ ts + (s-m_{1}^{2})^{2} \biggr] \cdot \biggr( \frac{4E_{2}sin\psi}{|\vec{p}_{1}|} \biggr)^{2}
\end{aligned}
\end{equation}

\begin{equation}
\begin{aligned}\label{eq:GFun}
G(\omega)=-a^{2}\omega + b\omega + c
\end{aligned}
\end{equation}

\item[(4)] then, one can follow the procedure of Ref.~\cite{HQSteph08QED} for the analytical evaluation of $\omega$ integral.
The obtained results for $\kappa_{T}$ and $\kappa_{L}$ are shown in Eq.~\ref{eq:KappaT_Hard} and \ref{eq:KappaL_Hard}, respectively.

\end{itemize}
%
%
\bibliography{Shuang_p6}

\begin{thebibliography}{79}%
\makeatletter
\providecommand \@ifxundefined [1]{%
 \@ifx{#1\undefined}
}%
\providecommand \@ifnum [1]{%
 \ifnum #1\expandafter \@firstoftwo
 \else \expandafter \@secondoftwo
 \fi
}%
\providecommand \@ifx [1]{%
 \ifx #1\expandafter \@firstoftwo
 \else \expandafter \@secondoftwo
 \fi
}%
\providecommand \natexlab [1]{#1}%
\providecommand \enquote  [1]{``#1''}%
\providecommand \bibnamefont  [1]{#1}%
\providecommand \bibfnamefont [1]{#1}%
\providecommand \citenamefont [1]{#1}%
\providecommand \href@noop [0]{\@secondoftwo}%
\providecommand \href [0]{\begingroup \@sanitize@url \@href}%
\providecommand \@href[1]{\@@startlink{#1}\@@href}%
\providecommand \@@href[1]{\endgroup#1\@@endlink}%
\providecommand \@sanitize@url [0]{\catcode `\\12\catcode `\$12\catcode
  `\&12\catcode `\#12\catcode `\^12\catcode `\_12\catcode `\%12\relax}%
\providecommand \@@startlink[1]{}%
\providecommand \@@endlink[0]{}%
\providecommand \url  [0]{\begingroup\@sanitize@url \@url }%
\providecommand \@url [1]{\endgroup\@href {#1}{\urlprefix }}%
\providecommand \urlprefix  [0]{URL }%
\providecommand \Eprint [0]{\href }%
\providecommand \doibase [0]{http://dx.doi.org/}%
\providecommand \selectlanguage [0]{\@gobble}%
\providecommand \bibinfo  [0]{\@secondoftwo}%
\providecommand \bibfield  [0]{\@secondoftwo}%
\providecommand \translation [1]{[#1]}%
\providecommand \BibitemOpen [0]{}%
\providecommand \bibitemStop [0]{}%
\providecommand \bibitemNoStop [0]{.\EOS\space}%
\providecommand \EOS [0]{\spacefactor3000\relax}%
\providecommand \BibitemShut  [1]{\csname bibitem#1\endcsname}%
\let\auto@bib@innerbib\@empty
\bibitem [{\citenamefont {Bzdak}\ \emph {et~al.}(2020)\citenamefont {Bzdak},
  \citenamefont {Esumi}, \citenamefont {Koch}, \citenamefont {Liao},
  \citenamefont {Stephanov},\ and\ \citenamefont {Xu}}]{2019pkr}%
  \BibitemOpen
  \bibfield  {author} {\bibinfo {author} {\bibfnamefont {A.}~\bibnamefont
  {Bzdak}}, \bibinfo {author} {\bibfnamefont {S.}~\bibnamefont {Esumi}},
  \bibinfo {author} {\bibfnamefont {V.}~\bibnamefont {Koch}}, \bibinfo {author}
  {\bibfnamefont {J.}~\bibnamefont {Liao}}, \bibinfo {author} {\bibfnamefont
  {M.}~\bibnamefont {Stephanov}}, \ and\ \bibinfo {author} {\bibfnamefont
  {N.}~\bibnamefont {Xu}},\ }\href {\doibase 10.1016/j.physrep.2020.01.005}
  {\bibfield  {journal} {\bibinfo  {journal} {Phys. Rept.}\ }\textbf {\bibinfo
  {volume} {853}},\ \bibinfo {pages} {1} (\bibinfo {year} {2020})},\ \Eprint
  {http://arxiv.org/abs/1906.00936} {arXiv:1906.00936 [nucl-th]} \BibitemShut
  {NoStop}%
\bibitem [{\citenamefont {{E.~Shuryak}}(1980)}]{ShuryakPR80}%
  \BibitemOpen
  \bibfield  {author} {\bibinfo {author} {\bibnamefont {{E.~Shuryak}}},\ }\href
  {\doibase 10.1016/0370-1573(80)90105-2} {\bibfield  {journal} {\bibinfo
  {journal} {{Phys.~Rep.}}\ }\textbf {\bibinfo {volume} {61}},\ \bibinfo
  {pages} {71} (\bibinfo {year} {1980})}\BibitemShut {NoStop}%
\bibitem [{\citenamefont {Braun-Munzinger}\ and\ \citenamefont
  {Wambach}(2009)}]{2009zz}%
  \BibitemOpen
  \bibfield  {author} {\bibinfo {author} {\bibfnamefont {P.}~\bibnamefont
  {Braun-Munzinger}}\ and\ \bibinfo {author} {\bibfnamefont {J.}~\bibnamefont
  {Wambach}},\ }\href {\doibase 10.1103/RevModPhys.81.1031} {\bibfield
  {journal} {\bibinfo  {journal} {Rev. Mod. Phys.}\ }\textbf {\bibinfo {volume}
  {81}},\ \bibinfo {pages} {1031} (\bibinfo {year} {2009})},\ \Eprint
  {http://arxiv.org/abs/0801.4256} {arXiv:0801.4256 [hep-ph]} \BibitemShut
  {NoStop}%
\bibitem [{\citenamefont {Braun-Munzinger}\ \emph {et~al.}(2016)\citenamefont
  {Braun-Munzinger}, \citenamefont {Koch}, \citenamefont {Sch\"afer},\ and\
  \citenamefont {Stachel}}]{2015hba}%
  \BibitemOpen
  \bibfield  {author} {\bibinfo {author} {\bibfnamefont {P.}~\bibnamefont
  {Braun-Munzinger}}, \bibinfo {author} {\bibfnamefont {V.}~\bibnamefont
  {Koch}}, \bibinfo {author} {\bibfnamefont {T.}~\bibnamefont {Sch\"afer}}, \
  and\ \bibinfo {author} {\bibfnamefont {J.}~\bibnamefont {Stachel}},\ }\href
  {\doibase 10.1016/j.physrep.2015.12.003} {\bibfield  {journal} {\bibinfo
  {journal} {Phys. Rept.}\ }\textbf {\bibinfo {volume} {621}},\ \bibinfo
  {pages} {76} (\bibinfo {year} {2016})},\ \Eprint
  {http://arxiv.org/abs/1510.00442} {arXiv:1510.00442 [nucl-th]} \BibitemShut
  {NoStop}%
\bibitem [{\citenamefont {Gyulassy}\ and\ \citenamefont
  {L.~McLerran}(2005)}]{Gyulassy05}%
  \BibitemOpen
  \bibfield  {author} {\bibinfo {author} {\bibfnamefont {M.}~\bibnamefont
  {Gyulassy}}\ and\ \bibinfo {author} {\bibfnamefont {L.}~\bibnamefont
  {L.~McLerran}},\ }\href {\doibase 10.1016/j.nuclphysa.2004.10.034} {\bibfield
   {journal} {\bibinfo  {journal} {Nucl.~Phys.~A}\ }\textbf {\bibinfo {volume}
  {750}},\ \bibinfo {pages} {30} (\bibinfo {year} {2005})}\BibitemShut
  {NoStop}%
\bibitem [{\citenamefont {{E.~Shuryak}}(2005)}]{Shuryak05}%
  \BibitemOpen
  \bibfield  {author} {\bibinfo {author} {\bibnamefont {{E.~Shuryak}}},\ }\href
  {\doibase 10.1016/j.nuclphysa.2004.10.022} {\bibfield  {journal} {\bibinfo
  {journal} {Nucl.~Phys.~A}\ }\textbf {\bibinfo {volume} {750}},\ \bibinfo
  {pages} {64} (\bibinfo {year} {2005})}\BibitemShut {NoStop}%
\bibitem [{\citenamefont {{B.~Muller, J.~Schukraft, and
  B.~Wyslouch}}(2012)}]{Muller12}%
  \BibitemOpen
  \bibfield  {author} {\bibinfo {author} {\bibnamefont {{B.~Muller,
  J.~Schukraft, and B.~Wyslouch}}},\ }\href {\doibase
  10.1146/annurev-nucl-102711-094910} {\bibfield  {journal} {\bibinfo
  {journal} {Ann.~Rev.~Nucl.~Part.~Sci.}\ }\textbf {\bibinfo {volume} {62}},\
  \bibinfo {pages} {361} (\bibinfo {year} {2012})}\BibitemShut {NoStop}%
\bibitem [{\citenamefont {Zhou}\ \emph {et~al.}(2014)\citenamefont {Zhou},
  \citenamefont {Xu}, \citenamefont {Xu},\ and\ \citenamefont
  {Zhuang}}]{Zhou:2014kka}%
  \BibitemOpen
  \bibfield  {author} {\bibinfo {author} {\bibfnamefont {K.}~\bibnamefont
  {Zhou}}, \bibinfo {author} {\bibfnamefont {N.}~\bibnamefont {Xu}}, \bibinfo
  {author} {\bibfnamefont {Z.}~\bibnamefont {Xu}}, \ and\ \bibinfo {author}
  {\bibfnamefont {P.}~\bibnamefont {Zhuang}},\ }\href {\doibase
  10.1103/PhysRevC.89.054911} {\bibfield  {journal} {\bibinfo  {journal} {Phys.
  Rev. C}\ }\textbf {\bibinfo {volume} {89}},\ \bibinfo {pages} {054911}
  (\bibinfo {year} {2014})},\ \Eprint {http://arxiv.org/abs/1401.5845}
  {arXiv:1401.5845 [nucl-th]} \BibitemShut {NoStop}%
\bibitem [{\citenamefont {Tang}\ \emph {et~al.}(2014)\citenamefont {Tang},
  \citenamefont {Xu}, \citenamefont {Zhou},\ and\ \citenamefont
  {Zhuang}}]{Tang:2014tga}%
  \BibitemOpen
  \bibfield  {author} {\bibinfo {author} {\bibfnamefont {Z.}~\bibnamefont
  {Tang}}, \bibinfo {author} {\bibfnamefont {N.}~\bibnamefont {Xu}}, \bibinfo
  {author} {\bibfnamefont {K.}~\bibnamefont {Zhou}}, \ and\ \bibinfo {author}
  {\bibfnamefont {P.}~\bibnamefont {Zhuang}},\ }\href {\doibase
  10.1088/0954-3899/41/12/124006} {\bibfield  {journal} {\bibinfo  {journal}
  {J. Phys. G}\ }\textbf {\bibinfo {volume} {41}},\ \bibinfo {pages} {124006}
  (\bibinfo {year} {2014})},\ \Eprint {http://arxiv.org/abs/1409.5559}
  {arXiv:1409.5559 [nucl-th]} \BibitemShut {NoStop}%
\bibitem [{\citenamefont {Andronic}\ \emph {et~al.}(2016)\citenamefont
  {Andronic} \emph {et~al.}}]{Andronic:2015wma}%
  \BibitemOpen
  \bibfield  {author} {\bibinfo {author} {\bibfnamefont {A.}~\bibnamefont
  {Andronic}} \emph {et~al.},\ }\href {\doibase 10.1140/epjc/s10052-015-3819-5}
  {\bibfield  {journal} {\bibinfo  {journal} {Eur. Phys. J. C}\ }\textbf
  {\bibinfo {volume} {76}},\ \bibinfo {pages} {107} (\bibinfo {year} {2016})},\
  \Eprint {http://arxiv.org/abs/1506.03981} {arXiv:1506.03981 [nucl-ex]}
  \BibitemShut {NoStop}%
\bibitem [{\citenamefont {{X.~Dong and V.~Greco}}(2019)}]{GrecoPIPNP19}%
  \BibitemOpen
  \bibfield  {author} {\bibinfo {author} {\bibnamefont {{X.~Dong and
  V.~Greco}}},\ }\href {\doibase 10.1016/j.ppnp.2018.08.001} {\bibfield
  {journal} {\bibinfo  {journal} {Prog. Part. Nucl. Phs.}\ }\textbf {\bibinfo
  {volume} {104}},\ \bibinfo {pages} {97} (\bibinfo {year} {2019})}\BibitemShut
  {NoStop}%
\bibitem [{\citenamefont {Dong}\ \emph {et~al.}(2019)\citenamefont {Dong},
  \citenamefont {Lee},\ and\ \citenamefont {Rapp}}]{Dong:2019byy}%
  \BibitemOpen
  \bibfield  {author} {\bibinfo {author} {\bibfnamefont {X.}~\bibnamefont
  {Dong}}, \bibinfo {author} {\bibfnamefont {Y.}~\bibnamefont {Lee}}, \ and\
  \bibinfo {author} {\bibfnamefont {R.}~\bibnamefont {Rapp}},\ }\href {\doibase
  10.1146/annurev-nucl-101918-023806} {\bibfield  {journal} {\bibinfo
  {journal} {Ann. Rev. Nucl. Part. Sci.}\ }\textbf {\bibinfo {volume} {69}},\
  \bibinfo {pages} {417} (\bibinfo {year} {2019})},\ \Eprint
  {http://arxiv.org/abs/1903.07709} {arXiv:1903.07709 [nucl-ex]} \BibitemShut
  {NoStop}%
\bibitem [{\citenamefont {Zhao}\ \emph {et~al.}(2020)\citenamefont {Zhao},
  \citenamefont {Zhou}, \citenamefont {Chen},\ and\ \citenamefont
  {Zhuang}}]{Zhao:2020jqu}%
  \BibitemOpen
  \bibfield  {author} {\bibinfo {author} {\bibfnamefont {J.}~\bibnamefont
  {Zhao}}, \bibinfo {author} {\bibfnamefont {K.}~\bibnamefont {Zhou}}, \bibinfo
  {author} {\bibfnamefont {S.}~\bibnamefont {Chen}}, \ and\ \bibinfo {author}
  {\bibfnamefont {P.}~\bibnamefont {Zhuang}},\ }\href {\doibase
  10.1016/j.ppnp.2020.103801} {\bibfield  {journal} {\bibinfo  {journal} {Prog.
  Part. Nucl. Phys.}\ }\textbf {\bibinfo {volume} {114}},\ \bibinfo {pages}
  {103801} (\bibinfo {year} {2020})},\ \Eprint
  {http://arxiv.org/abs/2005.08277} {arXiv:2005.08277 [nucl-th]} \BibitemShut
  {NoStop}%
\bibitem [{\citenamefont {Rapp}\ and\ \citenamefont {van
  Hees}(2010)}]{HQQGPRapp10}%
  \BibitemOpen
  \bibfield  {author} {\bibinfo {author} {\bibfnamefont {R.}~\bibnamefont
  {Rapp}}\ and\ \bibinfo {author} {\bibfnamefont {H.}~\bibnamefont {van
  Hees}},\ }in\ \href {\doibase 10.1142/9789814293297_0003} {\emph {\bibinfo
  {booktitle} {{Quark-Gluon Plasma 4 (World Sciencefic, Singapore, 2010)}}}}\
  (\bibinfo {year} {2010})\ pp.\ \bibinfo {pages} {111--206}\BibitemShut
  {NoStop}%
\bibitem [{\citenamefont {{F.~Prino and R.~Rapp}}(2016)}]{RalfSummary16}%
  \BibitemOpen
  \bibfield  {author} {\bibinfo {author} {\bibnamefont {{F.~Prino and
  R.~Rapp}}},\ }\href {\doibase 10.1088/0954-3899/43/9/093002} {\bibfield
  {journal} {\bibinfo  {journal} {J.~Phys.~G}\ }\textbf {\bibinfo {volume}
  {43}},\ \bibinfo {pages} {093002} (\bibinfo {year} {2016})}\BibitemShut
  {NoStop}%
\bibitem [{\citenamefont {{J.~F.~Liao and E.~Shuryak}}(2009)}]{JFLPRL09}%
  \BibitemOpen
  \bibfield  {author} {\bibinfo {author} {\bibnamefont {{J.~F.~Liao and
  E.~Shuryak}}},\ }\href {\doibase 10.1103/PhysRevLett.102.202302} {\bibfield
  {journal} {\bibinfo  {journal} {Phys.~Rev.~Lett.}\ }\textbf {\bibinfo
  {volume} {102}},\ \bibinfo {pages} {202302} (\bibinfo {year}
  {2009})}\BibitemShut {NoStop}%
\bibitem [{\citenamefont {{J.~C.~Xu, J.~F.~Liao and
  M.~Gyulassy}}(2015)}]{JFLCPL15}%
  \BibitemOpen
  \bibfield  {author} {\bibinfo {author} {\bibnamefont {{J.~C.~Xu, J.~F.~Liao
  and M.~Gyulassy}}},\ }\href {\doibase 10.1088/0256-307X/32/9/092501}
  {\bibfield  {journal} {\bibinfo  {journal} {Chin.~Phys.~Lett.}\ }\textbf
  {\bibinfo {volume} {32}},\ \bibinfo {pages} {092501} (\bibinfo {year}
  {2015})}\BibitemShut {NoStop}%
\bibitem [{\citenamefont {{J.~C.~Xu, J.~F.~Liao and
  M.~Gyulassy}}(2016)}]{CUJET3JHEP16}%
  \BibitemOpen
  \bibfield  {author} {\bibinfo {author} {\bibnamefont {{J.~C.~Xu, J.~F.~Liao
  and M.~Gyulassy}}},\ }\href {\doibase 10.1007/JHEP02(2016)169} {\bibfield
  {journal} {\bibinfo  {journal} {Journal of High Energy Physics}\ }\textbf
  {\bibinfo {volume} {1602}},\ \bibinfo {pages} {169} (\bibinfo {year}
  {2016})}\BibitemShut {NoStop}%
\bibitem [{\citenamefont {{JET Collaboration}}(2014)}]{JETCoef14}%
  \BibitemOpen
  \bibfield  {author} {\bibinfo {author} {\bibnamefont {{JET Collaboration}}},\
  }\href {\doibase 10.1103/PhysRevC.90.014909} {\bibfield  {journal} {\bibinfo
  {journal} {Phys.~Rev.~C}\ }\textbf {\bibinfo {volume} {90}},\ \bibinfo
  {pages} {014909} (\bibinfo {year} {2014})}\BibitemShut {NoStop}%
\bibitem [{\citenamefont {{S.~Z.~Shi, J.~F.~Liao, and
  M.~Gyulassy}}(2018)}]{CUJET3CPC18}%
  \BibitemOpen
  \bibfield  {author} {\bibinfo {author} {\bibnamefont {{S.~Z.~Shi, J.~F.~Liao,
  and M.~Gyulassy}}},\ }\href {\doibase 10.1088/1674-1137/42/10/104104}
  {\bibfield  {journal} {\bibinfo  {journal} {Chin.~Phys.~C}\ }\textbf
  {\bibinfo {volume} {42}},\ \bibinfo {pages} {104104} (\bibinfo {year}
  {2018})}\BibitemShut {NoStop}%
\bibitem [{\citenamefont {{S.~Z.~Shi, J.~F.~Liao, and
  M.~Gyulassy}}(2019)}]{CUJET3Arxiv18}%
  \BibitemOpen
  \bibfield  {author} {\bibinfo {author} {\bibnamefont {{S.~Z.~Shi, J.~F.~Liao,
  and M.~Gyulassy}}},\ }\href {\doibase 10.1088/1674-1137/43/4/044101}
  {\bibfield  {journal} {\bibinfo  {journal} {Chin.~Phys.~C}\ }\textbf
  {\bibinfo {volume} {43}},\ \bibinfo {pages} {044101} (\bibinfo {year}
  {2019})}\BibitemShut {NoStop}%
\bibitem [{\citenamefont {Ramamurti}\ and\ \citenamefont
  {Shuryak}(2018)}]{Ramamurti:2017zjn}%
  \BibitemOpen
  \bibfield  {author} {\bibinfo {author} {\bibfnamefont {A.}~\bibnamefont
  {Ramamurti}}\ and\ \bibinfo {author} {\bibfnamefont {E.}~\bibnamefont
  {Shuryak}},\ }\href {\doibase 10.1103/PhysRevD.97.016010} {\bibfield
  {journal} {\bibinfo  {journal} {Phys. Rev. D}\ }\textbf {\bibinfo {volume}
  {97}},\ \bibinfo {pages} {016010} (\bibinfo {year} {2018})},\ \Eprint
  {http://arxiv.org/abs/1708.04254} {arXiv:1708.04254 [hep-ph]} \BibitemShut
  {NoStop}%
\bibitem [{\citenamefont {{J.~E.~Bernhard, J.~S.~Moreland and
  S.~A.~Bass}}(2019)}]{BassNaturePhy19}%
  \BibitemOpen
  \bibfield  {author} {\bibinfo {author} {\bibnamefont {{J.~E.~Bernhard,
  J.~S.~Moreland and S.~A.~Bass}}},\ }\href {\doibase
  https://doi.org/10.1038/s41567-019-0611-8} {\bibfield  {journal} {\bibinfo
  {journal} {Nat.~Phys.}\ }\textbf {\bibinfo {volume} {15}},\ \bibinfo {pages}
  {1113} (\bibinfo {year} {2019})}\BibitemShut {NoStop}%
\bibitem [{\citenamefont {{S.~S.~Cao, G.~Y.~Qin, and
  S.~A.~Bass}}(2015)}]{CaoPRC15}%
  \BibitemOpen
  \bibfield  {author} {\bibinfo {author} {\bibnamefont {{S.~S.~Cao, G.~Y.~Qin,
  and S.~A.~Bass}}},\ }\href {\doibase 10.1103/PhysRevC.92.024907} {\bibfield
  {journal} {\bibinfo  {journal} {Phys.~Rev.~C}\ }\textbf {\bibinfo {volume}
  {92}},\ \bibinfo {pages} {024907} (\bibinfo {year} {2015})}\BibitemShut
  {NoStop}%
\bibitem [{\citenamefont {Cao}\ \emph {et~al.}(2016)\citenamefont {Cao},
  \citenamefont {Luo}, \citenamefont {Qin},\ and\ \citenamefont
  {Wang}}]{LBTPRC16}%
  \BibitemOpen
  \bibfield  {author} {\bibinfo {author} {\bibfnamefont {S.}~\bibnamefont
  {Cao}}, \bibinfo {author} {\bibfnamefont {T.}~\bibnamefont {Luo}}, \bibinfo
  {author} {\bibfnamefont {G.-Y.}\ \bibnamefont {Qin}}, \ and\ \bibinfo
  {author} {\bibfnamefont {X.-N.}\ \bibnamefont {Wang}},\ }\href {\doibase
  10.1103/PhysRevC.94.014909} {\bibfield  {journal} {\bibinfo  {journal} {Phys.
  Rev. C}\ }\textbf {\bibinfo {volume} {94}},\ \bibinfo {pages} {014909}
  (\bibinfo {year} {2016})},\ \Eprint {http://arxiv.org/abs/1605.06447}
  {arXiv:1605.06447 [nucl-th]} \BibitemShut {NoStop}%
\bibitem [{\citenamefont {{M.~Greif, F.~Senzel, H.~Kremer, K.~Zhou, C.~Greiner,
  and Z.~Xu}}(2017)}]{XuPRC17}%
  \BibitemOpen
  \bibfield  {author} {\bibinfo {author} {\bibnamefont {{M.~Greif, F.~Senzel,
  H.~Kremer, K.~Zhou, C.~Greiner, and Z.~Xu}}},\ }\href {\doibase
  10.1103/PhysRevC.95.054903} {\bibfield  {journal} {\bibinfo  {journal}
  {Phys.~Rev.~C}\ }\textbf {\bibinfo {volume} {95}},\ \bibinfo {pages} {054903}
  (\bibinfo {year} {2017})}\BibitemShut {NoStop}%
\bibitem [{\citenamefont {{S.~Li, C.~W.~Wang, X.~B.~Yuan, and
  S.~Q.~Feng}}(2018)}]{CTGUHybrid1}%
  \BibitemOpen
  \bibfield  {author} {\bibinfo {author} {\bibnamefont {{S.~Li, C.~W.~Wang,
  X.~B.~Yuan, and S.~Q.~Feng}}},\ }\href {\doibase 10.1103/PhysRevC.98.014909}
  {\bibfield  {journal} {\bibinfo  {journal} {Phys.~Rev.~C}\ }\textbf {\bibinfo
  {volume} {98}},\ \bibinfo {pages} {014909} (\bibinfo {year}
  {2018})}\BibitemShut {NoStop}%
\bibitem [{\citenamefont {{S.~Li and C.~W.~Wang}}(2018)}]{CTGUHybrid2}%
  \BibitemOpen
  \bibfield  {author} {\bibinfo {author} {\bibnamefont {{S.~Li and
  C.~W.~Wang}}},\ }\href {\doibase 10.1103/PhysRevC.98.034914} {\bibfield
  {journal} {\bibinfo  {journal} {Phys.~Rev.~C}\ }\textbf {\bibinfo {volume}
  {98}},\ \bibinfo {pages} {034914} (\bibinfo {year} {2018})}\BibitemShut
  {NoStop}%
\bibitem [{\citenamefont {Prado}\ \emph {et~al.}(2020)\citenamefont {Prado},
  \citenamefont {Xing}, \citenamefont {Cao}, \citenamefont {Qin},\ and\
  \citenamefont {Wang}}]{Prado:2019ste}%
  \BibitemOpen
  \bibfield  {author} {\bibinfo {author} {\bibfnamefont {C.~A.~G.}\
  \bibnamefont {Prado}}, \bibinfo {author} {\bibfnamefont {W.-J.}\ \bibnamefont
  {Xing}}, \bibinfo {author} {\bibfnamefont {S.}~\bibnamefont {Cao}}, \bibinfo
  {author} {\bibfnamefont {G.-Y.}\ \bibnamefont {Qin}}, \ and\ \bibinfo
  {author} {\bibfnamefont {X.-N.}\ \bibnamefont {Wang}},\ }\href {\doibase
  10.1103/PhysRevC.101.064907} {\bibfield  {journal} {\bibinfo  {journal}
  {Phys. Rev. C}\ }\textbf {\bibinfo {volume} {101}},\ \bibinfo {pages}
  {064907} (\bibinfo {year} {2020})},\ \Eprint
  {http://arxiv.org/abs/1911.06527} {arXiv:1911.06527 [nucl-th]} \BibitemShut
  {NoStop}%
\bibitem [{\citenamefont {Wang}\ \emph {et~al.}(2020)\citenamefont {Wang},
  \citenamefont {Dai}, \citenamefont {Zhang},\ and\ \citenamefont
  {Wang}}]{Wang:2020ukj}%
  \BibitemOpen
  \bibfield  {author} {\bibinfo {author} {\bibfnamefont {S.}~\bibnamefont
  {Wang}}, \bibinfo {author} {\bibfnamefont {W.}~\bibnamefont {Dai}}, \bibinfo
  {author} {\bibfnamefont {B.-W.}\ \bibnamefont {Zhang}}, \ and\ \bibinfo
  {author} {\bibfnamefont {E.}~\bibnamefont {Wang}},\ }\href {\doibase
  10.1088/1674-1137/abf4f5} {\  (\bibinfo {year} {2020}),\
  10.1088/1674-1137/abf4f5},\ \Eprint {http://arxiv.org/abs/2012.13935}
  {arXiv:2012.13935 [nucl-th]} \BibitemShut {NoStop}%
\bibitem [{\citenamefont {{M.~He, R.~J.~Fries, and
  R.~Rapp}}(2013)}]{HFModelHee13}%
  \BibitemOpen
  \bibfield  {author} {\bibinfo {author} {\bibnamefont {{M.~He, R.~J.~Fries,
  and R.~Rapp}}},\ }\href {\doibase 10.1103/PhysRevLett.110.112301} {\bibfield
  {journal} {\bibinfo  {journal} {Phys.~Rev.~Lett.}\ }\textbf {\bibinfo
  {volume} {110}},\ \bibinfo {pages} {112301} (\bibinfo {year}
  {2013})}\BibitemShut {NoStop}%
\bibitem [{\citenamefont {{T.~Song, H.~Berrehrah, D.~Cabrera, W.~Cassing, and
  E.~Bratkovskaya}}(2016)}]{PHSDPRC16}%
  \BibitemOpen
  \bibfield  {author} {\bibinfo {author} {\bibnamefont {{T.~Song, H.~Berrehrah,
  D.~Cabrera, W.~Cassing, and E.~Bratkovskaya}}},\ }\href {\doibase
  10.1103/PhysRevC.93.034906} {\bibfield  {journal} {\bibinfo  {journal}
  {Phys.~Rev.~C}\ }\textbf {\bibinfo {volume} {93}},\ \bibinfo {pages} {034906}
  (\bibinfo {year} {2016})}\BibitemShut {NoStop}%
\bibitem [{\citenamefont {{W.~M.~Alberico, A.~Beraudo, A.~De~Pace, A.~Molinari,
  M.~Monteno, M.~Nardi, and F.~Prino}}(2011)}]{POWLANGEPJC11}%
  \BibitemOpen
  \bibfield  {author} {\bibinfo {author} {\bibnamefont {{W.~M.~Alberico,
  A.~Beraudo, A.~De~Pace, A.~Molinari, M.~Monteno, M.~Nardi, and F.~Prino}}},\
  }\href {\doibase 10.1140/epjc/s10052-011-1666-6} {\bibfield  {journal}
  {\bibinfo  {journal} {Eur.~Phys.~J.~C}\ }\textbf {\bibinfo {volume} {71}},\
  \bibinfo {pages} {1666} (\bibinfo {year} {2011})}\BibitemShut {NoStop}%
\bibitem [{\citenamefont {{O.~Fochler, Z.~Xu, and
  C.~Greiner}}(2010)}]{BAMPS10}%
  \BibitemOpen
  \bibfield  {author} {\bibinfo {author} {\bibnamefont {{O.~Fochler, Z.~Xu, and
  C.~Greiner}}},\ }\href {\doibase 10.1103/PhysRevC.82.024907} {\bibfield
  {journal} {\bibinfo  {journal} {Phys.~Rev.~C}\ }\textbf {\bibinfo {volume}
  {82}},\ \bibinfo {pages} {024907} (\bibinfo {year} {2010})}\BibitemShut
  {NoStop}%
\bibitem [{\citenamefont {Gossiaux}(2019)}]{Gossiaux:2019mjc}%
  \BibitemOpen
  \bibfield  {author} {\bibinfo {author} {\bibfnamefont {P.~B.}\ \bibnamefont
  {Gossiaux}},\ }\href {\doibase 10.1016/j.nuclphysa.2018.10.076} {\bibfield
  {journal} {\bibinfo  {journal} {Nucl. Phys. A}\ }\textbf {\bibinfo {volume}
  {982}},\ \bibinfo {pages} {113} (\bibinfo {year} {2019})},\ \Eprint
  {http://arxiv.org/abs/1901.01606} {arXiv:1901.01606 [nucl-th]} \BibitemShut
  {NoStop}%
\bibitem [{\citenamefont {Cao}(2021)}]{Cao:2021ces}%
  \BibitemOpen
  \bibfield  {author} {\bibinfo {author} {\bibfnamefont {S.}~\bibnamefont
  {Cao}},\ }\href {\doibase 10.1016/j.nuclphysa.2020.121984} {\bibfield
  {journal} {\bibinfo  {journal} {Nucl. Phys. A}\ }\textbf {\bibinfo {volume}
  {1005}},\ \bibinfo {pages} {121984} (\bibinfo {year} {2021})}\BibitemShut
  {NoStop}%
\bibitem [{\citenamefont {Beraudo}\ \emph {et~al.}(2018)\citenamefont {Beraudo}
  \emph {et~al.}}]{Rapp:2018qla}%
  \BibitemOpen
  \bibfield  {author} {\bibinfo {author} {\bibfnamefont {A.}~\bibnamefont
  {Beraudo}} \emph {et~al.},\ }\href {\doibase 10.1016/j.nuclphysa.2018.09.002}
  {\bibfield  {journal} {\bibinfo  {journal} {Nucl. Phys. A}\ }\textbf
  {\bibinfo {volume} {979}},\ \bibinfo {pages} {21} (\bibinfo {year} {2018})},\
  \Eprint {http://arxiv.org/abs/1803.03824} {arXiv:1803.03824 [nucl-th]}
  \BibitemShut {NoStop}%
\bibitem [{\citenamefont {Xu}\ \emph {et~al.}(2019)\citenamefont {Xu} \emph
  {et~al.}}]{XuCoefficient18}%
  \BibitemOpen
  \bibfield  {author} {\bibinfo {author} {\bibfnamefont {Y.}~\bibnamefont {Xu}}
  \emph {et~al.},\ }\href {\doibase 10.1103/PhysRevC.99.014902} {\bibfield
  {journal} {\bibinfo  {journal} {Phys. Rev. C}\ }\textbf {\bibinfo {volume}
  {99}},\ \bibinfo {pages} {014902} (\bibinfo {year} {2019})},\ \Eprint
  {http://arxiv.org/abs/1809.10734} {arXiv:1809.10734 [nucl-th]} \BibitemShut
  {NoStop}%
\bibitem [{\citenamefont {Cao}\ \emph {et~al.}(2019)\citenamefont {Cao} \emph
  {et~al.}}]{Cao:2018ews}%
  \BibitemOpen
  \bibfield  {author} {\bibinfo {author} {\bibfnamefont {S.}~\bibnamefont
  {Cao}} \emph {et~al.},\ }\href {\doibase 10.1103/PhysRevC.99.054907}
  {\bibfield  {journal} {\bibinfo  {journal} {Phys. Rev. C}\ }\textbf {\bibinfo
  {volume} {C99}},\ \bibinfo {pages} {054907} (\bibinfo {year} {2019})},\
  \Eprint {http://arxiv.org/abs/1809.07894} {arXiv:1809.07894 [nucl-th]}
  \BibitemShut {NoStop}%
\bibitem [{\citenamefont {{S.~Li and J.~F.~Liao}}(2020)}]{CTGUHybrid4}%
  \BibitemOpen
  \bibfield  {author} {\bibinfo {author} {\bibnamefont {{S.~Li and
  J.~F.~Liao}}},\ }\href {\doibase 10.1140/epjc/s10052-020-8243-9} {\bibfield
  {journal} {\bibinfo  {journal} {Eur. Phys. J. C}\ }\textbf {\bibinfo {volume}
  {80}},\ \bibinfo {pages} {671} (\bibinfo {year} {2020})}\BibitemShut
  {NoStop}%
\bibitem [{\citenamefont {{S.~Li, C.~W.~Wang, R.~Z.~Wan, and
  J.~F.~Liao}}(2019)}]{CTGUHybrid3}%
  \BibitemOpen
  \bibfield  {author} {\bibinfo {author} {\bibnamefont {{S.~Li, C.~W.~Wang,
  R.~Z.~Wan, and J.~F.~Liao}}},\ }\href {\doibase 10.1103/PhysRevC.99.054909}
  {\bibfield  {journal} {\bibinfo  {journal} {Phys.~Rev.~C}\ }\textbf {\bibinfo
  {volume} {99}},\ \bibinfo {pages} {054909} (\bibinfo {year}
  {2019})}\BibitemShut {NoStop}%
\bibitem [{\citenamefont {{I.~Karpenko, P.~Huoviven and
  M.~Bleicher}}(2014)}]{vhlle}%
  \BibitemOpen
  \bibfield  {author} {\bibinfo {author} {\bibnamefont {{I.~Karpenko,
  P.~Huoviven and M.~Bleicher}}},\ }\href {\doibase 10.1016/j.cpc.2014.07.010}
  {\bibfield  {journal} {\bibinfo  {journal} {Comput.~Phys.~Commun.}\ }\textbf
  {\bibinfo {volume} {185}},\ \bibinfo {pages} {3016} (\bibinfo {year}
  {2014})}\BibitemShut {NoStop}%
\bibitem [{\citenamefont {{E.~Braaten, K.~Cheung, and
  T.~C.~Yuan}}(1993)}]{FragBraaten93}%
  \BibitemOpen
  \bibfield  {author} {\bibinfo {author} {\bibnamefont {{E.~Braaten, K.~Cheung,
  and T.~C.~Yuan}}},\ }\href {\doibase 10.1103/PhysRevD.48.R5049} {\bibfield
  {journal} {\bibinfo  {journal} {Phys.~Rev.~D}\ }\textbf {\bibinfo {volume}
  {48}},\ \bibinfo {pages} {R5049} (\bibinfo {year} {1993})}\BibitemShut
  {NoStop}%
\bibitem [{\citenamefont {{M.~Cacciari, P.~Nason and
  R.~Vogt}}(2005)}]{FragFONLLPRL}%
  \BibitemOpen
  \bibfield  {author} {\bibinfo {author} {\bibnamefont {{M.~Cacciari, P.~Nason
  and R.~Vogt}}},\ }\href {\doibase 10.1103/PhysRevLett.95.122001} {\bibfield
  {journal} {\bibinfo  {journal} {Phys.~Rev.~Lett.}\ }\textbf {\bibinfo
  {volume} {95}},\ \bibinfo {pages} {122001} (\bibinfo {year}
  {2005})}\BibitemShut {NoStop}%
\bibitem [{\citenamefont {{C.~B.~Dover, U.~Heinz, E.~Schnedermann, and
  J.~Zim\'anyi}}(1991)}]{CoalOriginalDover91}%
  \BibitemOpen
  \bibfield  {author} {\bibinfo {author} {\bibnamefont {{C.~B.~Dover, U.~Heinz,
  E.~Schnedermann, and J.~Zim\'anyi}}},\ }\href {\doibase
  10.1103/PhysRevC.44.1636} {\bibfield  {journal} {\bibinfo  {journal}
  {Phys.~Rev.~C}\ }\textbf {\bibinfo {volume} {44}},\ \bibinfo {pages} {1636}
  (\bibinfo {year} {1991})}\BibitemShut {NoStop}%
\bibitem [{\citenamefont {{B.~Svetitsky}}(1988)}]{Benjamin88}%
  \BibitemOpen
  \bibfield  {author} {\bibinfo {author} {\bibnamefont {{B.~Svetitsky}}},\
  }\href {\doibase 10.1103/PhysRevD.37.2484} {\bibfield  {journal} {\bibinfo
  {journal} {Phys.~Rev.~D}\ }\textbf {\bibinfo {volume} {37}},\ \bibinfo
  {pages} {2484} (\bibinfo {year} {1988})}\BibitemShut {NoStop}%
\bibitem [{\citenamefont {{P.~B.~Gossiaux and J.~Aichelin}}(2008)}]{PBGPRC08}%
  \BibitemOpen
  \bibfield  {author} {\bibinfo {author} {\bibnamefont {{P.~B.~Gossiaux and
  J.~Aichelin}}},\ }\href {\doibase 10.1103/PhysRevC.78.014904} {\bibfield
  {journal} {\bibinfo  {journal} {Phys.~Rev.~C}\ }\textbf {\bibinfo {volume}
  {78}},\ \bibinfo {pages} {014904} (\bibinfo {year} {2008})}\BibitemShut
  {NoStop}%
\bibitem [{\citenamefont {{S.~Peign$\acute{\rm e}$ and
  A.~Peshier}}(2008{\natexlab{a}})}]{HQSteph08QED}%
  \BibitemOpen
  \bibfield  {author} {\bibinfo {author} {\bibnamefont {{S.~Peign$\acute{\rm
  e}$ and A.~Peshier}}},\ }\href {\doibase 10.1103/PhysRevD.77.014015}
  {\bibfield  {journal} {\bibinfo  {journal} {Phys.~Rev.~D}\ }\textbf {\bibinfo
  {volume} {77}},\ \bibinfo {pages} {014015} (\bibinfo {year}
  {2008}{\natexlab{a}})}\BibitemShut {NoStop}%
\bibitem [{\citenamefont {{S.~Peign$\acute{\rm e}$ and
  A.~Peshier}}(2008{\natexlab{b}})}]{HQSteph08QCD}%
  \BibitemOpen
  \bibfield  {author} {\bibinfo {author} {\bibnamefont {{S.~Peign$\acute{\rm
  e}$ and A.~Peshier}}},\ }\href {\doibase 10.1103/PhysRevD.77.114017}
  {\bibfield  {journal} {\bibinfo  {journal} {Phys.~Rev.~D}\ }\textbf {\bibinfo
  {volume} {77}},\ \bibinfo {pages} {114017} (\bibinfo {year}
  {2008}{\natexlab{b}})}\BibitemShut {NoStop}%
\bibitem [{\citenamefont {{E.~Braaten and T.~C.~Yuan}}(1991)}]{Braaten91PRL}%
  \BibitemOpen
  \bibfield  {author} {\bibinfo {author} {\bibnamefont {{E.~Braaten and
  T.~C.~Yuan}}},\ }\href {\doibase 10.1103/PhysRevLett.66.2183} {\bibfield
  {journal} {\bibinfo  {journal} {Phys.~Rev.~Lett.}\ }\textbf {\bibinfo
  {volume} {2183}},\ \bibinfo {pages} {66} (\bibinfo {year}
  {1991})}\BibitemShut {NoStop}%
\bibitem [{\citenamefont {{J.~P.~Blaizot and E.~Iancu}}(2002)}]{JeanPR02}%
  \BibitemOpen
  \bibfield  {author} {\bibinfo {author} {\bibnamefont {{J.~P.~Blaizot and
  E.~Iancu}}},\ }\href {\doibase 10.1016/S0370-1573(01)00061-8} {\bibfield
  {journal} {\bibinfo  {journal} {Phys.~Rep.}\ }\textbf {\bibinfo {volume}
  {359}},\ \bibinfo {pages} {355} (\bibinfo {year} {2002})}\BibitemShut
  {NoStop}%
\bibitem [{\citenamefont {{J.~Ghiglieri, G.~Moore, and
  D.~Teaney}}(2016)}]{JetMediumJHEP16}%
  \BibitemOpen
  \bibfield  {author} {\bibinfo {author} {\bibnamefont {{J.~Ghiglieri,
  G.~Moore, and D.~Teaney}}},\ }\href {\doibase 10.1007/JHEP03(2016)095}
  {\bibfield  {journal} {\bibinfo  {journal} {J. High Energy Phys.}\ }\textbf
  {\bibinfo {volume} {03}},\ \bibinfo {pages} {095} (\bibinfo {year}
  {2016})}\BibitemShut {NoStop}%
\bibitem [{\citenamefont {{S.~Li, W.~Xiong, and
  R.~Z.~Wan}}(2020)}]{CTGUHybrid5}%
  \BibitemOpen
  \bibfield  {author} {\bibinfo {author} {\bibnamefont {{S.~Li, W.~Xiong, and
  R.~Z.~Wan}}},\ }\href {\doibase 10.1140/epjc/s10052-020-08708-y} {\bibfield
  {journal} {\bibinfo  {journal} {Eur. Phys. J. C}\ }\textbf {\bibinfo {volume}
  {80}},\ \bibinfo {pages} {1113} (\bibinfo {year} {2020})}\BibitemShut
  {NoStop}%
\bibitem [{\citenamefont {{ALICE
  Collaboration}}(2016{\natexlab{a}})}]{ALICEDesonPbPb2760RAA}%
  \BibitemOpen
  \bibfield  {author} {\bibinfo {author} {\bibnamefont {{ALICE
  Collaboration}}},\ }\href {\doibase 10.1007/JHEP03(2016)081} {\bibfield
  {journal} {\bibinfo  {journal} {Journal of High Energy Physics}\ }\textbf
  {\bibinfo {volume} {03}},\ \bibinfo {pages} {081} (\bibinfo {year}
  {2016}{\natexlab{a}})}\BibitemShut {NoStop}%
\bibitem [{\citenamefont {{ALICE
  Collaboration}}(2016{\natexlab{b}})}]{ALICEDsPbPb2760RAA}%
  \BibitemOpen
  \bibfield  {author} {\bibinfo {author} {\bibnamefont {{ALICE
  Collaboration}}},\ }\href {\doibase 10.1007/JHEP03(2016)082} {\bibfield
  {journal} {\bibinfo  {journal} {Journal of High Energy Physics}\ }\textbf
  {\bibinfo {volume} {03}},\ \bibinfo {pages} {082} (\bibinfo {year}
  {2016}{\natexlab{b}})}\BibitemShut {NoStop}%
\bibitem [{\citenamefont {{ALICE
  Collaboration}}(2018{\natexlab{a}})}]{ALICEDesonPbPb5020RAA}%
  \BibitemOpen
  \bibfield  {author} {\bibinfo {author} {\bibnamefont {{ALICE
  Collaboration}}},\ }\href {\doibase 10.1007/JHEP10(2018)174} {\bibfield
  {journal} {\bibinfo  {journal} {Journal of High Energy Physics}\ }\textbf
  {\bibinfo {volume} {10}},\ \bibinfo {pages} {174} (\bibinfo {year}
  {2018}{\natexlab{a}})}\BibitemShut {NoStop}%
\bibitem [{\citenamefont {{ALICE Collaboration}}(2014)}]{ALICEDesonPbPb2760V2}%
  \BibitemOpen
  \bibfield  {author} {\bibinfo {author} {\bibnamefont {{ALICE
  Collaboration}}},\ }\href {\doibase 10.1103/PhysRevC.90.034904} {\bibfield
  {journal} {\bibinfo  {journal} {Phys.~Rev.~C}\ }\textbf {\bibinfo {volume}
  {90}},\ \bibinfo {pages} {034904} (\bibinfo {year} {2014})}\BibitemShut
  {NoStop}%
\bibitem [{\citenamefont {{ALICE
  Collaboration}}(2018{\natexlab{b}})}]{ALICEDesonPbPb5020V2}%
  \BibitemOpen
  \bibfield  {author} {\bibinfo {author} {\bibnamefont {{ALICE
  Collaboration}}},\ }\href {\doibase 10.1103/PhysRevLett.120.102301}
  {\bibfield  {journal} {\bibinfo  {journal} {Phys.~Rev.~Lett.}\ }\textbf
  {\bibinfo {volume} {120}},\ \bibinfo {pages} {102301} (\bibinfo {year}
  {2018}{\natexlab{b}})}\BibitemShut {NoStop}%
\bibitem [{\citenamefont {{CMS
  Collaboration}}(2018{\natexlab{a}})}]{CMSd05020V2}%
  \BibitemOpen
  \bibfield  {author} {\bibinfo {author} {\bibnamefont {{CMS Collaboration}}},\
  }\href {\doibase 10.1103/PhysRevLett.120.202301} {\bibfield  {journal}
  {\bibinfo  {journal} {Phys.~Rev.~Lett.}\ }\textbf {\bibinfo {volume} {120}},\
  \bibinfo {pages} {202301} (\bibinfo {year} {2018}{\natexlab{a}})}\BibitemShut
  {NoStop}%
\bibitem [{\citenamefont {{W.~Y.~Ke, Y.~R.~Xu, and
  S.~A.~Bass}}(2018)}]{Lido18}%
  \BibitemOpen
  \bibfield  {author} {\bibinfo {author} {\bibnamefont {{W.~Y.~Ke, Y.~R.~Xu,
  and S.~A.~Bass}}},\ }\href {\doibase 10.1103/PhysRevC.98.064901} {\bibfield
  {journal} {\bibinfo  {journal} {Phys.~Rev.~C}\ }\textbf {\bibinfo {volume}
  {98}},\ \bibinfo {pages} {064901} (\bibinfo {year} {2018})}\BibitemShut
  {NoStop}%
\bibitem [{\citenamefont {{F.~Scardina, S.~K.~Das, V.~Minissale, S.~Plumari,
  and V.~Greco}}(2017)}]{Catania_2PiTDs}%
  \BibitemOpen
  \bibfield  {author} {\bibinfo {author} {\bibnamefont {{F.~Scardina,
  S.~K.~Das, V.~Minissale, S.~Plumari, and V.~Greco}}},\ }\href {\doibase
  10.1103/PhysRevC.96.044905} {\bibfield  {journal} {\bibinfo  {journal}
  {Phys.~Rev.~C}\ }\textbf {\bibinfo {volume} {96}},\ \bibinfo {pages} {044905}
  (\bibinfo {year} {2017})}\BibitemShut {NoStop}%
\bibitem [{\citenamefont {{G.~Coci}}(2018)}]{Catania_Qhat}%
  \BibitemOpen
  \bibfield  {author} {\bibinfo {author} {\bibnamefont {{G.~Coci}}},\ }\href
  {http://www.infn.it/thesis/thesis_dettaglio.php?tid=13663} {\bibfield
  {journal} {\bibinfo  {journal} {PhD~Thesis}\ } (\bibinfo {year}
  {2018})}\BibitemShut {NoStop}%
\bibitem [{\citenamefont {{G.~D.~Moore and D.~Teaney}}(2005)}]{Moore04}%
  \BibitemOpen
  \bibfield  {author} {\bibinfo {author} {\bibnamefont {{G.~D.~Moore and
  D.~Teaney}}},\ }\href {\doibase 10.1103/PhysRevC.71.064904} {\bibfield
  {journal} {\bibinfo  {journal} {Phys.~Rev.~C}\ }\textbf {\bibinfo {volume}
  {71}},\ \bibinfo {pages} {064904} (\bibinfo {year} {2005})}\BibitemShut
  {NoStop}%
\bibitem [{\citenamefont {{D.~Banerjee, S.~Datta, R.~Gavai, and
  P.~Majumdar}}(2012)}]{LQCDbanerjee12}%
  \BibitemOpen
  \bibfield  {author} {\bibinfo {author} {\bibnamefont {{D.~Banerjee, S.~Datta,
  R.~Gavai, and P.~Majumdar}}},\ }\href {\doibase 10.1103/PhysRevD.85.014510}
  {\bibfield  {journal} {\bibinfo  {journal} {Phys.~Rev.~D}\ }\textbf {\bibinfo
  {volume} {85}},\ \bibinfo {pages} {014510} (\bibinfo {year}
  {2012})}\BibitemShut {NoStop}%
\bibitem [{\citenamefont {{H.~T.~Ding, A.~Francis, O.~Kaczmarek, F.~Karsch,
  H.~Satz, and W.~Soeldner}}(2012)}]{LQCDding12}%
  \BibitemOpen
  \bibfield  {author} {\bibinfo {author} {\bibnamefont {{H.~T.~Ding,
  A.~Francis, O.~Kaczmarek, F.~Karsch, H.~Satz, and W.~Soeldner}}},\ }\href
  {\doibase 10.1103/PhysRevD.86.014509} {\bibfield  {journal} {\bibinfo
  {journal} {Phys.~Rev.~D}\ }\textbf {\bibinfo {volume} {86}},\ \bibinfo
  {pages} {014509} (\bibinfo {year} {2012})}\BibitemShut {NoStop}%
\bibitem [{\citenamefont {{O.~Kaczmarek}}(2014)}]{LQCDolaf14}%
  \BibitemOpen
  \bibfield  {author} {\bibinfo {author} {\bibnamefont {{O.~Kaczmarek}}},\
  }\href {\doibase 10.1016/j.nuclphysa.2014.09.031} {\bibfield  {journal}
  {\bibinfo  {journal} {Nucl.~Phys.~A}\ }\textbf {\bibinfo {volume} {931}},\
  \bibinfo {pages} {633} (\bibinfo {year} {2014})}\BibitemShut {NoStop}%
\bibitem [{\citenamefont {{N.~Brambilla, V.~Leino, P.~Petreczky, and
  A.~Vairo}}(2020)}]{LQCDBrambilla20}%
  \BibitemOpen
  \bibfield  {author} {\bibinfo {author} {\bibnamefont {{N.~Brambilla,
  V.~Leino, P.~Petreczky, and A.~Vairo}}},\ }\href {\doibase
  10.1103/PhysRevD.102.074503} {\bibfield  {journal} {\bibinfo  {journal}
  {Phys.~Rev.~D}\ }\textbf {\bibinfo {volume} {102}},\ \bibinfo {pages}
  {074503} (\bibinfo {year} {2020})}\BibitemShut {NoStop}%
\bibitem [{\citenamefont {{L.~Altenkort, A.~M.~Eller, O.~Kaczmarek, L.~Mazur,
  G.~D.Moore, and H.~T. Shu}}(2020)}]{LQCDAltenkort20}%
  \BibitemOpen
  \bibfield  {author} {\bibinfo {author} {\bibnamefont {{L.~Altenkort,
  A.~M.~Eller, O.~Kaczmarek, L.~Mazur, G.~D.Moore, and H.~T. Shu}}},\ }\href
  {\doibase 10.1103/PhysRevD.103.014511} {\bibfield  {journal} {\bibinfo
  {journal} {Phys.~Rev.~D}\ }\textbf {\bibinfo {volume} {103}},\ \bibinfo
  {pages} {014511} (\bibinfo {year} {2020})}\BibitemShut {NoStop}%
\bibitem [{\citenamefont {{L.~Tolos and
  J.~M.~Torres-Rincon}}(2013)}]{2PiTDs4Dmeson}%
  \BibitemOpen
  \bibfield  {author} {\bibinfo {author} {\bibnamefont {{L.~Tolos and
  J.~M.~Torres-Rincon}}},\ }\href {\doibase 10.1103/PhysRevD.88.074019}
  {\bibfield  {journal} {\bibinfo  {journal} {Phys.~Rev.~D}\ }\textbf {\bibinfo
  {volume} {88}},\ \bibinfo {pages} {074019} (\bibinfo {year}
  {2013})}\BibitemShut {NoStop}%
\bibitem [{\citenamefont {Vertesi}(2019)}]{Vertesi:2019awk}%
  \BibitemOpen
  \bibfield  {author} {\bibinfo {author} {\bibfnamefont {R.}~\bibnamefont
  {Vertesi}} (\bibinfo {collaboration} {ALICE}),\ }in\ \href@noop {} {\emph
  {\bibinfo {booktitle} {{18th Conference on Elastic and Diffractive
  Scattering}}}}\ (\bibinfo {year} {2019})\ \Eprint
  {http://arxiv.org/abs/1910.01981} {arXiv:1910.01981 [nucl-ex]} \BibitemShut
  {NoStop}%
\bibitem [{\citenamefont {{CMS
  Collaboration}}(2018{\natexlab{b}})}]{CMSD0PbPb5020V2}%
  \BibitemOpen
  \bibfield  {author} {\bibinfo {author} {\bibnamefont {{CMS Collaboration}}},\
  }\href {\doibase 10.1103/PhysRevLett.120.202301} {\bibfield  {journal}
  {\bibinfo  {journal} {Phys.~Rev.~Lett.}\ }\textbf {\bibinfo {volume} {120}},\
  \bibinfo {pages} {202301} (\bibinfo {year} {2018}{\natexlab{b}})}\BibitemShut
  {NoStop}%
\bibitem [{\citenamefont {Cao}\ \emph {et~al.}(2020)\citenamefont {Cao},
  \citenamefont {Sun}, \citenamefont {Li}, \citenamefont {Liu}, \citenamefont
  {Xing}, \citenamefont {Qin},\ and\ \citenamefont {Ko}}]{Cao:2019iqs}%
  \BibitemOpen
  \bibfield  {author} {\bibinfo {author} {\bibfnamefont {S.}~\bibnamefont
  {Cao}}, \bibinfo {author} {\bibfnamefont {K.-J.}\ \bibnamefont {Sun}},
  \bibinfo {author} {\bibfnamefont {S.-Q.}\ \bibnamefont {Li}}, \bibinfo
  {author} {\bibfnamefont {S.~Y.~F.}\ \bibnamefont {Liu}}, \bibinfo {author}
  {\bibfnamefont {W.-J.}\ \bibnamefont {Xing}}, \bibinfo {author}
  {\bibfnamefont {G.-Y.}\ \bibnamefont {Qin}}, \ and\ \bibinfo {author}
  {\bibfnamefont {C.~M.}\ \bibnamefont {Ko}},\ }\href {\doibase
  10.1016/j.physletb.2020.135561} {\bibfield  {journal} {\bibinfo  {journal}
  {Phys. Lett. B}\ }\textbf {\bibinfo {volume} {807}},\ \bibinfo {pages}
  {135561} (\bibinfo {year} {2020})},\ \Eprint
  {http://arxiv.org/abs/1911.00456} {arXiv:1911.00456 [nucl-th]} \BibitemShut
  {NoStop}%
\bibitem [{\citenamefont {{H.~A.~Weldon}}(1983)}]{WeldonPRD83}%
  \BibitemOpen
  \bibfield  {author} {\bibinfo {author} {\bibnamefont {{H.~A.~Weldon}}},\
  }\href {\doibase 10.1103/PhysRevD.28.2007} {\bibfield  {journal} {\bibinfo
  {journal} {Phys.~Rev.~D}\ }\textbf {\bibinfo {volume} {28}},\ \bibinfo
  {pages} {2007} (\bibinfo {year} {1983})}\BibitemShut {NoStop}%
\bibitem [{\citenamefont {{E.~Braaten and
  M.~H.~Thoma}}(1991{\natexlab{a}})}]{HQQGPBraaten91QED}%
  \BibitemOpen
  \bibfield  {author} {\bibinfo {author} {\bibnamefont {{E.~Braaten and
  M.~H.~Thoma}}},\ }\href {\doibase 10.1103/PhysRevD.44.1298} {\bibfield
  {journal} {\bibinfo  {journal} {Phys.~Rev.~D}\ }\textbf {\bibinfo {volume}
  {44}},\ \bibinfo {pages} {1298} (\bibinfo {year}
  {1991}{\natexlab{a}})}\BibitemShut {NoStop}%
\bibitem [{\citenamefont {{E.~Braaten and
  M.~H.~Thoma}}(1991{\natexlab{b}})}]{HQQGPBraaten91QCD}%
  \BibitemOpen
  \bibfield  {author} {\bibinfo {author} {\bibnamefont {{E.~Braaten and
  M.~H.~Thoma}}},\ }\href {\doibase 10.1103/PhysRevD.44.R2625} {\bibfield
  {journal} {\bibinfo  {journal} {Phys.~Rev.~D}\ }\textbf {\bibinfo {volume}
  {44}},\ \bibinfo {pages} {R2625(R)} (\bibinfo {year}
  {1991}{\natexlab{b}})}\BibitemShut {NoStop}%
\bibitem [{\citenamefont {{O.~Kaczmarek and F.~Zantow}}(2005)}]{TwoLoopGPRD05}%
  \BibitemOpen
  \bibfield  {author} {\bibinfo {author} {\bibnamefont {{O.~Kaczmarek and
  F.~Zantow}}},\ }\href {\doibase 10.1103/PhysRevD.71.114510} {\bibfield
  {journal} {\bibinfo  {journal} {Phys.~Rev.~D}\ }\textbf {\bibinfo {volume}
  {71}},\ \bibinfo {pages} {114510} (\bibinfo {year} {2005})}\BibitemShut
  {NoStop}%
\bibitem [{\citenamefont {{J.~D.~Bjorken}}(1982)}]{BjorkenFerlab82}%
  \BibitemOpen
  \bibfield  {author} {\bibinfo {author} {\bibnamefont {{J.~D.~Bjorken}}},\
  }\href {https://lss.fnal.gov/archive/1982/pub/Pub-82-059-T.pdf} {\bibfield
  {journal} {\bibinfo  {journal} {Fermilab Report No. PUB-82/59-THY}\ }
  (\bibinfo {year} {1982})}\BibitemShut {NoStop}%
\bibitem [{\citenamefont {{A.~Beraduo, A.~De~Pace, W.~M.~Alberico, and
  A.~Molinari}}(2009)}]{POWLANG09}%
  \BibitemOpen
  \bibfield  {author} {\bibinfo {author} {\bibnamefont {{A.~Beraduo,
  A.~De~Pace, W.~M.~Alberico, and A.~Molinari}}},\ }\href {\doibase
  10.1016/j.nuclphysa.2009.09.002} {\bibfield  {journal} {\bibinfo  {journal}
  {Nucl.~Phys.~A}\ }\textbf {\bibinfo {volume} {831}},\ \bibinfo {pages} {59}
  (\bibinfo {year} {2009})}\BibitemShut {NoStop}%
\bibitem [{\citenamefont {{B.~L.~Combridge}}(1979)}]{Combridge79}%
  \BibitemOpen
  \bibfield  {author} {\bibinfo {author} {\bibnamefont {{B.~L.~Combridge}}},\
  }\href {\doibase 10.1016/0550-3213(79)90449-8} {\bibfield  {journal}
  {\bibinfo  {journal} {Nucl.~Phys.~B}\ }\textbf {\bibinfo {volume} {151}},\
  \bibinfo {pages} {429} (\bibinfo {year} {1979})}\BibitemShut {NoStop}%
\end{thebibliography}%
\end{document}